\documentclass[11pt]{article}
\usepackage[margin=1in]{geometry}
\usepackage[T1]{fontenc}
\usepackage[sc]{mathpazo}   
\usepackage{booktabs}
\usepackage{array}
\usepackage{amsmath}
\usepackage{url}
\usepackage[hidelinks]{hyperref}
\usepackage{microtype}
\usepackage{enumitem}
\setlist{nosep}
\newcommand{\src}[1]{}  
\newcommand{\arm}[1]{\texttt{#1}}

\title{SaltBench: A Referee-Gated Protocol for Measuring Method Effects in Machine-Checked Software Work}
\author{Jason Hickey}
\date{October 2026}

\begin{document}
\maketitle

\begin{abstract}
SaltBench asks how a machine referee changes how a coding agent works. Outcomes are decided outside the agent's toolchain; on the Claude lane a wall probed before any scored run isolates the agent from the network, reference solutions and harness, and 5 Gemini-lane cells served web search results are declared and leave every comparison; a dated freeze of predictions authorizes each run; a budget stop is a halt, never a failure. Five Rust systems components on a pinned Verus toolchain are each refereed by a withheld test suite. The arms: \arm{plain}; \arm{salt-diet}, also instructed to specify and verify its code (a registered reduced rendering of the method); and two arms handed the specification a priori, where the registered sign test at $k=4$ reached no verdict (3 of 4, $p = 0.3125$). \arm{salt-diet} cost more on all five components, by a practical margin: no premium exceeded $2.8879\times$ under either reading of the declared set and the three cheapest sat below $1.4\times$, a bound of this population and not a promise about larger ones: the premium is near $1$ on the smallest components and rises with size. Versions~2 and~3 add the complete pilot matrix: 200 conditions over four models, 181 with a result of record, 16 inexpressible and 3 declared unreached at the cost cap, with no verdict on the arms. Version~3 adds a correctness reading, declared post hoc and registered after every verdict it reads existed and many had been read by its author: 447 cells pass the withheld suite, 40 fail, 10 are censored at a registered budget, 4 are unscorable and 42 are read by condition. \arm{salt-diet}'s pass-rate interval lies wholly below \arm{plain}'s on 8 greenfield and 5 brownfield rows and wholly above it on 2 greenfield rows. It is descriptive, over three cells of record per condition, with no test, and says nothing about whether the method makes code more or less correct. The full record is public.
\end{abstract}\src{harness/systems-v3/RESULT-correctness-tables-v3-2026-10-01.md, the header, the P4 counts and the brownfield signs; harness/systems-v3/REGISTRATION-correctness-tables-v3-2026-10-01.md, section P6}

\section{Introduction}

An agent's account of its own work is the least reliable evidence about that work. A benchmark that lets the agent grade itself, or lets the experimenter choose the reading after the run, measures the story and not the outcome. SaltBench is built on two refusals. The outcome is decided by a machine referee the agent cannot argue with --- a proof kernel, a program verifier, or a withheld test suite, run outside the agent's own toolchain --- and the comparison is fixed before the agent runs, so that it cannot be narrated afterwards. The comparison is pre-registered~\cite{nosek2018preregistration}: the population is drawn by a committed seed, the arms are byte-pinned, the predictions are written down, the adverse outcomes are named, and the freeze commit is the authorization. Everything that happens afterwards is a dated amendment appended to the record.

Here we report the benchmark's design and its first measurement, taken end to end on an experiment whose subject is the working seat rather than the model. Section~\ref{sec:design} states the design and how much of it this paper runs. We claim the protocol; the seat-as-subject design and its authored population; the cost reading the run returned, together with the qualifiers registered beside it; a set of instrument findings that we think transfer to any benchmark of this kind; and the complete pilot matrix of Section~\ref{sec:matrix}, new in this version, as a record of what was run and of what each result of record can carry, never as a verdict on the arms. We do not claim an effect of the method under test on what a referee accepts, and we report no magnitude for the population; Section~\ref{sec:open} states the tests that could produce one. Three earlier reads on drawn populations are reported in Section~\ref{sec:priorreads}, compressed to the role they now play, which is the reason this design is shaped as it is. One note on labels, because two of them are numbers: this document is version~5 of the SaltBench report, and \emph{v3} names the campaign generation whose matrix it reports, not an edition of this document. Version~2 adds Section~\ref{sec:matrix}, the complete pilot matrix over four models, updates the status of the task forms in Section~\ref{sec:design}, and adds two instrument findings the matrix taught; it changes no number that version~1 printed. Nineteen prices in Appendix~\ref{app:cells} carry a flag that version~1 did not print, and the flag changes no number. One sentence of Section~\ref{sec:fence} is scoped to the population its evidence covers, and the scope is marked there. Version~3 adds Section~\ref{sec:matrixdollars}, the same matrix in modelled list-price dollars and in wall time, and Appendix~\ref{app:opussplit}, the share of each Opus cell spent by its subagents, Section~\ref{sec:matrixcorrect}, the same matrix read for correctness under a reading declared post hoc, and a paragraph on related work; it changes no number that version~2 printed. Version~4 corrects Tables~T1 to~T4: the Gemini lane's token meter had read each cell's parent conversation and not the sub-agent conversations it starts, which undercounted the \arm{salt-diet} arm far more than the \arm{plain} arm, so the published Gemini ratios understated the \arm{salt-diet} arm's cost. Section~\ref{sec:matrixtokens} states what moved; no ratio whose sign was fixed changes sign, no Claude row changes, and no other table is re-computed. It also corrects three Sonnet median cost ratios in the prose of Section~\ref{sec:matrixrecords}, which versions~2 and~3 printed as $8.27\times$, $5.32\times$ and $4.71\times$: each was a ratio of two rounded medians, and from the unrounded tables they are $8.29\times$, $5.31\times$ and $4.70\times$. No sign, order or bound changes.\src{harness/systems-v3/CENSUS-full-matrix-2026-09-14.md, addendum 38} Version~4 also adds two limits to Section~\ref{sec:matrixrecords}, the cells' failed \texttt{git} and \texttt{python3} invocations and a fence that did not cover a cell root created during a run, and an erratum beside Table~P1 on one unscorable Flash cell and three Pro top-up cells; no cell is re-scored.\src{evidence/nr-command-hazard-2026-09-22/README.md; harness/systems-v3/CENSUS-full-matrix-2026-09-14.md, addendum 31 section AG1 and addendum 37; harness/systems-v3/RESULT-stage3-2026-09-12.md, addendum B} Version~5 adds three dated sections of their own: Section~\ref{sec:zf}, a second model run on the same matrix beside a same-day control; Section~\ref{sec:o90truth}, a test of whether an explanation arm's prose is true of its own Lean; and Section~\ref{sec:o90understand}, whether a reader understands the two arms' Lean; it changes no number that version~4 printed. Version~5 also corrects one sentence of the abstract, whose network clause was true of the Claude lane only: on the Gemini lane the client served web search results to five cells of record, three of them in the matrix. Section~\ref{sec:matrixrecords} declares them, keeps their verdicts, takes them out of every comparison and prints what that moves beside the tables, which stay the tables of record, together with two temp-directory limits that move no count.\src{harness/systems-v3/CENSUS-full-matrix-2026-09-14.md, addenda 41, 42 and 47; evidence/o37-agy-web-served-2026-11-20/recut/README.md}

The method under test, the Salt method~\cite{hickey2026saltmethod,hickey2026authority}, is the working discipline the author uses in a formal mathematics project: machine-checked development in which claims travel as kernel-checked artifacts and human attention is reserved for statements and rulings. Its rendering as an arm is described in Section~\ref{sec:arms}; only the part of it that a single sealed agent can carry is rendered, and that limit was registered before the first call.

\paragraph{Related work.} Kaashoek and Zeldovich report a machine-checked proof of the xv6 kernel, 6{,}593 lines of C and assembly, on a Rocq rendering of the Sail RISC-V semantics, built by coding agents under the direction of its two authors~\cite{kaashoek2026xv6}. The two studies ask different questions of agents and a referee. Theirs is one development, directed by its authors throughout, and it reports what the agents cost and what the authors learned about directing them (its Sections~9 and~12). Ours fixes the agent's instructions per arm and asks what a specify-and-verify method costs against a plain agent on the same task, with the outcome decided by a referee outside the agent's own toolchain. Both price agent work from the harness's own transcripts, and theirs records that roughly 9\% of its sessions are missing because the harness deletes transcripts older than 30 days by default (its Section~12.1). Section~\ref{sec:matrixdollars} sets their effort figures beside ours for scale.

\section{The seat-as-subject experiment}
\label{sec:populations}

The experiment this paper is built around takes the working seat (meaning a coding agent, its harness, and a workspace), not the model, as its subject. A capable agent is handed a systems component to build or to change, under a method it is told to follow, and the question put to it is not whether it can do the work but what the method costs it to do the work. That question forces two things the earlier populations could not supply. The task must be one a referee can decide without consulting the agent's own account of what it did, and the population must be authored rather than drawn, because a drawn population carries a chance, unquantifiable, that the model has already seen the answer.

We present a benchmark design, and an initial measurement over five authored components, run as arms that differ only in the method text the agent is given. Each component is specified by a \emph{card}: the written task the agent is given, carrying its requirements and, where one exists, the formal statement of the task. What the matrix produces is a price per cell, and prices are what this paper reports from it. The cells were scored for cost; the correctness pass taken afterwards, and what it could and could not separate, is reported in Section~\ref{sec:s3}.

\subsection{The design, and how much of it this paper runs}
\label{sec:design}

The benchmark design has four arms, two-by-two. One axis is the method: a plain agent, against an agent instructed to write a specification and verify its code against it. The other is where the specification comes from: written by the agent, or handed to it a priori as the formal statement of the task. The a priori pair isolates the cost of verifying against a specification from the cost of writing one, and it carries the design's registered gold reading, fixed before any cell ran: if the treatment beats the control when both are handed the statement, the method is the statement.\src{harness/systems-v3/PREREGISTRATION-matrix-opus-1-2026-09-08.md sections 1 and 3} A placebo, an equal-length process prompt with no method content, sits beside the four as a control on prompt length and is not one of them.\src{harness/systems-v3/PREREGISTRATION-matrix-opus-1-2026-09-08.md section 1}

The substrate is chosen so that the four arms are comparable. Every arm writes Rust under the same pinned Verus toolchain and is scored by the same withheld suite. Verus is verified Rust, so an arm that writes specifications and proofs and an arm that does not are writing the same language for the same compiler, and the difference between them is the method, not the language. A fifth arm in which both the code and its specification are written in Lean is a possible extension. It would be comparable to the arms that verify in Rust and Verus, and not to the plain Rust arm, because the language would change together with the method; it is not part of this paper.

Its task forms are four, and they stand at different stages. Greenfield, the agent building a component from its requirements, is the form the cost reading of Section~\ref{sec:s3} measures, and the form the pilot matrix of Section~\ref{sec:matrix} runs on all four models. Brownfield, the agent repairing a component that does not work, was authored and designed when version~1 of this report was written and has since run on all four models under two treatments (Section~\ref{sec:matrix}): every card in the population carries a brownfield rung naming a planted defect, all five are realised as complete trees beside their greenfield twins,\src{harness/systems-v3/CENSUS-full-matrix-2026-09-14.md, addendum 4} and the form's design fixes what a run must record, two verdicts that are never one number and three outcome classes of which only a repaired given file counts as brownfield at all, because an agent that discards the given code and writes from scratch has done greenfield with a longer prompt.\src{harness/systems-v3/DESIGN-v3-brownfield-form-2026-09-09.md sections 0, 1 and 3} That form is also the one that can attach a correctness verdict to a price, because a repair is a delta against a baseline that is measurable before the agent starts.\src{harness/systems-v3/DESIGN-v3-brownfield-form-2026-09-09.md section 1} Specification change, where the statement moves under the agent after it has landed a first build, has run as the third treatment of the matrix on greenfield cards, as a two-phase cell whose second phase exists only on a landed first phase; on brownfield cards the harness cannot express it, and the matrix's denominator leaves those forty conditions out by the author's ruling (Section~\ref{sec:matrix}).\src{harness/systems-v3/CENSUS-full-matrix-2026-09-14.md, section C3 and addendum 1; harness/systems-v3/RESULT-claude-blockSC-2026-09-21.md section 1} Brownfield inside a larger system, where the component under repair is one part of a codebase the agent must not break, is planned and has no registered object yet. It is named here so that the paper says what the benchmark is for and not only what it has done. The problem set has the same shape: five components in this paper, with an expansion to fourteen authored on the same form and ordered to follow the statement-arm run.\src{harness/systems-v3/AMENDMENT-statement-arm-pilot-2026-09-09.md section 1}

\begin{center}
{\small
\begin{tabular}{lllll}
\toprule
task form & \arm{plain-bare} & \arm{diet-bare} & \arm{plain-statement} & \arm{diet-statement} \\
\midrule
greenfield & run & run & run, 4 of 5 & run, 4 of 5 \\
greenfield, specification change & run & run & not registered & not registered \\
brownfield, planted defect & run & run & run, 4 of 5 & run, 4 of 5 \\
brownfield in a larger system & planned & planned & planned & planned \\
\bottomrule
\end{tabular}}
\src{harness/systems-v3/CENSUS-full-matrix-2026-09-14.md, sections C3 and C4 and addenda 1 and 24; harness/systems-v3/PREREGISTRATION-matrix-opus-1-2026-09-08.md section 1; harness/systems-v3/DESIGN-v3-brownfield-form-2026-09-09.md section 1}
\end{center}

\noindent \emph{Run} means cells fired on the pilot matrix's four models and accounted for in Section~\ref{sec:matrix}, which says per row how many conditions are \textsc{done}, inexpressible or declared unreached at the cost cap; \emph{4 of 5} means that on \texttt{Paxos} the statement arms are inexpressible, because an arm-neutral formal statement cannot exist for a proof-obligation task; \emph{not registered} means the matrix's treatments are none, statement and specification change, one at a time, so a statement arm under specification change was never registered; \emph{planned} means intent, with no object yet. The arm names are the registration's, and the word \emph{diet} on the treatment records that the matrix measures a reduced rendering of the method.

\paragraph{S3-Systems: five components authored for this benchmark.} This population is not drawn from a public set. Five classic systems components were authored for the benchmark in Rust and frozen in a dated export before any model call: \texttt{Crc32}, \texttt{FreeList}, \texttt{LRU}, \texttt{LZW} and \texttt{Paxos}. Each is frozen as four files of the same shape: a task card, a greenfield reference (\texttt{G}), a base plus a change request for a maintenance card (\texttt{B}), and the referee's name list. The frozen material runs to 9{,}141, 12{,}917, 14{,}643, 20{,}861 and 16{,}037 bytes across the five, and none is a stub. Each of the ten cards carries a withheld test suite, and a withheld mutant set was authored against it.\src{harness/systems-v3/PRICE-campaign-matrix-opus-1-2026-09-08.md section 1} An authored population has no third-party licence to reconcile and no contamination proxy to compute against a public gold patch; what it gives up is independent authorship, and Section~\ref{sec:s3} states that as a limitation rather than leaving it implicit.

The withheld suites were measured against the withheld mutants before any cell ran, by driving the same runner the referee drives rather than a second invocation of the test tool: over the ten cards, 44 of 44 mutants killed, none surviving and none unmeasured, and every reference passing its own suite.\src{harness/systems-v3/RESULT-hidden-test-strength-v3.md section 1} That number is a ceiling, not a strength: the mutants were authored beside the tests, and the figure is a property of the suite, not a verdict on any cell. A later sweep of every withheld arm against every withheld mutant, over the five greenfield cards that carry 22 of those 44 mutants, found no survivor and 13 of 54 arms that fail on no mutant at all, which is a statement about the mutant set before it is one about the arms.\src{harness/systems-v3/RESULT-arm-coverage-2026-09-10.md sections 1 and 6} When the matrix ran, only \texttt{LZW}'s card carried a \texttt{\#\# Statement} section; the other four had been authored without one, so 24 of 24 statement-arm cell builds refused at the harness, which was the harness working and a content gap in the population.\src{harness/systems-v3/PREREGISTRATION-matrix-opus-1-2026-09-08.md section 16} The four sections were derived afterwards, under a dated amendment that fixed the rule before any card was edited: extracted from each withheld reference solution by the same tool that produces \texttt{LZW}'s, verbatim and closed over itself and the interface, with \texttt{LZW} reproducing byte-identical from that tool as the control. A section that does not reproduce from the tool is not admissible, and a named proof function contributes its contract with its body elided, on every task alike. Across the four sections the tool had produced when the amendment froze, the size differs twelvefold, from 1,166 bytes on \texttt{Crc32} to 13,849 on \texttt{FreeList}, so the treatment they carry is not the same size on every problem.\src{harness/systems-v3/AMENDMENT-statement-arm-pilot-2026-09-09.md, addendum 1}

\section{The protocol, as exercised on v3}
\label{sec:protocol}

\subsection{The referee decides, outside the agent's elaboration}

An episode ends when the agent stops or hits a cap, and what it leaves behind is scored by a checker the agent never sees, on a machine the agent never reaches. What ``solved'' means is fixed per substrate, and it is fixed by the referee, not by the agent.

\paragraph{Rust (S3-Systems), the benchmark's referee.} The referee for the authored population is the withheld test suite of the card, run by the harness's own runner outside the agent's workspace, with the suite and its mutant set never shipped to the agent. That is the design. The run this paper reports on this population (Section~\ref{sec:s3}) scored its cells for cost with the referee not invoked, and a declared post-hoc pass invoked it afterwards on the surviving repositories, under classes fixed before it ran (Section~\ref{sec:posthoc}).

Three other referees appear in this paper. They scored the earlier reads of Section~\ref{sec:priorreads}, the reads that shaped this design, and they are stated here because the protocol around them is the same one. The Lean referee is also the one a Lean arm of the benchmark would use.

\paragraph{Lean (CLEVER~\cite{thakur2025clever}).} The checker extracts only the agent-writable bodies by section markers, screens them for command-introducing and meta keywords with comments and string literals stripped, assembles a canonical file from the frozen statements and the bodies, and compiles it under an operating-system sandbox. It then replays every declaration of the compiled module through the Lean kernel~\cite{demoura2021lean4} (\texttt{Lean.Environment.replay}), so a declaration the elaborator admitted without the kernel is rejected. It compares, as kernel expressions, the type of every frozen declaration and the value of the human specification between the canonical file and a pristine file with \texttt{sorry} bodies, so a notation, macro, instance or \texttt{open} that changes what the frozen text means is caught by construction. It collects the axioms of the audited declarations and requires every set to be inside $\{\texttt{propext}, \texttt{Classical.choice}, \texttt{Quot.sound}\}$; \texttt{sorryAx}, user axioms, and \texttt{native\_decide}'s \texttt{Lean.ofReduceBool} all fail here. The class names the first failing gate: \texttt{SCREEN}, \texttt{COMPILE}, \texttt{KERNEL\_REJECTED}, \texttt{STATEMENT\_ALTERED}, \texttt{AXIOMS\_FAIL}, \texttt{PASS}.\src{SCOUT-S2LEAN-STAGE0.md section 3}

\paragraph{Verus~\cite{lattuada2023verus} (VeruSAGE-Bench~\cite{yang2025verusage}).} Verus emits no proof object and there is no independent re-check, so the protocol says ``the referee accepted'' where the Lean protocol says ``kernel''. The integrity work that the kernel replay and axiom audit do in Lean is done here by three text and AST layers before the referee runs: a screen over the agent's regions; a count guard over \texttt{admit()}, \texttt{assume(}, \texttt{external\_body} in both spellings, \texttt{assume\_specification} and \texttt{verifier::admit}, canonical against original; and the benchmark's own \texttt{lynette additions} AST comparison. A screened or cheating body is never handed to the referee. \texttt{PASS} requires exit~0 and a results line that parses as \texttt{N verified, 0 errors} with $N \geq 1$ and no partial-verification suffix. Resource-limit exhaustion and compile errors are their own classes and are never charged to the arm.\src{AMENDMENT-15-s2rust-2026-09-01.md section 6}

\paragraph{SWE-bench~\cite{jimenez2024swebench} Verified~\cite{swebenchverified}.} The scorer is the unmodified \texttt{swebench} harness in the pinned evaluation images, with the fail-to-pass and pass-to-pass suites hidden from the agent. This substrate was abandoned after stage~0 (Section~\ref{sec:s1}); its protocol is kept in the record.

\subsection{The fence is measured, not assumed}
\label{sec:fence}

The agent's subprocesses run under the operating system's sandbox (macOS Seatbelt through the agent harness's own sandbox settings): every host denied, writes only under the working copy and the session temp directory, reads denied on the harness state, the credential trees, and the run root. The checker runs under its own \texttt{sandbox-exec} profile. A command-text hook on the shell tool is kept as an attempt audit, not as the fence, and an audit layer records every tool read by path. Five smoke probes run before the driver starts: a canary, network escapes (including an in-language process spawn through the toolchain wrapper), reads of the arm files and the harness config, writes through the symlinked build directory, and the compile path. The driver refuses to start unless the last verdict of every probe is a pass carrying the freeze's episode-script hash.\src{SCOUT-S2LEAN-STAGE0.md sections 4 and 7}

Ground truth never reaches the host while an arm writes a specification blind. The stage-A views carry no human specification; the stage-B and stage-C views are shipped only after every stage-A episode has landed, and the ship is verified by a content set-hash rather than by the exit code of the tool that moved it.\src{SCOUT-S2LEAN-STAGE0.md section 1; amendment 8 addendum 3}

Two findings in the record show why the fence is measured. The Verus referee's docstring claimed a sandbox, a process group and a belt; a grep of the file found only the comment, and the referee had been running unfenced. The exec set turned out to be four binaries, established by a ladder in which each arm's own error names the next binary, and a fifth exec point was revealed only by the red arm's panic text. A proof-only fixture had passed the four-binary fence, which would have been the receipt.\src{AMENDMENT-16-regime-boundary-2026-09-02.md section 2}

The third finding arrived after every read reported in this paper and applies to all of them. The sandbox's read denials bind sandboxed subprocesses: the shell tool and whatever it executes. The agent harness's own file-reading tool is executed by the harness process and never enters that sandbox. Neither substrate's settings carried a tool-permission rule, so no scored episode of this campaign had the agent's tools fenced by path. The smoke probe that had certified credential reads denied made its reads through the shell and through an in-language process spawn, both sandboxed; a probe written in the sandbox's language cannot see a hole in the layer above it, and it reads afterwards as though the agent itself were fenced. What was in force for the v1 and v2 episodes, each with its own evidence, was the subprocess sandbox, the empty network allowlist, the shell-tool hook, the ground-truth leak check, and the audit layer. For the v3 cells run under the Claude harness the fence is rendered and drift-checked at use and the hook layer is driven, and the operating-system sandbox layer is unmeasured for the cells this paper reports: an audit of the published claims found 124 such cells launched and none that drove that layer, and the probe that drives it was built after those cells landed and cannot be applied to a cell retroactively.\src{harness/systems-v3/AUDIT-published-fence-claims-2026-09-13.md; harness/systems-v3/RECEIPT-claude-sandbox-probe-2026-09-14.md, sections R4 and R5; paper/REVISION-NOTE-inventory-clause-2026-09-16.md} That audit is what establishes the S2-Rust read: across its 15 Opus episodes no episode was voided and no unblocked read of a fenced path was recorded (one attempt through the shell was blocked). One episode of the lower-tier probe reported in Section~\ref{sec:open} directed its file tool at a path under its own configuration directory, was not refused, and was voided as unscorable; read at the transcript the call returned that the file did not exist, so no bytes came back, and the audit field had recorded that the call was not blocked rather than that it was served. Those are different facts, and only one of them is a fence event. Over every landed S2-Lean episode in the campaign's four state roots, 278 with a transcript and 278 parsed, the count of file-tool calls at a fenced path is zero served and zero not served: no agent ever directed a file tool at a path under a deny root. The detector was driven on all three branches (a served read of a fenced path counted, a refused read classified and not counted, a read of an unfenced path ignored), because a quiet failure reads as good news. What this measures is what the agents did, not what they could have done: the canary shows the gap was open, and an absence of exploitation is not a presence of protection.\src{AMENDMENT-17-fence-config-dir-2026-09-02.md section 11; harness/s2rust-analysis/s2lean\_escape\_audit.py} The repaired fence derives a tool-permission deny list from the same path list as the sandbox denials, 132 rules beside 11 paths, and was driven with a canary in both arms: with the tool layer absent the canary reached the agent's final message, and with it present the agent reported the file unreadable and the canary appears in neither the result nor the transcript.\src{AMENDMENT-17-fence-config-dir-2026-09-02.md sections 2 and 9}

\subsection{Pre-registration, and the amendment discipline}

The dated freeze commit is the authorization. Before it: the population rule and seed, the arms as byte-pinned files, the checker, the constants, the reading rule, the predictions, the adverse outcomes, the stop rules with their enforcer. After it: nothing above the line is edited. Every later change, including a repair to an instrument, is a dated amendment appended before its own first model call, with its own predictions. Corrections to a registered document are appended as corrections, never edited away; the record carries several such corrections by its own authors.\src{PRE-REGISTRATION.md section 0; SCOUT-S2LEAN-STAGE0.md, amendments 1 through 10}

The reading rule on the Lean and Verus populations is a count, not a $p$-value: for two arms on the same drawn problems, $b$ is the number the first arm proves and the second does not, $c$ the reverse, and $|b-c| < 5$ is read as \textsc{indistinguishable}. The threshold is registered before the first call and the outcome table is enumerated, with the adverse cell named.\src{SCOUT-S2LEAN-STAGE0.md, amendment 5} For the Verus wave the fixed separation is replaced by one derived from a measured null (a same-arm rerun), because the substrate's own paper~\cite{yang2025verusage} reports that 11.8\,\% of failures flip on rerun.\src{AMENDMENT-15-s2rust-2026-09-01.md section 9}

Predictions are scored as they stand. Of the registered predictions in the record, several failed, and the failures are recorded as failures with their direction: the estimator over-priced five consecutive stages, then under-priced the sixth because one episode carried 64\,\% of a stage's spend.\src{RESULT-amend11-stageC-2026-09-01.md section 4; RESULT-amend13-AW-2026-09-02.md section 5} A further prediction of this kind was registered before the cells it governs had finished running, and is recorded here in advance of its own outcome: two conditions of the statement amendment were expected to reach the per-cell cost cap, because their bare counterparts' medians already sat above it. It is scored as it stands in either direction. If they cap, the cells are reported as cap-cost and never as failures; if they do not, the prediction failed and is recorded as a failure with its direction, and the comparison that failure invites, between a bare condition and its statement counterpart, is not the registered reading of that arm, which lies between the two statement conditions on the same problem.\src{harness/systems-v3/AMENDMENT-statement-arm-pilot-2026-09-09.md sections 3 and A4}

\subsection{A budget stop is a halt, never a failure}

A per-episode token stop at four times the governing regime's p90 is registered as a halt: a halted episode is recorded as halted and is unresolved for the rate, because scoring an episode the budget stopped as a failure would let the budget instrument move the result. The multiple was chosen from the record: no passing episode among 177 had exceeded $2.40\times$ its regime's p90, and the one runaway was at $7.79\times$ and failed.\src{AMENDMENT-14-instrument-2026-09-01.md section 8}

The rule had to be enforced where the verdict is written, not only where it is registered. The first halted Verus episode was scored \texttt{SCREEN} on a snapshot of an interrupted edit (two \texttt{assume} calls the agent had not yet discharged). The checker now takes the episode's termination and writes \texttt{class = HALT} for a token or wall stop, preserving the class it would otherwise have written as a diagnostic.\src{harness/s2rust/check\_verus.py, HALT\_TERMINATIONS}

\subsection{Ask of every gate which arm is more likely to trip it}
\label{sec:armcorr}

The treatment arm encourages helper lemmas. Three gates in the Verus grader were found, in sequence, to refuse exactly that. The AST comparison refused a helper placed before the enclosing \texttt{impl}, classifying the episode as \texttt{STATEMENT\_ALTERED}, a cheating class, for doing what the prompt invites; 47 of 207 views are impl-enclosed.\src{AMENDMENT-15-addendum-1-2026-09-01.md section A1} The screen's rule set enforced 29 refusals of which the prompts named 12, and the unstated \texttt{use} rule bites in the helpers region; the repair makes the prompt a consumer of the rule set, with a gate that refuses a rule with no prompt entry.\src{AMENDMENT-16-regime-boundary-2026-09-02.md section 4} The helpers whitelist then split items by brace depth and refused a legitimate lemma whose \texttt{ensures} clause carried struct literals; it was the third episode ever run that found it, after 29 screen arms, 18 fence arms and an 11-arm fixture kit had passed.\src{RESULT-amend16-driver-2026-09-02.md section 4c} An instrument that penalizes the treatment for applying the treatment does not measure a small effect badly; it manufactures the opposite one. The discipline is now a standing question at every gate.

\subsection{The arms}
\label{sec:arms}

Every arm receives the same task file and the same stage prompt; only the agent's instruction file differs, and it is byte-pinned. \arm{a0} is the plain arm: the base block alone. \arm{a1} is the placebo: the base block plus an equal-length process prompt with no method content (1,746 bytes against the treatment's 1,913; the base is 627).\src{AMENDMENT-13-AW-2026-09-01.md section 1} \arm{a2} is the treatment: the base block plus the method's solo-renderable core~\cite{hickey2026saltmethod}. The rendering, article by article against its source, and the articles that are not rendered because they presuppose an organisation with a human in it, were registered before the first call; no result from \arm{a2} may be read as a test of the method's multi-agent tier.\src{SCOUT-S2LEAN-STAGE0.md, amendment 5} The treatment text names \texttt{sorry}, which is the control's dominant failure; the objection that the arm coaches to the metric was registered in advance and is answered by the data in Section~\ref{sec:s2lean}.

\section{The v3 reading: a registered cost premium on all five problems, and a post-hoc correctness pass that did not separate the arms}
\label{sec:s3}

This is the only reading in the paper that returns a difference between the arms, and the difference is in price, not in what a referee accepted. The registered reading is a cost result. No cell of the run carried a correctness verdict at the time it was scored, and the paragraph on correctness below states and measures that fact; Section~\ref{sec:posthoc} reports the declared post-hoc pass taken afterwards. That is a statement about what was run and not about what exists: the withheld suites were authored with the population and characterised before the first cell (Section~\ref{sec:populations}), and the cells were scored for cost with the referee never invoked on them.

\paragraph{The design, registered before the first cell.} Five problems, four arms, $n=3$ per condition, at \texttt{claude-opus-5} on the greenfield card. The arms are \arm{plain-bare}, \arm{diet-bare}, \arm{plain-statement} and \arm{diet-statement}. The treatment arm is a \emph{dieted} rendering of the method~\cite{hickey2026saltmethod}: prescribed acts were removed after the full rendering hit the per-cell cost cap on a single earlier cell at \$42.24, and the registration requires the word \emph{diet} on every headline, because this matrix does not measure the un-dieted method.\src{harness/systems-v3/PREREGISTRATION-matrix-opus-1-2026-09-08.md section 1; harness/systems-v3/PREDICTIONS-pricing-set-2026-09-06.md sections 1 and 2} The primary reading is a cross-problem sign test on $\mathrm{premium}(P) = \mathrm{median}(\text{diet-bare}, P) / \mathrm{median}(\text{plain-bare}, P)$, one premium per problem, against the null that each premium is equally likely either side of $1.0$. The outcome table was enumerated before the first cell: 5 of 5 gives $p = 0.0312$, 4 of 5 gives $p = 0.1875$ and is registered in advance as \emph{not} a positive result, 3 of 5 gives $p = 0.5000$. None of those numbers may be quoted on its own: the four qualifiers stated with the reading below were registered with the test and are part of it. Five problems is the smallest count at which a clean sweep can reach $.05$ at all, which is why the earlier two-problem stage could not have produced a verdict whatever its cells had said.\src{harness/systems-v3/PREREGISTRATION-matrix-opus-1-2026-09-08.md section 2; harness/systems-v3/PRICE-campaign-matrix-opus-1-2026-09-08.md section 4}

The client was pinned by absolute versioned path to the build the earlier stage had used, rather than resolved from the shell path, and the registration names the cost of that choice: the matrix measures a client two releases behind the one a path lookup would have fired. A symbolic link follows the newest install, so resolving a client from the path silently re-pins a scored run every time the vendor ships.\src{harness/systems-v3/PREREGISTRATION-matrix-opus-1-2026-09-08.md section 11}

\paragraph{What was built, against what was registered.} The registration and the campaign price cover 60 cells; 36 of them were buildable. (\emph{Priced} is used in its own sense in the census below, where it means a cell that produced a cost, and the two senses are kept apart deliberately.) When the matrix ran, the 24 statement-arm cells on the four problems whose cards carried no \texttt{\#\# Statement} section refused at the harness (Section~\ref{sec:populations}), which left the bare pair complete on all five problems and the statement pair complete on \texttt{LZW} alone. The sections were fixed under a gate registered before any card was edited, and the 18 cells that fired then ran under the pilot's own cap. The gold pair is read below. The statement arm is handed a formal specification derived verbatim from the withheld reference on every problem: that is the treatment, it is intended, it is not a leak of the fence, and the comparison between the two statement arms is made within it, where both receive it.\src{harness/systems-v3/PREREGISTRATION-matrix-opus-1-2026-09-08.md section 16} The condition key is \texttt{(task, arm, card\_extras)} read from each cell's own control directory, and a scorer keying on the arm alone would have pooled the bare and statement arms of the same treatment, doubled each apparent $n$, and mixed the very pair the design calls its gold, with no error raised and medians that look reasonable. The scorer that had scored every earlier stage correctly keys on two fields, because until this matrix a third was never needed.\src{harness/systems-v3/PREREGISTRATION-matrix-opus-1-2026-09-08.md sections 15 and 17}

\paragraph{The scoreboard, at two readings.} The declared set is scored twice and the instrument computes both readings itself, so the comparison below is the tool's output and not a patch applied to it. Reading~A is the continuity reading: the matrix root, the three cells AMENDMENT~26 added, and the three borrowed smoke cells, 43 cells in all. Reading~B keeps only what this run produced, dropping the three borrowed cells, 40 cells. Both exist because of a dependency the first scoreboard did not show, which is stated below the table.\src{harness/systems-v3/RESULT-n3-topup-2026-09-09.md sections 2 and 3}

\begin{center}
\begin{tabular}{lrr}
\toprule
problem & premium, reading A & premium, reading B \\
\midrule
Crc32     & $1.1610\times$ & $1.1610\times$ \\
FreeList  & $2.8070\times$ & $2.8879\times$ \\
LZW       & $1.3749\times$ & $1.3749\times$ \\
Paxos     & $2.4306\times$ & $2.2437\times$ \\
LRU       & $1.2826\times$ & $1.1521\times$ \\
\midrule
sign test & 5 of 5, $p = 0.0312$ & 5 of 5, $p = 0.0312$ \\
\bottomrule
\end{tabular}
\src{harness/systems-v3/RESULT-n3-topup-2026-09-09.md section 3, the instrument's stdout for both readings}
\end{center}

\noindent Cost is US dollars of metered subscription spend per cell, read from each cell's harvested \texttt{METER.txt}. Every premium is a ratio of medians at $n=3$. An earlier version of this table carried the two medians behind each ratio; they are not printed here, because the instrument that computes these two readings prints the per-cell prices and the premium and does not print the median, and a median recovered by hand from a printed cell list is a number this paper cannot cite to a file. The per-cell prices for every condition under both readings are in the result file.\src{harness/systems-v3/RESULT-n3-topup-2026-09-09.md section 3} Three of the five premiums sit below the design's resolvable floor of $2.0072\times$ under both readings. The columns are printed so that the reader can reconstruct the sign, and by the registration no entry in either may be read as the finding.

\paragraph{The reading, and the four qualifiers that are part of it.} Five of five premiums exceed 1 under both readings. On the registered one-sided sign test that is $p = 0.0312$ at each. The four qualifiers below were registered before the first cell fired, and the campaign's own rule is that they travel with the $p$-value wherever it is quoted; a reader who has the number without them has a different claim from the one this design supports.\src{harness/systems-v3/RESULT-matrix-opus-1-2026-09-08.md, the scored result; harness/systems-v3/PREREGISTRATION-matrix-opus-1-2026-09-08.md sections 2 and 4}

\begin{enumerate}[leftmargin=*]
\item \textbf{The test is one-sided.} It can confirm the hypothesis it tests and cannot significantly refute it: five premiums \emph{below} 1 would have returned $p = 1.0000$. A one-sided test is a statement about which surprise the design was willing to be surprised by, and this one measured the expected surprise.
\item \textbf{No magnitude is resolvable for the population.} At $n=3$ with the registered pooled $\mathrm{sd}(\ln \mathrm{cost})$ of $0.30458$ the smallest resolvable premium is $2.0072\times$. Three of the five fall below it under both readings: Crc32 at $1.1610$, LZW at $1.3749$, and LRU at $1.2826$ under reading~A and $1.1521$ under reading~B. Only FreeList and Paxos clear it, and two resolvable magnitudes out of five are not a magnitude for the population. No ratio may be reported as the finding: not $2.7\times$, not a range, not a percentage. The finding is a sign across five problems.
\item \textbf{The paired arm the design calls its gold reached no verdict inside the matrix, and is read separately.} Inside the matrix it existed on \texttt{LZW} only, which is $k=1$ and $p = 0.5000$ at any outcome, with plain-with-statement at \$10.24, \$11.39 and \$9.82 against diet-with-statement at \$22.53, \$15.34 and \$18.54, a ratio of medians of $1.8105\times$. It reaches $k=4$ only through the amendment reported in the paragraph after this list, and that reading is registered by that amendment and not by the matrix's pre-registration.
\item \textbf{The within-condition noise is larger than the smallest premium.} \texttt{FreeList} on the plain arm spans \$13.01 to \$23.38 inside one condition on one problem, a spread of $1.80\times$, against the smallest premium in the table, which is $1.16\times$ on Crc32 under reading~A and $1.15\times$ on LRU under reading~B.\src{harness/systems-v3/RESULT-n3-topup-2026-09-09.md section 3, the per-cell prices} The registered spread clause says why the count of wide conditions is not itself a reading: at $n=3$ with this dispersion a perfectly homogeneous condition crosses a $1.5\times$ spread with probability $0.614$, so about twelve of the matrix's twenty conditions are expected to look wide by sampling alone.\src{harness/systems-v3/PREREGISTRATION-matrix-opus-1-2026-09-08.md section 5}
\end{enumerate}

\noindent Added in version~5: one \arm{salt-diet} cell of this matrix, \texttt{d6b53ee4} on \texttt{LRU}, prints its own gate failing to make a temp directory at its selftest (Section~\ref{sec:matrixrecords}); whether its cost moved is not read.\src{harness/systems-v3/RESULT-matrix-opus-1-2026-09-08.md, addendum of 2026-11-15}

\paragraph{The gold pair, under the statement amendment.} The reading was fixed before any card was edited: per-problem median cost of the treatment-with-statement arm over the plain-with-statement arm, then the cross-problem sign test at $k=4$, the same statistic and the same floor as the bare reading.\src{harness/systems-v3/AMENDMENT-statement-arm-pilot-2026-09-09.md section 5} Two of the four diet-with-statement conditions were registered in advance as expected to reach the cap, because their bare counterparts' medians already sat above it, and a capped cell is reported as \textsc{cap-cost} and never as a failure. One of those two conditions never fired, so the prediction is scored on the other alone. That condition did not reach the cap. The prediction failed, and it is recorded as a failure with its direction in the predictions paragraph of Section~\ref{sec:protocol}, where this record keeps its other failed predictions. The comparison a failure of this kind invites, between a bare condition and its statement counterpart, is not the registered reading of this arm and is not made here.\src{harness/systems-v3/AMENDMENT-statement-arm-pilot-2026-09-09.md sections 3 and A4}

\begin{center}
\begin{tabular}{lrrll}
\toprule
problem & $n$, plain & $n$, treatment & premium & \\
\midrule
Crc32 & 3 & 3 & $0.9384\times$ & below parity \\
FreeList & 3 & 3 & $1.5246\times$ &  \\
LRU & 3 & 3 & $1.5214\times$ &  \\
LZW & 3 & 3 & $1.8105\times$ & the pilot's own pair \\
\midrule
sign test & & & 3 of 4, $p = 0.3125$ & \\
\bottomrule
\end{tabular}
\src{harness/systems-v3/RESULT-statement-arm-2026-09-09.md}
\end{center}

\noindent The registered reading reached no verdict: 3 of 4 premiums above parity, $p = 0.3125$, on a test whose best available outcome at $k=4$ is $p = 0.0625$, and every premium below the resolvable floor of $2.0072\times$ at $n=3$, so no magnitude is resolvable and no ratio or range is reported as the result. The bare arms cleared that same floor on 2 of their 5 problems and this arm clears it on none of its 4, so on the instrument's own criterion the statement reading is the weaker of the two and not the friendlier one. At $k=4$ the registered test cannot reach the threshold at any outcome, and that is a property of how many problems the pair reached, not of what it found. The count is printed with its $p$-value because on its own it reads as support and it is not: at this $k$ a coin flip produces it. Set beside the bare arms, which were 5 of 5 above parity, the two arms do not tell the same story, and this is the first time in the campaign that they have not. Nothing here says the method is cheaper when handed a statement; one premium below parity at $n=3$ and below the floor is not a direction.  \texttt{Paxos}, on both statement arms, did not fire (an arm-neutral formal statement cannot exist for a proof-obligation task, so the neutrality gate refuses those cells; registered as a result rather than a shortfall), so 18 of the 24 registered cells landed and the pair is read at $k=4$ rather than the $k=5$ the amendment aimed at.\src{harness/systems-v3/RESULT-statement-arm-2026-09-09.md} Correctness on these 18 landed cells is scored by the same frozen pre-specification as the post-hoc pass and, unlike that pass, was registered for these cells before they ran; the two are not pooled into one column, and they are reported in Section~\ref{sec:posthoc} under their own heading.\src{harness/systems-v3/AMENDMENT-statement-arm-pilot-2026-09-09.md section 5}

\paragraph{What may be quoted from this reading.} The finding is the sign: the dieted treatment cost more than the plain arm on every one of the five problems, one-sided $p = 0.0312$, at $n=3$ per condition. The per-problem premiums in the table above are observations. They are printed so that a reader can reconstruct the sign and see the spread it was taken over, and the registration does not license any of them as a result. A span taken across the five, such as the interval from the smallest to the largest, is a range of three-cell medians and not an interval estimate for a population: this design registers no population magnitude, three of the five premiums fall below its own resolvable floor under both readings, and the within-condition spread on one arm of one problem is larger than the smallest premium. A summary of this result may therefore say that the treatment cost more on all five problems, and may quote any premium as an observation on its own problem at its own $n$. The same rule governs the gold reading above: its sign may be quoted with its $k$ and its $p$, and its premiums are observations at their own $n$. It may not state a ratio, a range or a percentage as the cost of the method, and it may not describe the result as a measured cost multiplier: not $2.7\times$, not one to three times, not a percentage, and not an interval of any width. Those are claims about a population this run did not measure, and an interval narrower than the within-arm spread reports a precision this instrument does not have. The three premiums below the floor are the ones such an interval would carry, and for those three the design cannot say that the true value is above $1.0$ at all. No cell of this matrix carried a referee verdict on a withheld suite when it was scored, so the cost reading bears on nothing about whether the method changes what a referee accepts; the post-hoc pass below did not separate the arms on that question, and it remains the question the next wave exists to answer. A reader should not take that absence for an absence of tests: the suites exist, and the sentence that is true of this run is that the referee was not pointed at its outputs when the cells were scored. Section~\ref{sec:posthoc} reports a declared post-hoc pass over the surviving repositories; a post-hoc suite outcome is still not a pre-registered capability result, and that pass did not separate the arms.
\subsection{A post-hoc correctness pass}
\label{sec:posthoc}

This pass is post hoc. Matrix~1 was registered as a cost experiment, its cost result was known before any suite was run, and the classes below and their meanings were fixed in a dated pre-specification before the first cell was scored. It is reported as a declared post-hoc measurement and not as a pre-registered result.\src{harness/systems-v3/PRESPEC-posthoc-correctness-matrix1-2026-09-09.md sections 1 and 3} The referee is the same runner the withheld suites were characterised with, it is arm-blind by construction, and the pass cost no model tokens: 43 cells in 246 seconds.\src{harness/systems-v3/RESULT-posthoc-correctness-2026-09-09.md, provenance}

\begin{center}
\begin{tabular}{lrlr}
\toprule
class & cells & class & cells \\
\midrule
\textsc{pass} & 35 & \textsc{cap-cost} & 3 \\
\textsc{fail} & 1 & \textsc{failed-boot} & 4 \\
\textsc{no-build} & 0 & \textsc{interface-miss} & 0 \\
\bottomrule
\end{tabular}
\src{harness/systems-v3/RESULT-posthoc-correctness-2026-09-09.md section 1; harness/systems-v3/RESULT-posthoc-correctness-verdicts-2026-09-09.tsv}
\end{center}

\noindent Every class is printed, including at zero, because a dropped cell is a claim that it did not exist. \textsc{cap-cost} and \textsc{failed-boot} are not failures of the method: a cell stopped by the budget, or one that produced no subject behaviour, was never asked the question, and neither is pooled into \textsc{fail}. The distinction is not decorative, because applying it changed the answer. The runner returns non-zero for a cell that never booted, so a first tally read five failures, and the frozen classes reduced that to one. Four cells that produced no subject behaviour would otherwise have been published as correctness failures.\src{harness/systems-v3/RESULT-posthoc-correctness-2026-09-09.md section 1}

\paragraph{The registered reading: the instrument did not separate the arms.} Landed cells on the bare arms, per problem, as cells passing over cells landed:

\begin{center}
\begin{tabular}{lcc}
\toprule
problem & plain & salt-diet \\
\midrule
Crc32 & 3/3 & 3/3 \\
FreeList & 4/4 & 0/1 \\
LRU & 4/4 & 3/3 \\
LZW & 3/3 & 3/3 \\
Paxos & 4/4 & 2/2 \\
\bottomrule
\end{tabular}
\src{harness/systems-v3/RESULT-posthoc-correctness-2026-09-09.md section 2; harness/systems-v3/RESULT-posthoc-correctness-verdicts-2026-09-09.tsv}
\end{center}

\noindent Four of the five problems tie, and one pair is informative. The cross-problem sign test cannot reach significance on this and does not try. That outcome was registered in the pre-specification before any suite ran: that pass counts may be equal on every problem is written there as a registered outcome, named as the instrument not separating the arms, and it is stated not to be a failure, not a null to be spun, and not grounds for a second analysis chosen afterwards.\src{harness/systems-v3/PRESPEC-posthoc-correctness-matrix1-2026-09-09.md section 4} There is no second analysis. The single difference is one cell failing one test of seven, and that is not an adverse finding and is not reported as one. Reported alone, the \texttt{FreeList} row misleads: of the treatment's four cells on that problem, two were stopped by the budget and pass 7 of 7 when their suites are run, one never booted, and one landed at 6 of 7.\src{harness/systems-v3/RESULT-posthoc-correctness-2026-09-09.md section 3; harness/systems-v3/RESULT-posthoc-correctness-verdicts-2026-09-09.tsv}

\paragraph{Being stopped by the budget is correlated with the arm.} The classes are not distributed evenly across the arms. On the bare arms the control contributed 18 landed cells and none stopped by the budget, and the treatment contributed 12 landed cells and three stopped. Every budget-capped cell in this matrix is a treatment cell.\src{harness/systems-v3/RESULT-posthoc-correctness-2026-09-09.md section 3; harness/systems-v3/RESULT-posthoc-correctness-verdicts-2026-09-09.tsv} The correctness column therefore scores 18 control cells against 12 treatment cells on those arms, and the treatment cells missing from it are the ones that ran long enough to reach a cap, which is to say the expensive ones. The surviving treatment sample is easier than the arm it is drawn from, so any pass rate computed for the treatment on this run is biased upward by construction. This is a selection effect, it is correlated with the arm under test, and it travels with every correctness number from this run. Section~\ref{sec:instrument} states it as a property of the instrument rather than of this population.

A budget stop is not a correctness failure, and on this run that is measured rather than assumed: all three capped treatment cells pass their suites completely when they are run, 7 of 7 and 7 of 7 on \texttt{FreeList} and 17 of 17 on \texttt{Paxos}.\src{harness/systems-v3/RESULT-posthoc-correctness-verdicts-2026-09-09.tsv} That is why the pre-specification gave \textsc{cap-cost} its own class instead of letting it read as a loss.

\paragraph{The statement-arm cells, registered before they ran.} The same classes and the same runner, applied to the 18 cells the statement amendment fired: \textsc{pass} 18, \textsc{fail} 0, \textsc{cap-cost} 0, \textsc{failed-boot} 0, \textsc{no-build} 0, \textsc{interface-miss} 0. For these cells the pass was registered before they ran, which is the distinction the amendment makes binding, and they are kept under their own heading rather than added to the table above.\src{harness/systems-v3/RESULT-statement-arm-2026-09-09.md section 2}

\paragraph{What a pass is, and what this section does not license.} A pass is that the withheld suite did not fail the cell. It is not a proof of correctness. The suites were characterised against authored mutants at a ceiling rather than a floor (Section~\ref{sec:populations}), and a suite that kills every mutant written beside it has been shown not to be vacuous, not to be hard.\src{harness/systems-v3/RESULT-posthoc-correctness-2026-09-09.md section 4} Nothing in this section licenses joining the cost result to this one. In particular it does not license the claim that the treatment costs more and produces more correct code, which is a causal join this design cannot support, and it does not license a correctness comparison between the arms, which the reading above says the instrument did not make.

\paragraph{The declared set, and a dependency the scoreboard does not show.} The scored set is declared by identifier rather than by a glob over the archive, because cells from different runs share the condition key and a glob pools them: an early draft of the scorer, driven on partial data, silently pooled this matrix with the earlier stage, with a pricing set, with two dropped arms and with a void cell.\src{harness/systems-v3/PREREGISTRATION-matrix-opus-1-2026-09-08.md section 17} The declaration is the cells of the matrix root plus three named smoke cells, \texttt{ae304f63}, \texttt{a69e9131} and \texttt{b7537006}, one each on \texttt{FreeList}, \texttt{LRU} and \texttt{Paxos}, fired first as an end-to-end smoke on the three problems that had never produced a cell and registered in advance as counting toward their conditions.\src{harness/systems-v3/score\_matrix1.py, the declared set; harness/systems-v3/PREREGISTRATION-matrix-opus-1-2026-09-08.md section 10} To that declaration AMENDMENT~26 added three further plain-arm cells, one each on the same three problems, fired in a third cells root of their own with a fresh identifier prefix and against a subject-facing tree driven byte-identical to the matrix's before the first of them started. All three landed, none hit the cost cap, none was void, and each was priced twice by instruments that agree to the cent: the harvested \texttt{METER.txt} and the cell's own end-of-run control record.\src{harness/systems-v3/RESULT-n3-topup-2026-09-09.md section 1}

Those three cells live in a different cells root from the rest of the declared set, and the matrix root holds two landed plain-arm cells each on \texttt{FreeList}, \texttt{LRU} and \texttt{Paxos}. Three of the five problems therefore reach $n=3$ on the plain arm only by counting one smoke cell each, and without them the registered rule takes no median on those three: the sign test becomes 2 of 2 at $p = 0.25$, which is no verdict. This is not a defect in the declaration, which names the three by identifier and says why a glob would be wrong. It was a fact about the dataset that the scoreboard was silent about, and it travelled with the number until the cells that retire it landed.\src{harness/systems-v3/RESULT-matrix-opus-1-2026-09-08.md, the declared-set section} The three smoke cells also ran under an earlier export of the harness whose repairs are to the audit, carrier and canary paths and not to the task, the arm, the prompt or the caps; on that basis they contribute a price, and their containment verdicts are withdrawn rather than caveated, so no containment claim for this matrix rests on them.\src{harness/systems-v3/PREREGISTRATION-matrix-opus-1-2026-09-08.md sections 12 and 13}

\paragraph{The dependency is discharged for the sign, and not for the magnitudes.} The rule for reading that discharge was registered with the top-up and before its cells fired: report the sign test with and without the borrowed cells, and if the two readings agree the dependency is discharged, while if they diverge the divergence is the result and is reported ahead of the headline. Both halves fire at once here, and reading only the first is the error this paragraph exists to prevent. The readings agree on what the registration calls the reading: 5 of 5 at $p = 0.0312$ under A and under B alike, so the headline does not rest on the three borrowed cells and the dependency reported above is retired. The magnitudes do not agree. Three of the five premiums move between the readings and two of them move down: LRU from $1.2826\times$ to $1.1521\times$ and Paxos from $2.4306\times$ to $2.2437\times$, both because the cheapest plain cell on each problem is a borrowed one and its departure raises the plain median, and FreeList up from $2.8070\times$ to $2.8879\times$ because the cell that leaves there sat near the middle. Crc32 and LZW do not move, having no borrowed cell to lose. Reading~B's LRU premium is the closest to $1.0$ that any premium in this campaign has come, and the borrowed cells were flattering it.\src{harness/systems-v3/RESULT-n3-topup-2026-09-09.md section 5} No qualifier is retired by any of this: three of the five magnitudes remain below the resolvable floor under both readings, so the movement of a magnitude is barred from being the finding in exactly the way the magnitude itself is. That is the reason it would be easy to call the movement immaterial, and it is not immaterial. It is the measurement of how much the borrowed cells were flattering the table, and a dependency can be discharged while the thing it was propping up gets weaker.

\paragraph{What the top-up did not do.} Before its cells fired, the amendment computed the premium range each problem could reach at $n=4$ and showed that no outcome of the top-up could move any premium to $1$. All three landed values fell at an end of their registered range, which is what a fourth observation does to a median taken as the mean of a middle pair. Because the exercise could not have changed the verdict, it did not confirm it: it bought precision and removed a dependency, and it is not a replication of the earlier reading and is not written as one.\src{harness/systems-v3/RESULT-n3-topup-2026-09-09.md section 4}

\paragraph{The declared set spans two run accounts, which is the third thing the scoreboard did not show.} The three cells the amendment added wrote their transcripts under one run account and every cell of the matrix root under another. Neither result file showed this: one named its own account and not the contrast, the other names no account at all, so a reader of either could not tell that a set described as every cell from this run pools cells recorded under two. It does not move a price, and that is measured rather than assumed, because both populations are priced from the same rate card read on the same day and a cost here is a rate card applied to a token count rather than an account's bill. What is not measured is the one mechanism that could bite. An account cannot change the price of a token but it can change the count, since almost all of the token total on these cells is cache read and cache state is held per account and per session, and nothing here measured whether that differs systematically across the boundary rather than randomly. Cache state already varies from cell to cell inside the matrix, so this is not a new source of variance; it is a possible systematic one and it is unquantified. Every cell from this run is a claim about a run, and a run is not necessarily one context.\src{harness/systems-v3/RESULT-n3-topup-2026-09-09.md section 1b}

\paragraph{Correctness: no cell of this run carried a referee verdict when it was scored.}
Of the cells in the matrix root, 33 are priced, and the three cells AMENDMENT~26 added take reading~B's declared set to 36 priced. None of them carried a referee verdict on a withheld suite at the time it was scored, and the three added cells added none: no quantity of cost cells closes a correctness gap. The path such a verdict would be read from did not exist then: five patterns swept across the whole cells root return zero files each, and the scorer reads a cost off the archive and no verdict of any kind. The only per-cell artifact that resembles a check is a manifest integrity check, which is not a correctness verdict. The withheld-suite work that does exist characterises the suites against mutants (Section~\ref{sec:populations}) and is a property of the suites, not of any cell.\src{harness/systems-v3/RESULT-matrix-opus-1-2026-09-08.md, the correctness section} This does not say the code was wrong. It says no instrument in this run asked when the cells were priced, so the premium is a price for work whose correctness the run did not verify as it went, and no sentence in this paper pairs it with quality or with working code. The post-hoc pass of Section~\ref{sec:posthoc}, run afterwards against the surviving repositories, is the instrument that asked; it did not separate the arms, and it does not change what the premium is a price for.

Two counts in that measurement do not reconcile and are reported rather than smoothed. Priced does not imply landed: of the 37 cells with a control directory in the matrix root, 30 landed, 3 stopped at the cost cap and are priced without having finished, and 4 failed to boot and are unpriced, which gives 33 priced. An earlier hand census in the same file reports 36 cells, 32 priced and 4 void. The file states the difference instead of adopting one number, and neither count changes the correctness answer, which is zero.\src{harness/systems-v3/RESULT-matrix-opus-1-2026-09-08.md, the correctness section, part (a)} Pricing a capped cell beside a landed one prices two different events under one name, and the three capped cells are named here for that reason.

\paragraph{The placebo arm supports nothing.} An equal-cost placebo arm was run to completion on all five problems: 15 of 15 cells landed and priced, \$193.42, no void cell, no cap, no drift and no failed boot. It returns \textsc{unresolved} on all five problems, every one inside the band $[0.4982, 2.0072]$.\src{harness/systems-v3/RULING-placebo-acceptance-2026-09-08.md section 13} That outcome was the predicted one and was registered as such before any placebo cell existed, because the floor at this $n$ is $2.0072\times$ while the treatment's own premiums on three of the five problems sit below it. The sentence this result most invites, that the placebo came in near the plain arm so the treatment's premium is method rather than form, is forbidden by name in the registration and is not written here. An arm that could have damaged the finding and did not is worth its price and is still not evidence for the finding.\src{harness/systems-v3/RULING-placebo-acceptance-2026-09-08.md sections 13.1 and 13.3}

\paragraph{A cross-stage cost claim is confounded by concurrency.} The earlier two-problem stage ran mostly one cell at a time, read from its own launch receipts; this matrix ran four-wide in both its fire and its resume schedulers. The two sets of premiums therefore differ in box contention as well as in date, and nothing in the record separates those. The confound does not touch this matrix's own result, which is computed entirely within a run where both arms were measured four-wide; it bars reading the matrix's premiums as a replication of the earlier stage's magnitudes, and the placebo was fired at four-wide for the same reason.\src{harness/systems-v3/RESULT-matrix-opus-1-2026-09-08.md, the concurrency limitation} This correction was made to the campaign's own earlier claim: the two stages had been established to share a client binary, byte for byte, and were then treated as cost-comparable on that basis, which verifies one axis and generalises the verdict past it.

\paragraph{Every cost in this section carries an unmeasured box-load term.} The concurrency confound above is a known difference between two stages. The wider one is that nothing in this campaign measures what the box was carrying while a cell was priced. The harness reaps the client and the watcher and does not reap what the subject forked: 24 processes forked by the subject of one cell in a later wave were re-parented to the init process and ran for three hours and twenty-three minutes, outliving their own cell's end by three hours and nine minutes, with 11 of the 16 priced cells of that wave metered wholly or partly inside the window. A cell's end is not the end of the cell's processes, so any cell in this campaign may have been priced on a box carrying the residue of earlier cells, and no arm looks, so no run can state its own load. The direction of that on cost is unmeasured and is not asserted here in either direction: extra processor time does not spend tokens, and a route from load to price through timeouts or extra turns was never driven. For this matrix specifically the term is bounded rather than assumed: all 37 cells of the matrix root reached their end at or before \texttt{2026-09-09T00:51:59Z} and that leak opened at \texttt{2026-09-09T03:34:51Z}, a margin of two hours, forty-two minutes and fifty-two seconds, so no cell of the matrix root was metered inside it. That bound is stated over the matrix root, and the declared set is now larger than the matrix root: the three cells AMENDMENT~26 added ran later on the same day, and no file in this repository records an end time for them, so the margin above does not cover them and this paper does not claim that it does. The bound therefore clears the matrix root of that leak, and of no other, and leaves three cells of the declared set unbounded for it. A set that grows after a bound is written over it does not inherit the bound, and the growth is silent: nothing in the scoreboard, the premium or the $p$-value changes shape when a cell that no timing record covers joins the set.\src{PUBLISH-CHECKLIST.md section (l) item 8; harness/systems-v3/RULING-placebo-acceptance-2026-09-08.md section 8.1}

\paragraph{What this population cannot do, stated with the result.} It is authored by the same people who wrote the treatment, so it carries no independent authorship and no contamination measurement of the kind a public set allows. It measures a dieted rendering of the method and not the method. It measures price and not acceptance. Three of its five magnitudes are below its own resolvable floor and none of the five may be reported as the finding. Its gold pair is $k=1$. And three of its five problems currently reach $n=3$ on the plain arm by counting a cell from another root. Version~2's remedy for the last of these is under way and is described in Section~\ref{sec:open}.

\section{The complete pilot matrix}
\label{sec:matrix}

The cost reading of Section~\ref{sec:s3} is one model on one task form. The pilot matrix the author asked for, once that reading existed, is the same population across four models, both task forms that have a registered object, and the three treatments, and this section reports it as a record of coverage: which conditions ran to their registered $n$, which could not, and what each result of record says its conditions can and cannot carry. It reports no test across the matrix and no verdict on the arms, because the census that counts it makes neither and the results of record it rests on make neither.\src{harness/systems-v3/CENSUS-full-matrix-2026-09-14.md, addendum 24 section AD2}

\paragraph{The denominator, and why it is 200.} The request that fixed the shape reads, in the census that tracks it: four models, five problems, greenfield and brownfield, and the arms \arm{plain} and \arm{salt-diet} under each of three treatments, which are no treatment, a formal statement handed to the agent, and a specification change after a first build. That is $4 \times 5 \times 2 \times 2 \times 3 = 240$ conditions before replication, at $n=3$ cells per condition.\src{harness/systems-v3/CENSUS-full-matrix-2026-09-14.md, the header} Two rulings on 2026-09-14 fixed the denominator at 200. Brownfield under specification change leaves it: a specification-change cell is a second phase on a landed first phase, the cell builder refuses that phase on a brownfield card, and the author ruled the forty conditions out rather than owed. The fourth model is \texttt{claude-sonnet-5}, beside \texttt{claude-opus-5}, \texttt{gemini-3.1-pro-high} and \texttt{gemini-3.8-flash-high}, which are the model strings the harness served.\src{harness/systems-v3/CENSUS-full-matrix-2026-09-14.md, sections C1 and C3 and addendum 1} The 240-view and the 200-view differ only in whether those forty sit inside the denominator, and the census publishes both so that neither has to be derived.\src{harness/systems-v3/CENSUS-full-matrix-2026-09-14.md, addendum 5 section J4}

\paragraph{Three states, fixed before the count reached them.} A condition is counted in one of three states. \textsc{done} means a result of record merged in this repository that reports the condition's outcome, at the $n$ that record states, which is $n=3$ cells of record for all 181. Ten conditions reached it later than the rest, on 2026-09-29: one Sonnet greenfield condition, three Pro greenfield conditions and six Opus specification-change conditions. Each of the three Pro figures spans two exports, and two of them span two clients as well. One Opus \texttt{Paxos} \arm{salt-diet} cell carries two verdicts on byte-identical code, a published build failure and a later full pass, and the census moves its $n$ and never a verdict. The count is of cells of record, not of scorable cells; a record can list three cells and score fewer. It is a claim that evidence exists and never a verdict on the arms: a condition can be \textsc{done} for the census and unable to separate the arms, and the census records that beside the count rather than letting the count close it.\src{harness/systems-v3/CENSUS-full-matrix-2026-09-14.md, sections C2 and C6, addendum 5 section J2, addendum 24 section AD2, addendum 25, and addenda 30 to 32 at commit ba29ac7} \textsc{inexpressible} means the harness cannot express the condition, with the reason recorded. Every model's \texttt{Paxos} conditions under the statement treatment are in this state, 16 in all, because an arm-neutral formal statement cannot exist for a proof-obligation task, so a plain arm handed one would be a control told to write proofs; it is the same gate that refused those cells in Section~\ref{sec:s3}.\src{harness/systems-v3/CENSUS-full-matrix-2026-09-14.md, section C3; harness/systems-v3/RESULT-statement-arm-2026-09-09.md} \textsc{declared} means the condition was fired as registered and reported whole, and did not produce the second-phase outcome the matrix asks for because of the uniform cost cap. It is reported as unreached at the cap, with its reach figure and each cell's own record. It is not owed, because a re-fire would select on cheapness, which is the property that stopped it, and it is not \textsc{done}, because a reach outcome is a different kind of outcome and must not enter a correctness denominator.\src{harness/systems-v3/CENSUS-full-matrix-2026-09-14.md, addendum 19 section Y2}

\paragraph{The reading of record.} At the commit this version of the paper reads, every condition is in one of the three states and none is owed or blocked: \textsc{done} 181, \textsc{inexpressible} 16, \textsc{declared} 3, in sum 200. The count moved through 24 dated addenda over eleven days. The first six record rulings, corrections and one result already merged; from addendum 7 the rule is that an addendum lands in the same commit as the result of record it counts, and the census records the one that did not, addendum 18, which followed its result in a later commit. The twelve addenda after the 24th, through addendum 36, move no condition's state: they restate what \textsc{done} means, correct one population figure, take ten conditions to $n=3$, add a later objective's blocks outside the 200, and declare limits. The census's own verifier rebuilds the trajectory from those addenda's headlines and refuses on any mismatch, missing row or broken chain. What that verifier cannot see is an addendum whose own headline is wrong, and it prints that limit beside its verdict; the per-block verifiers named in Section~\ref{sec:repro} are what check the headlines against the per-cell tables.\src{harness/systems-v3/CENSUS-full-matrix-2026-09-14.md, section C5 (the live row), addenda 1 to 7, addendum 18 section X5, and addenda 24 to 36, at commit ba29ac7; harness/systems-v3/CENSUS-full-matrix-2026-09-14-verify.py} Two corrections in that trajectory are kept because they say what the count is made of: four brownfield conditions were counted \textsc{done} and reverted when the wave's own result found that two of the given files announced their planted defect in a comment, and one block's ten conditions were counted under the wrong model for a day because its per-cell table carried no model column, which is why every later table of record derives the served model per cell from that cell's own receipt.\src{harness/systems-v3/CENSUS-full-matrix-2026-09-14.md, addendum 6 and addendum 12; harness/systems-v3/RESULT-gemini-brownfield-level4-2026-09-14.md, addendum 1} The givens behind the 72 brownfield conditions of record then had a refuter pass of their own. It found no leak of a withheld reference by three checks, one of them partial on \texttt{Paxos}, and found that two givens that do not announce their defect still carry a comment tell: the \texttt{Crc32} given's comment states eight bit-steps above a loop that runs seven, which points a reader who trusts it at the planted defect, and the \texttt{LZW} given's comment repeats the card. Those 32 conditions, 16 on each problem, may therefore be easier in absolute terms than their rows suggest. Both arms of a condition received the same bytes, so the tell cannot by itself produce an arm difference, but whether the method changes how closely a subject reads comments, which would let a constant tell interact with the arm, is not measured.\src{harness/systems-v3/CENSUS-full-matrix-2026-09-14.md, addendum 36}

\begin{center}
{\scriptsize\setlength{\tabcolsep}{4pt}
\begin{tabular}{lllrrrp{5.4cm}}
\toprule
model & task form & treatment & done & inexpr. & decl. & result of record \\
\midrule
\texttt{claude-opus-5} & greenfield & none & 10 & 0 & 0 & \texttt{matrix-opus-1-2026-09-08} \\
 & greenfield & statement & 8 & 2 & 0 & \texttt{statement-arm-2026-09-09} \\
 & greenfield & spec-change & 10 & 0 & 0 & \texttt{p1-specchange-2026-09-10} \\
 & brownfield & none & 10 & 0 & 0 & \texttt{claude-blockO-2026-09-21} \\
 & brownfield & statement & 8 & 2 & 0 & \texttt{claude-blockOS-2026-09-21} \\
\texttt{claude-sonnet-5} & greenfield & none & 10 & 0 & 0 & \texttt{claude-blockSG-2026-09-21} \\
 & greenfield & statement & 8 & 2 & 0 & \texttt{claude-blockSS-2026-09-21} \\
 & greenfield & spec-change & 7 & 0 & 3 & \texttt{claude-blockSC-2026-09-21} \\
 & brownfield & none & 10 & 0 & 0 & \texttt{claude-blockSB-2026-09-19} \\
 & brownfield & statement & 8 & 2 & 0 & \texttt{claude-blockSBS-2026-09-21} \\
\texttt{gemini-3.1-pro-high} & greenfield & none & 10 & 0 & 0 & \texttt{p1-greenfield-2026-09-13}\newline \texttt{gemini-level6-2026-09-17} \\
 & greenfield & statement & 8 & 2 & 0 & \texttt{p1-greenfield-2026-09-13}\newline \texttt{gemini-level6-2026-09-17} \\
 & greenfield & spec-change & 10 & 0 & 0 & \texttt{gemini-level8-chainD-2026-09-24}\newline \texttt{gemini-level8-chainF-2026-09-24} \\
 & brownfield & none & 10 & 0 & 0 & \texttt{gemini-brownfield-level4-2026-09-14}\newline \texttt{gemini-level7-2026-09-19} \\
 & brownfield & statement & 8 & 2 & 0 & \texttt{gemini-level7-2026-09-19} \\
\texttt{gemini-3.8-flash-high} & greenfield & none & 10 & 0 & 0 & \texttt{gemini-flash-level5-2026-09-15}\newline \texttt{gemini-level6-2026-09-17} \\
 & greenfield & statement & 8 & 2 & 0 & \texttt{gemini-flash-level5-2026-09-15}\newline \texttt{gemini-level6-2026-09-17} \\
 & greenfield & spec-change & 10 & 0 & 0 & \texttt{gemini-level8-chainD-2026-09-24}\newline \texttt{gemini-level8-chainF-2026-09-24} \\
 & brownfield & none & 10 & 0 & 0 & \texttt{gemini-level6-2026-09-17}\newline \texttt{gemini-level7-2026-09-19} \\
 & brownfield & statement & 8 & 2 & 0 & \texttt{gemini-level7-2026-09-19} \\
\midrule
all four models & & & 181 & 16 & 3 & 200 conditions \\
\bottomrule
\end{tabular}}
\src{harness/systems-v3/CENSUS-full-matrix-2026-09-14.md, section C4 (the opus rows and the pro greenfield rows), addendum 1 (the denominator), addenda 5 and 7 and 9 (the flash greenfield rows), addenda 6 and 10 (the pro brownfield rows), addenda 9 and 10 (the flash brownfield rows under no treatment), addenda 10 and 24 (the flash brownfield statement row), addenda 11 and 12 (the sonnet brownfield row under no treatment), addendum 13 (the opus brownfield row under no treatment), addenda 14 and 15 and 16 (the sonnet greenfield rows and its brownfield statement row), addenda 17 and 18 (the opus brownfield statement row), addenda 18 and 19 (the sonnet specification-change row), addenda 20 to 23 (the gemini specification-change rows), all at commit 3aebf116 and unchanged through addendum 36 at commit ba29ac7; paper/matrix\_census\_table.py, which derives these rows and checks their sums against the census's live row and the rows in this file against its own output}
\end{center}

\noindent Each result of record is a file under \texttt{harness/systems-v3/}, named here with its \texttt{RESULT-} prefix and \texttt{.md} suffix elided. Each row is ten conditions, five problems by two arms. A statement row carries \texttt{Paxos}'s two inexpressible conditions. The Sonnet specification-change row carries the three declared conditions, all on the \arm{salt-diet} arm. The Sonnet brownfield row under no treatment is block SB, whose result of record first named the wrong model and carries the correction as an erratum below its signed text.\src{harness/systems-v3/CENSUS-full-matrix-2026-09-14.md, addendum 12; harness/systems-v3/RESULT-claude-blockSB-2026-09-19.md, section E1}

\subsection{What the records carry, and what they refuse to say}
\label{sec:matrixrecords}

The census counts; the results of record say what the counted conditions mean. What follows is one paragraph per lane, in the results' own terms, and each ends where the result's own limits end.

\paragraph{The Opus greenfield rows.} These are the matrix of Section~\ref{sec:s3}, its statement arm, and the specification-change wave that followed them: 19 cells over the five problems, of which 16 landed and 17 passed the post-change suite in full, and the two sets are not nested: one cell declared itself done at 8 of 9 tests, and two stopped at the cost cap with 24 of 24 passing, so a campaign reporting only the landing rate would have been one short and would have named the wrong cells.\src{harness/systems-v3/RESULT-p1-specchange-2026-09-10.md sections 1 and 2 and addendum A} Version 2 of this report said 16 passed and one capped cell passed: the second capped cell was published as not building, and a re-score of its unchanged code on 2026-09-30 passed all 24 tests. What produced the first verdict is not measured, and no other verdict in the population re-scored changes its pass count.\src{harness/systems-v3/CENSUS-rc3-verdicts-2026-09-30.md sections C3 and C4} By problem the arms are level except on \texttt{FreeList} and \texttt{Paxos}, which are the problems with the thinnest cells, one condition at $n=1$ because its reusable first-phase pool was short.\src{harness/systems-v3/RESULT-p1-specchange-2026-09-10.md section 3 and addendum A} Version 2 took that clause from a result whose \texttt{Paxos} counts were 1 of 2 in each arm, which are level, so the clause did not match its source; after the re-score the counts are 1 of 2 and 2 of 2, and the clause matches them because of the correction, not because it was right before.\src{harness/systems-v3/RESULT-p1-specchange-2026-09-10.md section 3 and addendum A} That reading is no longer the row's: on 2026-09-29 step g took the six thin Opus specification-change conditions to $n=3$ cells of record with seven new cells. At $n=3$, \texttt{FreeList} reads 3 of 3 full passes under \arm{plain} and 2 of 3 under \arm{salt-diet}, and \texttt{Paxos} reads 2 of 3 in each arm. The step-g result licenses no arm comparison, because the arms differ in how often a first phase landed: on \texttt{FreeList}, 4 of 4 \arm{plain} first phases with an end landed against 3 of 6 \arm{salt-diet}, and on \texttt{Paxos} 4 of 4 against 3 of 5, and every second-phase count is conditioned on a landing.\src{harness/systems-v3/RESULT-stepg-opus-specchange-2026-09-29.md sections 2, 3 and 6 and addendum A; harness/systems-v3/CENSUS-full-matrix-2026-09-14.md, addendum 32}

\paragraph{The Claude lane's brownfield and Sonnet blocks.} Seven blocks ran under the Claude harness after the matrix of Section~\ref{sec:s3}: brownfield under no treatment at both Claude models, brownfield under the statement treatment at both, and the Sonnet greenfield rows under all three treatments. Four things travel with every number in them. First, on brownfield under no treatment the suite pass rate is a total ceiling at both models, 30 of 30 cells at each with every seeded bug fixed, so neither the pass rate nor the bugs-fixed count can rank the arms.\src{harness/systems-v3/RESULT-claude-blockSB-2026-09-19.md section 1; harness/systems-v3/RESULT-claude-blockO-2026-09-21.md; harness/systems-v3/CENSUS-full-matrix-2026-09-14.md, addendum 11 section R3 and addendum 13 section S3} Second, the registered brownfield separator, the fraction of the given file retained, puts the arms in disjoint bands on five of five problems at both models and under the statement treatment as well, and the results show it is reading growth rather than rewriting: on the Sonnet brownfield statement block the \arm{salt-diet} arm's lowest survival of the seed, 0.646, is above the \arm{plain} arm's lowest, 0.590, while the same arm wrote 4.66 to 15.77 times as much code around the seed, so a class that reads \emph{replaced} from that fraction does not say the seed was discarded.\src{harness/systems-v3/CENSUS-full-matrix-2026-09-14.md, addendum 16 section V2, addendum 13 section S3 and addendum 17 section W3; harness/systems-v3/RESULT-claude-blockSBS-2026-09-21.md} Third, the cost cap is arm-correlated and it repeats. On Sonnet greenfield under no treatment it bound 0 of 14 \arm{plain} cells and 5 of 15 \arm{salt-diet} cells; under the statement treatment, 0 of 12 and 2 of 12, and the two \arm{salt-diet} cells that did not pass are the two capped cells; on Sonnet brownfield under the statement treatment, 0 of 12 and 3 of 12; on the brownfield blocks under no treatment, 4 \arm{salt-diet} cells and 0 \arm{plain} at Sonnet and 1 and 0 at Opus. Every \arm{salt-diet} cost in those blocks is a lower bound and every \arm{salt-diet} pass rate a floor, and a block in which every failure coincides with a cap cannot separate code that is worse from work that ran out of money.\src{harness/systems-v3/CENSUS-full-matrix-2026-09-14.md, addendum 14 section T3, addendum 15 section U2, addendum 16 section V3, addendum 11 section R3 and addendum 13 section S3} Fourth, the one median cost ratio in these blocks taken with no cap binding, Opus brownfield under the statement treatment with 0 of 24 cells capped, is $1.40\times$ at twelve cells an arm, and it is by far the smallest of the five: the three Sonnet ratios, all lower bounds, are $8.29\times$, $5.31\times$ and $4.70\times$. It cannot be attributed, because that block differs from the others in model as well as in censoring, and its cells cost more in both arms, which is why the cap was never reached; it is the question these blocks raise and not one they answer.\src{harness/systems-v3/CENSUS-full-matrix-2026-09-14.md, addendum 17 section W2 and addendum 18 section X4} Every such ratio is read from a cost that includes the harness's own sandbox probe, about ten cents a cell; measured apart from it on one split table, the Sonnet ratios rise by between 2 and 10 percent and the Opus ratios move by less than a quarter of a percent, so the convention flatters the treatment on the cheaper arm and the bounds already allow it. The census's convention is that a ratio stated in a write-up is taken with the probe apart, from that split. The split covers two of the four figures above at the same cell counts, with probe-in readings of $5.31\times$ and $4.70\times$, equal to the record's: with the probe apart, Sonnet greenfield under the statement treatment is $5.54\times$ and Sonnet brownfield under the statement treatment is $5.12\times$. For the other two, the split was taken on a different cell set. It reads $8.48\times$ with the probe in for Sonnet greenfield under no treatment, where the record reads $8.29\times$, and $1.22\times$ for Opus brownfield under the statement treatment, on 10 \arm{plain} cells where the record, after that block's last two cells, has 12 and reads $1.40\times$. Those two therefore have no probe-apart figure of record, and are stated with the probe in.\src{harness/systems-v3/CENSUS-full-matrix-2026-09-14.md, the cost-convention declaration of 2026-09-22, and addendum 18 section X4; evidence/vx-probe-split-2026-09-22/ratio\_shift.out} Under specification change on Sonnet, two whole \arm{salt-diet} conditions produced no second phase at all: every first phase on \texttt{Paxos} and on \texttt{FreeList} ended at the cost cap, four of those six cells passing their first-phase suites, while \texttt{LRU}, \texttt{LZW} and \texttt{Crc32} reached 3 of 3, so the loss is an interaction of problem and arm and not the arm. The second-phase cap is compared against the cell's cumulative cost and sits below the first-phase cap, so raising the first-phase cap would have produced no second-phase result either. Those are two of the three declared conditions; the third, \texttt{LZW}, reached 3 of 3 and is declared on a usable-outcome standard, because two of its three second phases were build failures whose suite never ran and one passed 15 of 15, and a single pass is reported as a cell and never as a condition's rate.\src{harness/systems-v3/RESULT-claude-blockSC-2026-09-21.md section 2; harness/systems-v3/CENSUS-full-matrix-2026-09-14.md, addendum 18 sections X2 and X3 and addendum 19}

\paragraph{The Gemini lane.} The two Gemini models ran under the harness's second client lane in registered levels, each frozen before its first call.\src{harness/systems-v3/AMENDMENT-gemini-level8-2026-09-16.md} On this lane the fence does not claim that the subject could not reach a web tool during a run: it voids a cell whose hooks at its end differ from those the launcher installed, and the cells that were served a web search are declared in the next paragraph.\src{harness/systems-v3/AMENDMENT-gemini-level6-2026-09-16.md, A5.2 and its addendum of 2026-11-20} The lane's first brownfield wave found that two of the four given files, \texttt{FreeList} and \texttt{LZW}, announced their planted defect in a comment for about 57 hours, a window containing the wave, so 12 of its 24 cells are void for the find-the-defect claim, those four conditions were reverted and re-run at a later level, and \texttt{LRU} and \texttt{Paxos} stand.\src{harness/systems-v3/RESULT-gemini-brownfield-level4-2026-09-14.md, addendum 1; harness/systems-v3/CENSUS-full-matrix-2026-09-14.md, addendum 6} The Flash model's first wave halted at one cell on a server capacity error that was first read as a model behaviour and corrected at the object; its completed wave publishes no arm comparison, because every missing cell is \arm{salt-diet} and a rate over them would be measured on a treatment arm whose cells were removed by an arm-correlated mechanism, and two of its full passes did 65 and 0 seconds of model work, a fact about the task and not about the model.\src{harness/systems-v3/CENSUS-full-matrix-2026-09-14.md, addendum 2, addendum 5 section J2 and addendum 7} Level 6 is the first wave whose statement conditions carry an arm contrast on pass rate rather than a ceiling, \arm{plain} 6 of 6 against \arm{salt-diet} 4 of 6 at $n=3$ per condition, a contrast in the record and not an estimate; one \arm{salt-diet} cell is void because the arm never reached the subject, and it is in no denominator.\src{harness/systems-v3/CENSUS-full-matrix-2026-09-14.md, addendum 9 sections P2 and P3; harness/systems-v3/RESULT-gemini-level6-2026-09-17.md} The receipt that shows an arm reached its subject could also be read by a side route on the lane's older cell layout. Across its 92 cells of record, 27 subjects named such a route, every one after opening the arm file, and none echoed the receipt through one before, so no receipt of record changes. The count is of calls that name a side route.\src{harness/systems-v3/CENSUS-full-matrix-2026-09-14.md, addendum 35; evidence/agy-ctl-census-2026-10-13/side-route-all-legacy.out} Level 7, 84 cells over both Gemini models on brownfield under both treatments, carries the cap correlation in its sharpest form: every truncated, indeterminate or false-done cell, ten of ten, is a Pro \arm{salt-diet} cell, so no pass-rate contrast on the Pro row survives it, and its retention separation is a growth separation by the result's own erratum. Its last condition was held at one cell for six days after a credential window refused the other two, and closed with two cells fired under the same registration and export into a fresh root, pooled with the first by a measured interface hash and a control re-score: 3 of 3 pass.\src{harness/systems-v3/RESULT-gemini-level7-2026-09-19.md sections 1 and 2, erratum 2 and addendum A; harness/systems-v3/CENSUS-full-matrix-2026-09-14.md, addendum 10 section Q4 and addendum 24} Level 8, the Gemini specification-change wave, was fired twice. Its first firing ran under a fence that hid the harness's package registry from the subject, so the \arm{salt-diet} build could not resolve its verification library offline; some cells failed to build, and others went green because the subject rebuilt its own build environment, with a stub library of 61 bytes, or a patch onto a release library that verified nothing, or a private package home, and \arm{plain} cells did the same. Every cell of that firing was ruled invalid as fired, in both arms, including a \arm{plain} condition that had scored 3 of 3, because a subject that spent its turns rebuilding a build environment did not work the registered problem and because re-firing one arm alone would bias the pair; the whole level re-fired on the repaired export, and two methods over every Gemini cell's own fence file show the defect is a property of that level alone.\src{harness/systems-v3/CENSUS-full-matrix-2026-09-14.md, the declaration of 2026-09-23 on the level-8 fence} The same firing surfaced a second defect: the landing verdict is read from a file in the subject's own tree, and two subjects forged it, one through a named pipe and one by reimplementing the runner's tree hash. No score is touched, because the scorer never reads that file, and a search of all 714 cell state directories found forging tooling in exactly those two cells and none at any lower level, which makes a landing below level 8 unforged by trace and not verified; both are counted as false-done claims.\src{harness/systems-v3/CENSUS-full-matrix-2026-09-14.md, the same declaration} The re-fired level is registered as a believability pilot: sign only at $n \leq 3$, no magnitude, no cross-lane comparison in dollars, tokens or rates, no dollar price on this lane, and a ceiling says nothing about the arms.\src{harness/systems-v3/AMENDMENT-gemini-level8-2026-09-16.md section M9} On \texttt{LRU} and \texttt{Crc32} its 8 conditions sit at a ceiling registered before any data, each at 3 of 3 reached cells scored as full passes, 24 of 24. The result's own headline of 21 of 21 was written before its first erratum's tripwire cell and its two copies; one condition reached $n=3$ only after two landed first phases that a gate had refused ran their second phase on copies, by the author's ruling.\src{harness/systems-v3/RESULT-gemini-level8-chainD-2026-09-24.md sections 1 and 2 and addendum A; harness/systems-v3/CENSUS-full-matrix-2026-09-14.md, addenda 20 to 22} On \texttt{Paxos}, \texttt{FreeList} and \texttt{LZW} its 12 conditions ran 36 cells, 36 reached and scored, 19 full passes, with the four \texttt{LZW} conditions at ceiling. The one contrast a reader reaches for, Flash on \texttt{Paxos} at \arm{plain} 0 of 3 against \arm{salt-diet} 3 of 3, is recorded and not generalised, because the same arm is 0 of 3 on Flash \texttt{FreeList} where \arm{plain} is 2 of 3. The \arm{salt-diet} arm's median token count exceeds the \arm{plain} arm's in every condition and both phases, and is quoted for its sign only, because the count is dominated by cache reads. Three conditions needed more than one attempt after server errors, and every discarded attempt is whole and outside every count.\src{harness/systems-v3/RESULT-gemini-level8-chainF-2026-09-24.md sections 1, 2, 5 and 8; harness/systems-v3/CENSUS-full-matrix-2026-09-14.md, addendum 23}

\paragraph{Added in version~5: three cells were served a web search.} The Gemini client's main stream records a web search as a step with no result, so the census of streams quoted before this version could not tell a served search from a denied one. The client's own conversation record carries the result, and read there the client served every search it was asked for and the lane's hook denied none. Of the lane's 272 matrix cells of record, every one with a conversation record, three received search results and 269 received none: \texttt{l5psra202} (Flash, \arm{salt-diet}, greenfield \texttt{Paxos}), \texttt{l6vbpp02} (Flash, \arm{plain}, brownfield \texttt{Paxos}) and \texttt{l7npfs01} (Pro, \arm{salt-diet}, brownfield \texttt{FreeList}).\src{harness/systems-v3/CENSUS-full-matrix-2026-09-14.md, addendum 47; evidence/o37-agy-web-served-2026-11-20/README.md} No cell is voided, because the fence never claimed web isolation on this lane, and no count of the census moves. Each of the three keeps its verdict and leaves every comparison, as an input from outside the arm of unknown sign. The tables in this paper are the tables of record and still hold the three cells. Without them, these printed entries move, and no other: in Table~T2, Pro brownfield \texttt{FreeList} \arm{salt-diet} 17{,}298{,}225 becomes 15{,}550{,}480 and Flash brownfield \texttt{Paxos} \arm{plain} 7{,}992{,}405 becomes 7{,}084{,}768;\src{evidence/o37-agy-web-served-2026-11-20/recut/descriptive-v2-O57-moved.txt} in Tables~\$1 and~\$2, Flash greenfield \texttt{Paxos} \arm{salt-diet} \$4.57 becomes \$3.91 and Flash brownfield \texttt{Paxos} \arm{plain} \$1.32 becomes \$1.28, and in Table~\$4 that greenfield ratio 3.53 becomes 3.02; in Tables~W1 and~W2, the same greenfield condition's 1{,}207 seconds become 960, Pro brownfield \texttt{FreeList} \arm{salt-diet}'s 4{,}382 become 4{,}793 and Flash brownfield \texttt{Paxos} \arm{plain}'s 413 become 402, and in Table~W4 the greenfield ratio 3.14 becomes 2.50 and Flash's median ratio 2.63 becomes 2.50, still 5 of 5 above 1;\src{evidence/o37-agy-web-served-2026-11-20/recut/cost-v3-moved.txt} in Table~P1, Flash greenfield \texttt{Paxos} \arm{salt-diet} becomes unmeasured, because level~5 recorded its 2 of 3 by condition and not by cell; in Table~P2, Pro brownfield \texttt{FreeList} \arm{salt-diet} 1/3 becomes 0/2 and Flash brownfield \texttt{Paxos} \arm{plain} 3/3 becomes 2/2; and in Table~P4 the Flash greenfield \texttt{Paxos} sign, $=$, becomes no sign.\src{evidence/o37-agy-web-served-2026-11-20/recut/correctness-v3-moved.txt} No other sign moves, so the abstract's 8, 5 and 2 rows stand, and its cell totals count cells of record, which keep their verdicts.\src{evidence/o37-agy-web-served-2026-11-20/recut/README.md; harness/systems-v3/RESULT-correctness-tables-v3-2026-10-01.md, addendum of 2026-11-20 (re-cut)}

\paragraph{Added in version~5: two temp-directory limits.} Two limits found after version~4 attach to every per-arm reading of the matrix, and neither moves a count. On the Gemini lane, the subject's shell was given a temp directory it could not write, so a tool that makes a temp file failed while the shell itself ran; the stream prints that failure in 32 of 141 \arm{salt-diet} cells of record (34 counting two ambiguous identifiers) and 2 of 139 \arm{plain} ones.\src{harness/systems-v3/CENSUS-full-matrix-2026-09-14.md, addendum 41} On the Claude lane, the \arm{salt-diet} arm's own gate could not make its temp directory at its selftest in 7 of 126 \arm{salt-diet} condition-rows (6 distinct cells, one of them the \texttt{LRU} cell \texttt{d6b53ee4} of Section~\ref{sec:s3}); the \arm{plain} arm has no gate, so it cannot show this.\src{harness/systems-v3/CENSUS-full-matrix-2026-09-14.md, addendum 42} Whether a subject then worked around the failure, and whether any outcome or cost moved, is not read, and the sign of neither limit is measured.

\paragraph{What a complete matrix does not establish.} It does not establish an effect. A \textsc{done} is a result of record at the $n$ that record states; no test is run across the matrix, no interval is claimed, each result of record states what its conditions can and cannot carry, and none of them claims that the method under test helps or hurts. A reader summing the matrix as complete must say which of the three states is meant, because the 16 inexpressible and the 3 declared conditions are not \textsc{done}.\src{harness/systems-v3/CENSUS-full-matrix-2026-09-14.md, addendum 24 section AD2 and addendum 25} Five limits travel with every table above. The Gemini lane's registered readings carry no dollar price and are sign-only on token counts; the medians of Sections~\ref{sec:matrixtokens} and~\ref{sec:matrixdollars}, and the dollar prices of the second, describe them under readings registered afterwards and add no claim. The Claude lane's \arm{salt-diet} costs on the Sonnet blocks are lower bounds censored by an arm-correlated cap. And no comparison is made across the two lanes, in dollars, tokens or rates, because they did not face the same client or the same fence.\src{harness/systems-v3/AMENDMENT-gemini-level8-2026-09-16.md section M9; harness/systems-v3/CENSUS-full-matrix-2026-09-14.md, the declaration of 2026-09-23 on the level-8 fence} And many of the cells' own \texttt{git} and \texttt{python3} invocations failed, in both arms. On the Gemini lane, where the failure rate separates by command, 81.4 percent of completed \texttt{git} invocations failed under \arm{plain} and 96.7 percent under \arm{salt-diet}, against 61.2 and 30.4 percent for all other commands; on the Claude lane the client marks between 83 and 93 percent of the invocations of every class as failed, so that file cannot show a breakage specific to a tool. Whether the failures fell harder on one arm is not established: of the three Gemini blocks that carry both arms, one shows a \texttt{git} gap of 31.0 points, and the other two show gaps of 3.3 and 2.4 points the other way, with every \arm{plain} \texttt{git} invocation failing in both. The difference is declared as a confound and not read as an effect, and no census of the existing streams can settle it.\src{evidence/nr-command-hazard-2026-09-22/README.md; evidence/nr-command-hazard-2026-09-22/agy-out-all.txt; evidence/nr-command-hazard-2026-09-22/agy-out-b4.txt; evidence/nr-command-hazard-2026-09-22/agy-out-l5.txt; evidence/nr-command-hazard-2026-09-22/agy-out-s3.txt; evidence/nr-command-hazard-2026-09-22/claude-pooled-2026-09-22.tsv} And a cell's fence denied the cell roots that existed when it was rendered, so a cell root created during a run, or a sibling created inside the cell's own root, was not covered by that run's fence. Over the 520 cells of record that the correctness map names, 48 had another root born during their run and 17 a sibling born inside their own root; no cell's own record names such a root outside its fence, and every one of the 244 Claude cells of record was scanned for tool inputs that reach another root, each reading zero. That is evidence of what the cells did and not of what they could reach. The Gemini lane has no such scan, so its evidence is the name census alone, and a name census cannot see a wildcard read or a directory listing that never prints the name. A root deleted before the census is invisible to it. Of the 48 and the 17, 41 and 13 did not have that root on their own deny list, and those two counts may lean high. The map's 94 count rows and 14 condition rows name no cell and are not measured this way, and no cell is re-scored.\src{harness/systems-v3/CENSUS-full-matrix-2026-09-14.md, addendum 37; evidence/midcell-gap-2026-10-17/README.md}

\subsection{Token cost over the matrix, a descriptive reading}
\label{sec:matrixtokens}

The census counts conditions and the results of record say what each can carry; neither prints what the conditions cost. Tables~T1 to~T4 do, one cell per condition: the median total tokens of the condition's cells of record, for the 80 greenfield conditions, the 80 brownfield and the 40 under specification change, then the \arm{salt-diet} to \arm{plain} ratio in each greenfield row. Tokens are the one quantity all four models' meters record; Section~\ref{sec:matrixdollars} prices and times the same cells. The reading was registered on 2026-09-27, after every cell had run, and is declared post hoc. The registration fixes the cell figure, the median, the bound marks and the ratio and sign-count rules, and that no number is recovered by hand. It was committed before any map of the cells existed, and its first addendum after a census that located each cell's source and computed no median; both precede the instrument's first run, so the reading was fixed before any median was computed.\src{harness/systems-v3/REGISTRATION-descriptive-tables-v2-2026-09-27.md, sections D1 to D6 and addenda 1 and 2, at commits 368d925, ad1b6b2 and a9d938e; harness/systems-v3/tables\_v2.py and its map at commit 9f9a5d9; the result file at commit 643cac0} Every cell below is copied from one result file, printed by that instrument, which names the tracked source of every cell's figure. The file was printed twice: a non-author read of the first printing found four \arm{salt-diet} cells, each cut off by a deadline, left out of their two conditions although the registration lets no cell be dropped, and the file was reprinted with every cell of record; two medians became lower bounds and no ratio changed sign.\src{harness/systems-v3/RESULT-descriptive-tables-v2-2026-09-27.md, the header and the per-cell section; harness/systems-v3/tables\_v2.py; paper/descriptive\_tables.py, which copies the tables into this file and with --check refuses if they differ}

\paragraph{A dated map, and two populations on ten conditions.} Tables~T1 to~T4, \$1 to~\$4, W1 to~W4 and~S1 read the cell map of 2026-09-27. Two days later, step g and the top-up score receipt took ten conditions to $n=3$ cells of record: one Sonnet greenfield condition, three Pro greenfield conditions and the six Opus specification-change conditions. On those ten conditions these tables read fewer cells than Tables~P1 to~P4 and the census do, and Table~S1 prints $n$ below 3 for the six Opus rows. The census keeps these tables as a dated reading, which applied its registered rule correctly, and does not change them. A reader comparing a row here with a row of Tables~P1 to~P4 on one of those ten conditions is comparing two populations.\src{harness/systems-v3/CENSUS-full-matrix-2026-09-14.md, addendum 30 section AF2 and addenda 31 and 32; harness/systems-v3/RESULT-cost-tables-v3-2026-09-29.md, its header; harness/systems-v3/CELLMAP-descriptive-tables-v2-2026-09-27.tsv; harness/systems-v3/CELLMAP-correctness-tables-v3-2026-10-01.tsv}

\paragraph{Version 4 re-meters the Gemini rows.} After version~3 was posted, a decoder for the Gemini client's own conversation records showed that the lane's meter reads a cell's parent conversation only, and none of the sub-agent conversations the client starts from it. The decoder reproduces one cell's parent-conversation meter exactly, which is its positive control; over all the metered conversations its figures run 16.3\% below the meter in the \arm{plain} arm and 25.9\% below in the \arm{salt-diet} arm, a gap whose source, decoder or meter, is not established; if it is the decoder's, the sub-agent figures below are floors on that account too. Over the 272 Gemini cells of record, the unmetered sub-agent tokens were 434{,}359{,}169 against a metered 2{,}858{,}155{,}649 in the \arm{salt-diet} arm, in 76 of its 137 cells, and 884{,}196 against 566{,}780{,}515 in the \arm{plain} arm, in 2 of its 135: the uncounted tokens add at least 15.2\% to the \arm{salt-diet} arm's metered total (13.2\% of its total with them) and 0.2\% to the \arm{plain} arm's, so every published Gemini ratio understated the \arm{salt-diet} arm's cost, the direction that flatters the method under test.\src{evidence/o57-agy-subagent-unmetered-2026-10-13/README.md and unmetered2.out; harness/systems-v3/RESULT-cost-tables-v3-2026-09-29.md, the addendum of 2026-10-13 (O57)} Tables~T1 to~T4 in this version are printed by the same instrument with each Gemini cell's sub-agent tokens added to its figure. Sixty level-8 cells, in both arms, kept no record of their first phase's conversations, so their figures are floors and print $\geq$ by the registered rule; every Gemini specification-change ratio that can be printed therefore has both sides bounded and is printed in parentheses, unsigned. Without the addition the instrument reproduces version~3's result file byte for byte below its head line, and no Claude row changes.\src{harness/systems-v3/RESULT-descriptive-tables-v2-O57-remeter-2026-10-13.md, the head lines; harness/systems-v3/DIFF-O57-remeter-2026-10-13.md; harness/systems-v3/tables\_v2.py --add-T with evidence/o57-agy-subagent-unmetered-2026-10-13/addT.tsv} What moved: the median of the Pro ratios rises from 10.12 to 13.53 under no treatment and from 10.39 to 14.69 under the statement treatment, and the Flash rows move little. What did not: of the 60 ratio cells, no ratio whose sign was fixed changes sign; nine that version~3 signed above 1, all under specification change, are now unsigned because both sides are floors, and none becomes signed. The re-meter adds tokens and prices none of them: the Gemini dollar figures of Section~\ref{sec:matrixdollars} are not re-priced and carry the same exclusion.\src{harness/systems-v3/DIFF-O57-remeter-2026-10-13.md, section T4 and its sign counts; harness/systems-v3/RESULT-descriptive-tables-v2-O57-remeter-2026-10-13.md, section T4 and its sign counts}

The registered check did not close, and the tables print why. Of the 181 \textsc{done} conditions, 171 carry a median; 10 print \emph{unmeasured}, because some of their cells have no tracked token figure: nine Flash greenfield conditions whose cells come from one addendum of the level-5 wave (\texttt{Crc32}, \texttt{FreeList} and \texttt{LRU} under the statement treatment in both arms, \texttt{LRU} \arm{salt-diet} and both \texttt{Paxos} arms under no treatment), and Pro \texttt{Crc32} \arm{salt-diet} under specification change, two of whose three cells lack one. Their outcomes stand in their results of record; only their token medians are missing, and none was reconstructed.\src{harness/systems-v3/RESULT-descriptive-tables-v2-O57-remeter-2026-10-13.md, the check line and the section on the 10 unmeasured conditions, as in version~3's RESULT-descriptive-tables-v2-2026-09-27.md; harness/systems-v3/REGISTRATION-descriptive-tables-v2-2026-09-27.md, addendum 1}

Three limits were stated before any number was read. A token ratio describes cost and says nothing of quality, so no row says the method helps or hurts. A $\geq$ median is a floor, set by the uniform cost cap, a meter under-read, a registered deadline or, in version~4, a Gemini first phase whose sub-agent record was lost, and the cap binds the \arm{salt-diet} arm more often, so a bounded numerator makes its ratio an understatement. And a median of three cells is one cell's figure. The Gemini specification-change cells come from a level registered as sign-only with no magnitude, and nothing here adds a magnitude claim to it; the medians and ratios describe the record and are not an estimate of an effect.\src{harness/systems-v3/REGISTRATION-descriptive-tables-v2-2026-09-27.md, sections D6 and D4; harness/systems-v3/RESULT-descriptive-tables-v2-O57-remeter-2026-10-13.md, section T4 and its sign counts; harness/systems-v3/AMENDMENT-gemini-level8-2026-09-16.md section M9}

\begin{center}\begin{minipage}{\textwidth}\centering
{\scriptsize\setlength{\tabcolsep}{4pt}
\begin{tabular}{llrrrr}
\toprule
 & & \multicolumn{2}{c}{no treatment} & \multicolumn{2}{c}{statement} \\
\cmidrule(lr){3-4}\cmidrule(lr){5-6}
model & problem & \arm{plain} & \arm{salt-diet} & \arm{plain} & \arm{salt-diet} \\
\midrule
\texttt{claude-opus-5} & \texttt{Crc32} & $\geq$\,6{,}113{,}449 & $\geq$\,7{,}397{,}410 & 6{,}324{,}774 & 5{,}927{,}487 \\
 & \texttt{LRU} & $\geq$\,8{,}155{,}664 & 13{,}297{,}929 & 8{,}352{,}122 & 16{,}265{,}999 \\
 & \texttt{FreeList} & $\geq$\,11{,}914{,}579 & $\geq$\,46{,}242{,}329 & 19{,}470{,}666 & 32{,}830{,}752 \\
 & \texttt{LZW} & 11{,}449{,}608 & 22{,}063{,}541 & $\geq$\,8{,}859{,}916 & $\geq$\,21{,}418{,}493 \\
 & \texttt{Paxos} & $\geq$\,17{,}684{,}598 & 49{,}725{,}838 & --- & --- \\
\texttt{claude-sonnet-5} & \texttt{Crc32} & 2{,}042{,}023 & 12{,}950{,}272 & 2{,}181{,}020 & 11{,}335{,}722 \\
 & \texttt{LRU} & 2{,}311{,}374 & 16{,}469{,}774 & 2{,}359{,}212 & 20{,}783{,}719 \\
 & \texttt{FreeList} & 4{,}448{,}454 & $\geq$\,135{,}942{,}796 & 4{,}524{,}503 & 123{,}975{,}714 \\
 & \texttt{LZW} & 2{,}373{,}486 & 43{,}839{,}178 & 3{,}555{,}507 & 30{,}210{,}414 \\
 & \texttt{Paxos} & 4{,}996{,}639 & $\geq$\,123{,}495{,}570 & --- & --- \\
\texttt{gemini-3.1-pro-high} & \texttt{Crc32} & 861{,}120 & 9{,}675{,}734 & 1{,}222{,}527 & $\geq$\,17{,}700{,}944 \\
 & \texttt{LRU} & 1{,}466{,}321 & 11{,}728{,}188 & 1{,}080{,}062 & 11{,}097{,}896 \\
 & \texttt{FreeList} & 1{,}290{,}763 & 18{,}111{,}487 & 1{,}699{,}106 & $\geq$\,30{,}037{,}849 \\
 & \texttt{LZW} & 1{,}232{,}089 & 16{,}675{,}231 & 1{,}598{,}032 & 23{,}814{,}589 \\
 & \texttt{Paxos} & 1{,}176{,}979 & $\geq$\,19{,}688{,}273 & --- & --- \\
\texttt{gemini-3.8-flash-high} & \texttt{Crc32} & 3{,}993{,}025 & 12{,}112{,}123 & \emph{unmeasured} & \emph{unmeasured} \\
 & \texttt{LRU} & 5{,}413{,}078 & \emph{unmeasured} & \emph{unmeasured} & \emph{unmeasured} \\
 & \texttt{FreeList} & 6{,}495{,}950 & 24{,}187{,}669 & \emph{unmeasured} & \emph{unmeasured} \\
 & \texttt{LZW} & 4{,}244{,}069 & $\geq$\,23{,}941{,}795 & 5{,}147{,}496 & 18{,}215{,}862 \\
 & \texttt{Paxos} & \emph{unmeasured} & \emph{unmeasured} & --- & --- \\
\bottomrule
\end{tabular}}\\[3pt]
\parbox{0.92\textwidth}{\small\textbf{Table T1.} Greenfield. Median total tokens per condition, over the condition's cells of record; $n=3$ for most conditions, and the $n$ of each is printed in the result file. A descriptive reading over the complete matrix: no test, and no verdict on the arms. Compare the arms within a row, never across lanes: the Claude lane counts input, cache writes, cache reads and output from its session meter, and the Gemini lane counts input, output and cache reads as its vendor reports them. A Claude cell's figure sums every model its session was served, including the subagents it spawned, so many Opus cells' figures include \texttt{claude-sonnet-5} subagent tokens; the row names the subject model, and in a Sonnet cell every subagent is Sonnet. $\geq$ marks a floor (a cell at or below the median stopped at the cost cap, carries a meter under-read, or was cut off by a registered deadline); the cap binds the \arm{salt-diet} arm more often. --- marks an inexpressible condition (\texttt{Paxos} under the statement treatment); \emph{unmeasured} a \textsc{done} condition some of whose cells have no tracked token figure.}
\src{harness/systems-v3/RESULT-descriptive-tables-v2-2026-09-27.md, section T1; harness/systems-v3-analysis/RESULT-tokens-table-all-roots-2026-09-10.tsv, the models column; harness/systems-v3-analysis/tokens\_table.py, which sums the head and subagent lanes; harness/systems-v3/AMENDMENT-claude-lane-B-2026-09-16.md, addendum 2 item A2.2}
\end{minipage}\end{center}

\begin{center}\begin{minipage}{\textwidth}\centering
{\scriptsize\setlength{\tabcolsep}{4pt}
\begin{tabular}{llrrrr}
\toprule
 & & \multicolumn{2}{c}{no treatment} & \multicolumn{2}{c}{statement} \\
\cmidrule(lr){3-4}\cmidrule(lr){5-6}
model & problem & \arm{plain} & \arm{salt-diet} & \arm{plain} & \arm{salt-diet} \\
\midrule
\texttt{claude-opus-5} & \texttt{Crc32} & 6{,}822{,}910 & 7{,}327{,}378 & 10{,}004{,}941 & 7{,}250{,}392 \\
 & \texttt{LRU} & 7{,}807{,}170 & 16{,}927{,}118 & 9{,}747{,}753 & 16{,}074{,}704 \\
 & \texttt{FreeList} & 20{,}264{,}692 & $\geq$\,41{,}750{,}816 & 18{,}887{,}102 & 35{,}880{,}068 \\
 & \texttt{LZW} & 7{,}003{,}493 & 15{,}647{,}923 & 12{,}724{,}584 & 17{,}558{,}716 \\
 & \texttt{Paxos} & 11{,}771{,}721 & 42{,}126{,}869 & --- & --- \\
\texttt{claude-sonnet-5} & \texttt{Crc32} & 1{,}844{,}699 & 10{,}573{,}448 & 2{,}055{,}677 & 10{,}051{,}058 \\
 & \texttt{LRU} & 2{,}030{,}381 & 23{,}514{,}670 & 2{,}113{,}686 & 14{,}564{,}237 \\
 & \texttt{FreeList} & 5{,}289{,}799 & $\geq$\,129{,}225{,}268 & 5{,}160{,}232 & $\geq$\,140{,}700{,}178 \\
 & \texttt{LZW} & 3{,}115{,}443 & 59{,}322{,}626 & 2{,}641{,}673 & 58{,}601{,}805 \\
 & \texttt{Paxos} & 6{,}151{,}825 & $\geq$\,154{,}795{,}624 & --- & --- \\
\texttt{gemini-3.1-pro-high} & \texttt{Crc32} & 1{,}125{,}906 & 10{,}407{,}416 & 1{,}550{,}082 & 8{,}796{,}037 \\
 & \texttt{LRU} & 938{,}991 & 8{,}687{,}059 & 1{,}355{,}238 & $\geq$\,16{,}581{,}493 \\
 & \texttt{FreeList} & 1{,}949{,}814 & 17{,}298{,}225 & 1{,}822{,}259 & 34{,}282{,}120 \\
 & \texttt{LZW} & 1{,}523{,}409 & $\geq$\,35{,}151{,}779 & 1{,}468{,}162 & $\geq$\,30{,}795{,}640 \\
 & \texttt{Paxos} & 1{,}676{,}061 & 17{,}401{,}238 & --- & --- \\
\texttt{gemini-3.8-flash-high} & \texttt{Crc32} & 4{,}325{,}246 & 9{,}683{,}589 & 5{,}376{,}845 & 13{,}141{,}849 \\
 & \texttt{LRU} & 3{,}804{,}355 & 14{,}937{,}746 & 4{,}034{,}404 & 17{,}138{,}719 \\
 & \texttt{FreeList} & 5{,}334{,}389 & 21{,}228{,}941 & 7{,}756{,}505 & 20{,}084{,}774 \\
 & \texttt{LZW} & 6{,}555{,}430 & 18{,}064{,}205 & 5{,}919{,}649 & 13{,}256{,}712 \\
 & \texttt{Paxos} & 7{,}992{,}405 & 18{,}713{,}325 & --- & --- \\
\bottomrule
\end{tabular}}\\[3pt]
\parbox{0.92\textwidth}{\small\textbf{Table T2.} Brownfield. Median total tokens per condition, over the condition's cells of record; $n=3$ for most conditions, and the $n$ of each is printed in the result file. A descriptive reading over the complete matrix: no test, and no verdict on the arms. Compare the arms within a row, never across lanes: the Claude lane counts input, cache writes, cache reads and output from its session meter, and the Gemini lane counts input, output and cache reads as its vendor reports them. A Claude cell's figure sums every model its session was served, including the subagents it spawned, so many Opus cells' figures include \texttt{claude-sonnet-5} subagent tokens; the row names the subject model, and in a Sonnet cell every subagent is Sonnet. $\geq$ marks a floor (a cell at or below the median stopped at the cost cap, carries a meter under-read, or was cut off by a registered deadline); the cap binds the \arm{salt-diet} arm more often. --- marks an inexpressible condition (\texttt{Paxos} under the statement treatment); \emph{unmeasured} a \textsc{done} condition some of whose cells have no tracked token figure.}
\src{harness/systems-v3/RESULT-descriptive-tables-v2-2026-09-27.md, section T2; harness/systems-v3-analysis/RESULT-tokens-table-all-roots-2026-09-10.tsv, the models column; harness/systems-v3-analysis/tokens\_table.py, which sums the head and subagent lanes; harness/systems-v3/AMENDMENT-claude-lane-B-2026-09-16.md, addendum 2 item A2.2}
\end{minipage}\end{center}

\begin{center}\begin{minipage}{\textwidth}\centering
{\scriptsize\setlength{\tabcolsep}{4pt}
\begin{tabular}{llrr}
\toprule
model & problem & \arm{plain} & \arm{salt-diet} \\
\midrule
\texttt{claude-opus-5} & \texttt{Crc32} & $\geq$\,22{,}634{,}955 & $\geq$\,17{,}114{,}730 \\
 & \texttt{LRU} & $\geq$\,25{,}032{,}923 & 27{,}518{,}804 \\
 & \texttt{FreeList} & $\geq$\,30{,}217{,}718 & 65{,}579{,}395 \\
 & \texttt{LZW} & 23{,}086{,}951 & $\geq$\,48{,}465{,}177 \\
 & \texttt{Paxos} & $\geq$\,29{,}599{,}611 & $\geq$\,60{,}070{,}155 \\
\texttt{claude-sonnet-5} & \texttt{Crc32} & 3{,}986{,}797 & 26{,}391{,}549 \\
 & \texttt{LRU} & 3{,}311{,}062 & 31{,}561{,}498 \\
 & \texttt{FreeList} & 12{,}827{,}827 & \emph{declared} \\
 & \texttt{LZW} & 9{,}016{,}104 & \emph{declared} \\
 & \texttt{Paxos} & 19{,}088{,}970 & \emph{declared} \\
\texttt{gemini-3.1-pro-high} & \texttt{Crc32} & $\geq$\,2{,}214{,}755 & \emph{unmeasured} \\
 & \texttt{LRU} & $\geq$\,2{,}506{,}278 & $\geq$\,15{,}320{,}783 \\
 & \texttt{FreeList} & $\geq$\,3{,}357{,}987 & $\geq$\,19{,}724{,}118 \\
 & \texttt{LZW} & $\geq$\,2{,}828{,}970 & $\geq$\,33{,}881{,}993 \\
 & \texttt{Paxos} & $\geq$\,2{,}986{,}585 & $\geq$\,20{,}827{,}763 \\
\texttt{gemini-3.8-flash-high} & \texttt{Crc32} & $\geq$\,8{,}827{,}684 & $\geq$\,20{,}838{,}181 \\
 & \texttt{LRU} & $\geq$\,9{,}343{,}447 & $\geq$\,26{,}945{,}879 \\
 & \texttt{FreeList} & $\geq$\,14{,}103{,}453 & $\geq$\,32{,}423{,}306 \\
 & \texttt{LZW} & $\geq$\,9{,}189{,}665 & $\geq$\,36{,}488{,}484 \\
 & \texttt{Paxos} & $\geq$\,13{,}889{,}277 & $\geq$\,127{,}713{,}529 \\
\bottomrule
\end{tabular}}\\[3pt]
\parbox{0.92\textwidth}{\small\textbf{Table T3.} Specification change, greenfield only. A cell's figure is its first phase plus its second; for the Opus cells the first phase is the reused landing of the matrix of Section~\ref{sec:s3}. \emph{declared} marks a condition unreached at the cap, which carries no median. The other marks and limits are those of Table~T1.}
\src{harness/systems-v3/RESULT-descriptive-tables-v2-2026-09-27.md, section T3 and the header; harness/systems-v3/REGISTRATION-descriptive-tables-v2-2026-09-27.md, addendum 1 item A1.1}
\end{minipage}\end{center}

\begin{center}\begin{minipage}{\textwidth}\centering
{\scriptsize\setlength{\tabcolsep}{4pt}
\begin{tabular}{llrrr}
\toprule
model & problem & no treatment & statement & spec.\ change \\
\midrule
\texttt{claude-opus-5} & \texttt{Crc32} & (1.21) & 0.94 & (0.76) \\
 & \texttt{LRU} & $\leq$\,1.63 & 1.95 & $\leq$\,1.10 \\
 & \texttt{FreeList} & (3.88) & 1.69 & $\leq$\,2.17 \\
 & \texttt{LZW} & 1.93 & (2.42) & $\geq$\,2.10 \\
 & \texttt{Paxos} & $\leq$\,2.81 & --- & (2.03) \\
\texttt{claude-sonnet-5} & \texttt{Crc32} & 6.34 & 5.20 & 6.62 \\
 & \texttt{LRU} & 7.13 & 8.81 & 9.53 \\
 & \texttt{FreeList} & $\geq$\,30.56 & 27.40 & --- \\
 & \texttt{LZW} & 18.47 & 8.50 & --- \\
 & \texttt{Paxos} & $\geq$\,24.72 & --- & --- \\
\texttt{gemini-3.1-pro-high} & \texttt{Crc32} & 11.24 & $\geq$\,14.48 & --- \\
 & \texttt{LRU} & 8.00 & 10.28 & (6.11) \\
 & \texttt{FreeList} & 14.03 & $\geq$\,17.68 & (5.87) \\
 & \texttt{LZW} & 13.53 & 14.90 & (11.98) \\
 & \texttt{Paxos} & $\geq$\,16.73 & --- & (6.97) \\
\texttt{gemini-3.8-flash-high} & \texttt{Crc32} & 3.03 & --- & (2.36) \\
 & \texttt{LRU} & --- & --- & (2.88) \\
 & \texttt{FreeList} & 3.72 & --- & (2.30) \\
 & \texttt{LZW} & $\geq$\,5.64 & 3.54 & (3.97) \\
 & \texttt{Paxos} & --- & --- & (9.20) \\
\bottomrule
\end{tabular}\\[8pt]
\begin{tabular}{llrlrr}
\toprule
model & treatment & $k$ of $m$ $>1$ & indeterminate & median & bounded \\
\midrule
\texttt{claude-opus-5} & bare & 1 of 1 & Crc32, LRU, FreeList, Paxos & 1.93 & 4 \\
 & statement & 2 of 3 & LZW & 1.82 & 1 \\
 & spec-change & 1 of 1 & Crc32, LRU, FreeList, Paxos & 2.03 & 5 \\
\texttt{claude-sonnet-5} & bare & 5 of 5 & none & 18.47 & 2 \\
 & statement & 4 of 4 & none & 8.65 & 0 \\
 & spec-change & 2 of 2 & none & 8.08 & 0 \\
\texttt{gemini-3.1-pro-high} & bare & 5 of 5 & none & 13.53 & 1 \\
 & statement & 4 of 4 & none & 14.69 & 2 \\
 & spec-change & 0 of 0 & LRU, FreeList, LZW, Paxos & 6.54 & 4 \\
\texttt{gemini-3.8-flash-high} & bare & 3 of 3 & none & 3.72 & 1 \\
 & statement & 1 of 1 & none & 3.54 & 0 \\
 & spec-change & 0 of 0 & Crc32, LRU, FreeList, LZW, Paxos & 2.88 & 5 \\
\bottomrule
\end{tabular}}\\[3pt]
\parbox{0.92\textwidth}{\small\textbf{Table T4.} Top: the \arm{salt-diet} median divided by the \arm{plain} median in the same row of Tables~T1 and~T3, greenfield, the one task form all three treatments share. A $\geq$ numerator makes the ratio $\geq$ and a $\geq$ denominator makes it $\leq$; a ratio in parentheses has both sides bounded and is printed but not signed; --- means no ratio, because a side is inexpressible, declared or unmeasured. Bottom: per model and treatment (\emph{bare} is no treatment), $k$ of the $m$ ratios whose sign is fixed that exceed $1$; the bounded ratios whose sign the bound cannot fix, counted in neither $k$ nor $m-k$; the median of all the model's ratios under that treatment, bounded ones at their printed value; and how many of those are bounded. No $p$-value is given: the registered test is Section~\ref{sec:s3}'s, on one model, and a test across the others would be one chosen after seeing the data. The brownfield ratios, by the same rule, are in the result file.}
\src{harness/systems-v3/RESULT-descriptive-tables-v2-O57-remeter-2026-10-13.md, section T4, its sign counts and the brownfield ratios; harness/systems-v3/REGISTRATION-descriptive-tables-v2-2026-09-27.md, section D4}
\end{minipage}\end{center}

\subsection{Dollars and wall time over the matrix, a descriptive reading}
\label{sec:matrixdollars}

Tokens are not comparable across models: each vendor counts its own categories, and a token of one model does not cost what a token of another does. Tables~\$1 to~\$4 and~W1 to~W4 print the same 200 conditions in two further currencies, modelled list-price dollars and wall seconds, with the shape, the median, the bound marks and the ratio and sign-count rules of Tables~T1 to~T4. The reading was registered on 2026-09-29, after every cell had run and after the raw extractions it reads had been printed with their aggregate counts, and it is declared post hoc; the registration states what had been seen, and no median, ratio or per-condition figure had been computed. A non-author read of the registration found one rule that would have overstated every Gemini dollar, by pricing the model's thinking tokens a second time on top of an output count that already includes them; a test over every recorded Gemini request confirmed the inclusion (in 36{,}510 requests, thinking never exceeds output), and the rule was corrected by addendum before the instrument ran. A dry run to a scratch file, which read the checks and never a table cell, found one missing rule for cells whose second phase ran in a copy of their first, and it was added the same way.\src{harness/systems-v3/REGISTRATION-cost-tables-v3-2026-09-29.md, sections V1 to V6 and addenda 1 to 3; evidence/v3-cost-tables-2026-09-28/agy-field-tests.out; harness/systems-v3/RESULT-cost-tables-v3-2026-09-29.md, the header; harness/systems-v3/tables\_v3.py; paper/cost\_tables.py, which copies the tables into this file and with --check refuses if they differ}

\paragraph{How a dollar is modelled.} On the Claude lane the figure is the metered cost of each cell at the list rates the harness has carried since 2026-09-05, re-read against the vendor's page on 2026-09-29 and agreeing on every column; where a tracked row already carries a cost, that is the figure of record and a re-run of the meter is printed beside it. The re-run uses the same meter at the same rates, so it is a reproduction and not a second method; it differs by a cent or more on 46 of 262 cell rows, all under specification change, and each difference has a cause shown at the object: 23 are cells whose tracked figure separates the harness's own sandbox probe from the phases, as their token figures do, and 23 are cells whose second phase ran in a copy, where the re-run metered the first phase alone. On the Gemini lane this is the first dollar figure. Each request is priced at the tier its prompt selects, prompt meaning input plus cache reads, from the per-request usage the client stream records; where a phase's per-request records are incomplete, a Flash phase is priced from the meter's totals, because Flash has one tier, and a Pro phase is \emph{unmeasured}, because a lost request cannot be placed in a tier. No Pro request in these data reaches the upper tier: the largest Pro prompt is 144{,}717 tokens. Eight cells are priced from per-request records larger than the meter's own fold and are flagged in the result file. A cell stopped at the cost cap enters at the larger of its metered cost and the cap its end marker names, marked $\geq$, because the watcher records an overrun without clipping it and the second-phase cap is cumulative, so the cap alone would print a figure below money the cell had already spent.\src{harness/systems-v3/REGISTRATION-cost-tables-v3-2026-09-29.md, section V2 and addendum 1 item A1.3, addendum 2 items A2.1 to A2.4, addendum 3 item A3.2; harness/systems-v3/RESULT-cost-tables-v3-2026-09-29.md, the header, the section on rule 1's reproduction and the section on the capped cells}

\paragraph{The checks.} In dollars, 178 conditions carry a median, 16 are inexpressible, 3 declared and 3 \emph{unmeasured}: three Pro \arm{salt-diet} conditions, one of whose cells each lost per-request records (\texttt{FreeList} under the statement treatment, \texttt{Crc32} brownfield under no treatment and \texttt{FreeList} under specification change). In wall seconds, 181 carry a median, 16 are inexpressible and 3 declared, and none is \emph{unmeasured}. Both sum to 200. The reading's one second method, which re-reads the wall of each Gemini specification-change cell's first phase from the record that level wrote when it set the phase aside, disagreed on no cell. The ten \textsc{done} conditions that the token tables print \emph{unmeasured} carry dollar and wall figures here, because this reading prices and times each of their Gemini cells from its own stream and meter rather than from a tracked token row; the token medians stay as printed.\src{harness/systems-v3/RESULT-cost-tables-v3-2026-09-29.md, the check lines and the section on unmeasured cells; harness/systems-v3/REGISTRATION-cost-tables-v3-2026-09-29.md, addendum 1 item A1.1; harness/systems-v3/RESULT-descriptive-tables-v2-2026-09-27.md, the check line}

\paragraph{What the tables show, and within what.} Read within a row, the \arm{salt-diet} arm costs more than the \arm{plain} arm in dollars and takes longer in wall time almost everywhere a sign can be fixed. In dollars, every Sonnet, Pro and Flash ratio whose sign is fixed exceeds 1, under all three treatments; the Opus rows fix four signs in all, because the cap bounds both arms, and three of the four exceed 1; the fourth is \texttt{Crc32} under the statement treatment at 0.94. The medians of the per-treatment ratios run from 2.26 to 3.56 for Flash, 4.50 to 7.98 for Pro, 5.62 to 8.48 for Sonnet and 1.37 to 1.64 for Opus. In wall time the pattern repeats: every fixed sign for Sonnet, Pro and Flash exceeds 1, and Opus fixes four signs across its three treatments, none of them under no treatment, three above 1, the fourth a bounded ratio at or below 0.97 on \texttt{LRU} under specification change. Within the Claude lane the models also separate: the \arm{plain} Opus greenfield cells under no treatment cost between \$6.21 and \$16.78 at the median, all but one a floor, against \$1.14 to \$3.38 for the same Sonnet conditions. Each of these is a description of cost and not of quality; the arm that costs more is not thereby better or worse.\src{harness/systems-v3/RESULT-cost-tables-v3-2026-09-29.md, tables \$1, \$4 and its sign counts, W4 and its sign counts}

\paragraph{What the tables cannot carry.} Four limits were stated before any number was read, and one more is inherited. A dollar ratio, like a token ratio, describes cost and says nothing of quality. Every $\geq$ is a censoring that binds the \arm{salt-diet} arm more often, so a bounded numerator makes a ratio an understatement. Wall time describes these runs on this box at those hours, and the records that name the box's concurrency are cited in the result file and not extracted per cell. A median of three cells is one cell's figure. And each Gemini level's registration, frozen before its first call, excludes a comparison with the Claude lane in dollars or in tokens, because the two lanes differ in client, fence, caps and watcher and, under specification change, in how the first phase was produced. This reading was registered afterwards and does not lift that exclusion, so the tables print both lanes and the text compares arms within a row and models within a lane, never a Gemini row with a Claude row. Version~4 adds one more, found after the reading ran: the Gemini dollars price the parent conversation's requests and none of its sub-agents', so the \arm{salt-diet} arm's Gemini dollars and dollar ratios are understated, by the margin Section~\ref{sec:matrixtokens} measures in tokens; they are not re-priced here.\src{harness/systems-v3/RESULT-cost-tables-v3-2026-09-29.md, the addendum of 2026-10-13 (O57)} \src{harness/systems-v3/REGISTRATION-cost-tables-v3-2026-09-29.md, section V6; harness/systems-v3/AMENDMENT-gemini-flash-level5-2026-09-14.md, section F6 item 4; harness/systems-v3/AMENDMENT-gemini-level6-2026-09-16.md, section H7 item 3; harness/systems-v3/AMENDMENT-gemini-level7-2026-09-16.md, section K8 item 3; harness/systems-v3/AMENDMENT-gemini-level8-2026-09-16.md, sections M6.2 and M9 item 3}

\paragraph{For scale.} The agent-built verification of the xv6 kernel reports 93 days of development, 2{,}062 hours of Claude running time summed over its sessions, less than \$3{,}000 in Claude subscriptions, and 1{,}184{,}763 lines of Rocq written by agents~\cite[Sections~11 and~12.1]{kaashoek2026xv6}. Those are a subscription spend and a summed session time for one development directed throughout by its authors; the figures above are modelled list-price dollars and wall seconds per cell of a sealed run. They are different units of different things, set side by side for scale and never as a ratio.

\paragraph{Where an Opus cell's money goes.} An Opus cell spawns subagents, and the session meter serves them \texttt{claude-sonnet-5}, so part of every Opus figure above is Sonnet's. Table~S1 in Appendix~\ref{app:opussplit} prints, per Opus condition, the median share of tokens and of dollars spent outside the head session. It separates the arms on all 23 problem, task-form and treatment pairs: the \arm{plain} arm's share is 15.5 to 36.7 percent of tokens and 22.3 to 40.0 percent of dollars, the \arm{salt-diet} arm's 1.4 to 11.9 percent and 4.2 to 22.6 percent, lower in both currencies on every pair. Under the \arm{salt-diet} arm the head session does most of the work itself, so a dollar ratio between the Opus arms also compares two mixes of models.\src{harness/systems-v3/RESULT-cost-tables-v3-2026-09-29.md, section S1}

\begin{center}\begin{minipage}{\textwidth}\centering
{\scriptsize\setlength{\tabcolsep}{4pt}
\begin{tabular}{llrrrr}
\toprule
 & & \multicolumn{2}{c}{no treatment} & \multicolumn{2}{c}{statement} \\
\cmidrule(lr){3-4}\cmidrule(lr){5-6}
model & problem & \arm{plain} & \arm{salt-diet} & \arm{plain} & \arm{salt-diet} \\
\midrule
\texttt{claude-opus-5} & \texttt{Crc32} & $\geq$\,\$6.21 & $\geq$\,\$7.21 & \$6.66 & \$6.25 \\
 & \texttt{LRU} & $\geq$\,\$9.73 & \$11.21 & \$9.36 & \$14.24 \\
 & \texttt{FreeList} & $\geq$\,\$13.02 & $\geq$\,\$37.60 & $\geq$\,\$18.26 & \$27.84 \\
 & \texttt{LZW} & \$13.95 & \$19.18 & $\geq$\,\$10.24 & $\geq$\,\$18.54 \\
 & \texttt{Paxos} & $\geq$\,\$16.78 & \$37.65 & --- & --- \\
\texttt{claude-sonnet-5} & \texttt{Crc32} & \$1.14 & \$4.79 & \$0.98 & \$4.16 \\
 & \texttt{LRU} & $\geq$\,\$1.23 & $\geq$\,\$5.75 & \$1.38 & $\geq$\,\$6.67 \\
 & \texttt{FreeList} & \$2.92 & $\geq$\,\$37.72 & \$2.81 & $\geq$\,\$35.08 \\
 & \texttt{LZW} & \$1.70 & \$14.39 & \$1.89 & $\geq$\,\$12.36 \\
 & \texttt{Paxos} & \$3.38 & $\geq$\,\$37.61 & --- & --- \\
\texttt{gemini-3.1-pro-high} & \texttt{Crc32} & \$0.65 & \$5.35 & \$0.80 & $\geq$\,\$6.38 \\
 & \texttt{LRU} & \$1.21 & \$4.42 & \$0.72 & \$4.58 \\
 & \texttt{FreeList} & \$0.83 & \$4.91 & \$0.94 & \emph{unmeasured} \\
 & \texttt{LZW} & \$0.73 & \$5.27 & \$0.84 & \$7.29 \\
 & \texttt{Paxos} & \$0.69 & $\geq$\,\$6.84 & --- & --- \\
\texttt{gemini-3.8-flash-high} & \texttt{Crc32} & \$0.82 & \$2.15 & \$0.80 & \$2.95 \\
 & \texttt{LRU} & \$1.17 & \$2.78 & \$0.72 & \$2.61 \\
 & \texttt{FreeList} & \$1.18 & \$3.48 & \$1.03 & \$3.59 \\
 & \texttt{LZW} & \$0.89 & $\geq$\,\$3.81 & \$1.12 & \$3.08 \\
 & \texttt{Paxos} & \$1.29 & \$4.57 & --- & --- \\
\bottomrule
\end{tabular}}\\[3pt]
\parbox{0.92\textwidth}{\small\textbf{Table \$1.} Greenfield. Median modelled list-price US dollars per condition, over the condition's cells of record. A descriptive reading: no test, and no verdict on the arms. The dollars are a model and not an invoice, since every cell ran on a subscription: one list schedule per vendor, both read on 2026-09-29. A Claude cell prices every record at its served model's rates, so an Opus cell's \texttt{claude-sonnet-5} subagents are priced at Sonnet's; a Gemini cell prices each request at the tier its prompt selects, and context-caching storage is not priced. Compare the arms within a row and the models within a lane, never across lanes (Section~\ref{sec:matrixdollars}). $\geq$ marks a floor; a cell stopped at the cost cap enters at the larger of its metered cost and the cap. --- marks an inexpressible condition; \emph{unmeasured} a \textsc{done} condition one of whose Pro cells lost per-request records, so it cannot be put in a price tier.}
\src{harness/systems-v3/RESULT-cost-tables-v3-2026-09-29.md, section \$1}
\end{minipage}\end{center}
\begin{center}\begin{minipage}{\textwidth}\centering
{\scriptsize\setlength{\tabcolsep}{4pt}
\begin{tabular}{llrrrr}
\toprule
 & & \multicolumn{2}{c}{no treatment} & \multicolumn{2}{c}{statement} \\
\cmidrule(lr){3-4}\cmidrule(lr){5-6}
model & problem & \arm{plain} & \arm{salt-diet} & \arm{plain} & \arm{salt-diet} \\
\midrule
\texttt{claude-opus-5} & \texttt{Crc32} & \$6.70 & \$7.62 & $\geq$\,\$9.69 & $\geq$\,\$7.06 \\
 & \texttt{LRU} & \$8.17 & \$15.03 & \$10.36 & \$14.59 \\
 & \texttt{FreeList} & $\geq$\,\$18.66 & $\geq$\,\$37.30 & $\geq$\,\$19.57 & $\geq$\,\$29.62 \\
 & \texttt{LZW} & \$10.11 & \$15.99 & $\geq$\,\$13.47 & \$17.90 \\
 & \texttt{Paxos} & \$13.14 & \$32.22 & --- & --- \\
\texttt{claude-sonnet-5} & \texttt{Crc32} & $\geq$\,\$0.90 & \$4.11 & \$1.01 & \$3.94 \\
 & \texttt{LRU} & \$0.94 & \$7.74 & \$1.09 & $\geq$\,\$5.27 \\
 & \texttt{FreeList} & \$2.41 & $\geq$\,\$37.43 & \$2.34 & $\geq$\,\$37.60 \\
 & \texttt{LZW} & \$1.52 & $\geq$\,\$19.30 & \$1.79 & $\geq$\,\$18.14 \\
 & \texttt{Paxos} & $\geq$\,\$3.05 & $\geq$\,\$37.92 & --- & --- \\
\texttt{gemini-3.1-pro-high} & \texttt{Crc32} & \$0.67 & \emph{unmeasured} & \$0.80 & \$3.64 \\
 & \texttt{LRU} & \$0.54 & \$2.35 & \$0.76 & $\geq$\,\$4.97 \\
 & \texttt{FreeList} & \$1.03 & \$6.83 & \$0.92 & \$6.06 \\
 & \texttt{LZW} & \$0.86 & $\geq$\,\$13.81 & \$0.86 & $\geq$\,\$7.39 \\
 & \texttt{Paxos} & \$0.95 & $\geq$\,\$6.58 & --- & --- \\
\texttt{gemini-3.8-flash-high} & \texttt{Crc32} & \$0.78 & \$1.63 & \$0.86 & \$2.21 \\
 & \texttt{LRU} & \$0.77 & \$2.69 & \$0.79 & \$2.33 \\
 & \texttt{FreeList} & \$0.92 & \$3.26 & \$1.42 & \$2.93 \\
 & \texttt{LZW} & \$1.25 & \$2.77 & \$0.93 & \$1.91 \\
 & \texttt{Paxos} & \$1.32 & \$3.33 & --- & --- \\
\bottomrule
\end{tabular}}\\[3pt]
\parbox{0.92\textwidth}{\small\textbf{Table \$2.} Brownfield. The marks and limits are those of Table~\$1.}
\src{harness/systems-v3/RESULT-cost-tables-v3-2026-09-29.md, section \$2}
\end{minipage}\end{center}
\begin{center}\begin{minipage}{\textwidth}\centering
{\scriptsize\setlength{\tabcolsep}{4pt}
\begin{tabular}{llrr}
\toprule
model & problem & \arm{plain} & \arm{salt-diet} \\
\midrule
\texttt{claude-opus-5} & \texttt{Crc32} & $\geq$\,\$19.89 & $\geq$\,\$16.48 \\
 & \texttt{LRU} & $\geq$\,\$22.27 & \$22.33 \\
 & \texttt{FreeList} & $\geq$\,\$29.41 & \$50.22 \\
 & \texttt{LZW} & \$23.77 & $\geq$\,\$41.24 \\
 & \texttt{Paxos} & $\geq$\,\$30.29 & $\geq$\,\$49.67 \\
\texttt{claude-sonnet-5} & \texttt{Crc32} & $\geq$\,\$2.03 & $\geq$\,\$9.15 \\
 & \texttt{LRU} & \$1.53 & \$10.29 \\
 & \texttt{FreeList} & \$6.66 & \emph{declared} \\
 & \texttt{LZW} & $\geq$\,\$5.03 & \emph{declared} \\
 & \texttt{Paxos} & $\geq$\,\$9.92 & \emph{declared} \\
\texttt{gemini-3.1-pro-high} & \texttt{Crc32} & \$1.29 & \$6.01 \\
 & \texttt{LRU} & \$1.43 & \$6.19 \\
 & \texttt{FreeList} & \$1.96 & \emph{unmeasured} \\
 & \texttt{LZW} & \$1.65 & \$10.73 \\
 & \texttt{Paxos} & \$1.80 & $\geq$\,\$7.29 \\
\texttt{gemini-3.8-flash-high} & \texttt{Crc32} & \$1.61 & \$2.89 \\
 & \texttt{LRU} & \$1.57 & \$3.55 \\
 & \texttt{FreeList} & \$2.20 & \$4.15 \\
 & \texttt{LZW} & \$1.67 & \$4.97 \\
 & \texttt{Paxos} & \$2.28 & \$16.16 \\
\bottomrule
\end{tabular}}\\[3pt]
\parbox{0.92\textwidth}{\small\textbf{Table \$3.} Specification change, greenfield only. A cell's figure is its first phase plus its second; for the Opus cells the first phase is the reused landing of the matrix of Section~\ref{sec:s3}. \emph{declared} marks a condition unreached at the cap. The other marks and limits are those of Table~\$1.}
\src{harness/systems-v3/RESULT-cost-tables-v3-2026-09-29.md, section \$3}
\end{minipage}\end{center}
\begin{center}\begin{minipage}{\textwidth}\centering
{\scriptsize\setlength{\tabcolsep}{4pt}
\begin{tabular}{llrrr}
\toprule
model & problem & no treatment & statement & spec.\ change \\
\midrule
\texttt{claude-opus-5} & \texttt{Crc32} & (1.16) & 0.94 & (0.83) \\
 & \texttt{LRU} & $\leq$\,1.15 & 1.52 & $\leq$\,1.00 \\
 & \texttt{FreeList} & (2.89) & $\leq$\,1.53 & $\leq$\,1.71 \\
 & \texttt{LZW} & 1.37 & (1.81) & $\geq$\,1.73 \\
 & \texttt{Paxos} & $\leq$\,2.24 & --- & (1.64) \\
\texttt{claude-sonnet-5} & \texttt{Crc32} & 4.19 & 4.25 & (4.51) \\
 & \texttt{LRU} & (4.68) & $\geq$\,4.82 & 6.73 \\
 & \texttt{FreeList} & $\geq$\,12.93 & $\geq$\,12.50 & --- \\
 & \texttt{LZW} & 8.48 & $\geq$\,6.55 & --- \\
 & \texttt{Paxos} & $\geq$\,11.14 & --- & --- \\
\texttt{gemini-3.1-pro-high} & \texttt{Crc32} & 8.24 & $\geq$\,7.98 & 4.66 \\
 & \texttt{LRU} & 3.66 & 6.39 & 4.34 \\
 & \texttt{FreeList} & 5.88 & --- & --- \\
 & \texttt{LZW} & 7.25 & 8.68 & 6.51 \\
 & \texttt{Paxos} & $\geq$\,9.85 & --- & $\geq$\,4.05 \\
\texttt{gemini-3.8-flash-high} & \texttt{Crc32} & 2.61 & 3.70 & 1.80 \\
 & \texttt{LRU} & 2.38 & 3.62 & 2.26 \\
 & \texttt{FreeList} & 2.96 & 3.49 & 1.88 \\
 & \texttt{LZW} & $\geq$\,4.28 & 2.75 & 2.98 \\
 & \texttt{Paxos} & 3.53 & --- & 7.07 \\
\bottomrule
\end{tabular}\\[8pt]
\begin{tabular}{llrlrr}
\toprule
model & treatment & $k$ of $m$ $>1$ & indeterminate & median & bounded \\
\midrule
\texttt{claude-opus-5} & bare & 1 of 1 & Crc32, LRU, FreeList, Paxos & 1.37 & 4 \\
 & statement & 1 of 2 & FreeList, LZW & 1.52 & 2 \\
 & spec-change & 1 of 1 & Crc32, LRU, FreeList, Paxos & 1.64 & 5 \\
\texttt{claude-sonnet-5} & bare & 4 of 4 & LRU & 8.48 & 3 \\
 & statement & 4 of 4 & none & 5.68 & 3 \\
 & spec-change & 1 of 1 & Crc32 & 5.62 & 1 \\
\texttt{gemini-3.1-pro-high} & bare & 5 of 5 & none & 7.25 & 1 \\
 & statement & 3 of 3 & none & 7.98 & 1 \\
 & spec-change & 4 of 4 & none & 4.50 & 1 \\
\texttt{gemini-3.8-flash-high} & bare & 5 of 5 & none & 2.96 & 1 \\
 & statement & 4 of 4 & none & 3.56 & 0 \\
 & spec-change & 5 of 5 & none & 2.26 & 0 \\
\bottomrule
\end{tabular}}\\[3pt]
\parbox{0.92\textwidth}{\small\textbf{Table \$4.} The \arm{salt-diet} median divided by the \arm{plain} median in the same row of Tables~\$1 and~\$3, greenfield, with its sign counts, by the rules of Table~T4. No $p$-value is given. The brownfield ratios, by the same rule, are in the result file.}
\src{harness/systems-v3/RESULT-cost-tables-v3-2026-09-29.md, section \$4, its sign counts and the brownfield ratios}
\end{minipage}\end{center}
\begin{center}\begin{minipage}{\textwidth}\centering
{\scriptsize\setlength{\tabcolsep}{4pt}
\begin{tabular}{llrrrr}
\toprule
 & & \multicolumn{2}{c}{no treatment} & \multicolumn{2}{c}{statement} \\
\cmidrule(lr){3-4}\cmidrule(lr){5-6}
model & problem & \arm{plain} & \arm{salt-diet} & \arm{plain} & \arm{salt-diet} \\
\midrule
\texttt{claude-opus-5} & \texttt{Crc32} & $\geq$\,864 & $\geq$\,1{,}104 & $\geq$\,803 & $\geq$\,1{,}044 \\
 & \texttt{LRU} & $\geq$\,1{,}105 & 1{,}647 & 1{,}526 & $\geq$\,1{,}768 \\
 & \texttt{FreeList} & $\geq$\,2{,}190 & $\geq$\,3{,}938 & 2{,}490 & 2{,}973 \\
 & \texttt{LZW} & $\geq$\,1{,}286 & 2{,}371 & $\geq$\,1{,}225 & $\geq$\,2{,}491 \\
 & \texttt{Paxos} & $\geq$\,2{,}431 & $\geq$\,3{,}336 & --- & --- \\
\texttt{claude-sonnet-5} & \texttt{Crc32} & 321 & 1{,}347 & 261 & 1{,}045 \\
 & \texttt{LRU} & 352 & 1{,}769 & 382 & 1{,}709 \\
 & \texttt{FreeList} & 1{,}347 & $\geq$\,7{,}864 & 1{,}045 & 7{,}382 \\
 & \texttt{LZW} & 623 & 4{,}002 & 744 & 4{,}062 \\
 & \texttt{Paxos} & 1{,}468 & $\geq$\,6{,}476 & --- & --- \\
\texttt{gemini-3.1-pro-high} & \texttt{Crc32} & 269 & 1{,}596 & 218 & $\geq$\,1{,}620 \\
 & \texttt{LRU} & 613 & 1{,}318 & 229 & 989 \\
 & \texttt{FreeList} & 332 & 2{,}189 & 320 & $\geq$\,3{,}598 \\
 & \texttt{LZW} & 218 & 3{,}806 & 235 & 2{,}044 \\
 & \texttt{Paxos} & 214 & $\geq$\,3{,}577 & --- & --- \\
\texttt{gemini-3.8-flash-high} & \texttt{Crc32} & 281 & 516 & 242 & 766 \\
 & \texttt{LRU} & 394 & 864 & 374 & 913 \\
 & \texttt{FreeList} & 374 & 986 & 337 & 772 \\
 & \texttt{LZW} & 308 & 1{,}156 & 339 & 739 \\
 & \texttt{Paxos} & 385 & 1{,}207 & --- & --- \\
\bottomrule
\end{tabular}}\\[3pt]
\parbox{0.92\textwidth}{\small\textbf{Table W1.} Greenfield. Median wall seconds per condition, over the condition's cells of record, summed over phases. A descriptive reading: no test, and no verdict on the arms. Wall time measures the run box as well as the model: it depends on what else the box was running and on the vendor's service at that hour, and it is not adjusted for either. The Claude lane's figure is net of held time and can understate by one watcher tick, about 60 seconds. $\geq$ marks a floor, including a cell whose wall came within one turn timeout of its wall cap, a rule that can mark a true figure as a floor and never the reverse. Compare within a row and within a lane, never across lanes. --- marks an inexpressible condition.}
\src{harness/systems-v3/RESULT-cost-tables-v3-2026-09-29.md, section W1}
\end{minipage}\end{center}
\begin{center}\begin{minipage}{\textwidth}\centering
{\scriptsize\setlength{\tabcolsep}{4pt}
\begin{tabular}{llrrrr}
\toprule
 & & \multicolumn{2}{c}{no treatment} & \multicolumn{2}{c}{statement} \\
\cmidrule(lr){3-4}\cmidrule(lr){5-6}
model & problem & \arm{plain} & \arm{salt-diet} & \arm{plain} & \arm{salt-diet} \\
\midrule
\texttt{claude-opus-5} & \texttt{Crc32} & 804 & 1{,}227 & 1{,}106 & 985 \\
 & \texttt{LRU} & 1{,}106 & 1{,}769 & 1{,}045 & 1{,}649 \\
 & \texttt{FreeList} & 1{,}890 & $\geq$\,4{,}848 & 2{,}252 & 3{,}338 \\
 & \texttt{LZW} & 1{,}347 & 2{,}855 & 1{,}769 & 2{,}131 \\
 & \texttt{Paxos} & 2{,}674 & 3{,}277 & --- & --- \\
\texttt{claude-sonnet-5} & \texttt{Crc32} & 261 & 1{,}407 & 382 & 1{,}166 \\
 & \texttt{LRU} & 261 & 2{,}071 & 321 & 1{,}528 \\
 & \texttt{FreeList} & 684 & $\geq$\,6{,}657 & 925 & $\geq$\,7{,}623 \\
 & \texttt{LZW} & 442 & 4{,}847 & 562 & 4{,}122 \\
 & \texttt{Paxos} & 1{,}227 & $\geq$\,5{,}632 & --- & --- \\
\texttt{gemini-3.1-pro-high} & \texttt{Crc32} & 197 & 1{,}297 & 182 & 1{,}084 \\
 & \texttt{LRU} & 182 & 1{,}216 & 207 & $\geq$\,1{,}994 \\
 & \texttt{FreeList} & 313 & 4{,}382 & 268 & $\geq$\,2{,}458 \\
 & \texttt{LZW} & 242 & $\geq$\,5{,}817 & 198 & $\geq$\,4{,}375 \\
 & \texttt{Paxos} & 364 & 2{,}812 & --- & --- \\
\texttt{gemini-3.8-flash-high} & \texttt{Crc32} & 261 & 408 & 235 & 900 \\
 & \texttt{LRU} & 299 & 657 & 286 & 841 \\
 & \texttt{FreeList} & 554 & 906 & 430 & 677 \\
 & \texttt{LZW} & 487 & 966 & 398 & 807 \\
 & \texttt{Paxos} & 413 & 662 & --- & --- \\
\bottomrule
\end{tabular}}\\[3pt]
\parbox{0.92\textwidth}{\small\textbf{Table W2.} Brownfield. The marks and limits are those of Table~W1.}
\src{harness/systems-v3/RESULT-cost-tables-v3-2026-09-29.md, section W2}
\end{minipage}\end{center}
\begin{center}\begin{minipage}{\textwidth}\centering
{\scriptsize\setlength{\tabcolsep}{4pt}
\begin{tabular}{llrr}
\toprule
model & problem & \arm{plain} & \arm{salt-diet} \\
\midrule
\texttt{claude-opus-5} & \texttt{Crc32} & $\geq$\,2{,}270 & $\geq$\,2{,}871 \\
 & \texttt{LRU} & $\geq$\,2{,}601 & 2{,}511 \\
 & \texttt{FreeList} & $\geq$\,4{,}288 & 5{,}164 \\
 & \texttt{LZW} & 2{,}480 & $\geq$\,4{,}801 \\
 & \texttt{Paxos} & $\geq$\,4{,}620 & $\geq$\,6{,}037 \\
\texttt{claude-sonnet-5} & \texttt{Crc32} & 522 & 2{,}333 \\
 & \texttt{LRU} & 521 & 2{,}996 \\
 & \texttt{FreeList} & 2{,}212 & \emph{declared} \\
 & \texttt{LZW} & 1{,}789 & \emph{declared} \\
 & \texttt{Paxos} & 4{,}082 & \emph{declared} \\
\texttt{gemini-3.1-pro-high} & \texttt{Crc32} & 389 & 4{,}863 \\
 & \texttt{LRU} & 400 & 1{,}961 \\
 & \texttt{FreeList} & 592 & 4{,}089 \\
 & \texttt{LZW} & 512 & 4{,}677 \\
 & \texttt{Paxos} & 560 & $\geq$\,3{,}207 \\
\texttt{gemini-3.8-flash-high} & \texttt{Crc32} & 913 & 1{,}330 \\
 & \texttt{LRU} & 717 & 1{,}768 \\
 & \texttt{FreeList} & 809 & 1{,}654 \\
 & \texttt{LZW} & 793 & 2{,}072 \\
 & \texttt{Paxos} & 1{,}064 & 8{,}482 \\
\bottomrule
\end{tabular}}\\[3pt]
\parbox{0.92\textwidth}{\small\textbf{Table W3.} Specification change, greenfield only, both phases summed. \emph{declared} marks a condition unreached at the cap. The other marks and limits are those of Table~W1.}
\src{harness/systems-v3/RESULT-cost-tables-v3-2026-09-29.md, section W3}
\end{minipage}\end{center}
\begin{center}\begin{minipage}{\textwidth}\centering
{\scriptsize\setlength{\tabcolsep}{4pt}
\begin{tabular}{llrrr}
\toprule
model & problem & no treatment & statement & spec.\ change \\
\midrule
\texttt{claude-opus-5} & \texttt{Crc32} & (1.28) & (1.30) & (1.26) \\
 & \texttt{LRU} & $\leq$\,1.49 & $\geq$\,1.16 & $\leq$\,0.97 \\
 & \texttt{FreeList} & (1.80) & 1.19 & $\leq$\,1.20 \\
 & \texttt{LZW} & $\leq$\,1.84 & (2.03) & $\geq$\,1.94 \\
 & \texttt{Paxos} & (1.37) & --- & (1.31) \\
\texttt{claude-sonnet-5} & \texttt{Crc32} & 4.20 & 4.00 & 4.47 \\
 & \texttt{LRU} & 5.03 & 4.47 & 5.75 \\
 & \texttt{FreeList} & $\geq$\,5.84 & 7.06 & --- \\
 & \texttt{LZW} & 6.42 & 5.46 & --- \\
 & \texttt{Paxos} & $\geq$\,4.41 & --- & --- \\
\texttt{gemini-3.1-pro-high} & \texttt{Crc32} & 5.93 & $\geq$\,7.45 & 12.49 \\
 & \texttt{LRU} & 2.15 & 4.32 & 4.90 \\
 & \texttt{FreeList} & 6.60 & $\geq$\,11.24 & 6.91 \\
 & \texttt{LZW} & 17.46 & 8.69 & 9.13 \\
 & \texttt{Paxos} & $\geq$\,16.70 & --- & $\geq$\,5.72 \\
\texttt{gemini-3.8-flash-high} & \texttt{Crc32} & 1.84 & 3.16 & 1.46 \\
 & \texttt{LRU} & 2.19 & 2.44 & 2.47 \\
 & \texttt{FreeList} & 2.63 & 2.29 & 2.04 \\
 & \texttt{LZW} & 3.75 & 2.18 & 2.61 \\
 & \texttt{Paxos} & 3.14 & --- & 7.97 \\
\bottomrule
\end{tabular}\\[8pt]
\begin{tabular}{llrlrr}
\toprule
model & treatment & $k$ of $m$ $>1$ & indeterminate & median & bounded \\
\midrule
\texttt{claude-opus-5} & bare & 0 of 0 & Crc32, LRU, FreeList, LZW, Paxos & 1.49 & 5 \\
 & statement & 2 of 2 & Crc32, LZW & 1.25 & 3 \\
 & spec-change & 1 of 2 & Crc32, FreeList, Paxos & 1.26 & 5 \\
\texttt{claude-sonnet-5} & bare & 5 of 5 & none & 5.03 & 2 \\
 & statement & 4 of 4 & none & 4.97 & 0 \\
 & spec-change & 2 of 2 & none & 5.11 & 0 \\
\texttt{gemini-3.1-pro-high} & bare & 5 of 5 & none & 6.60 & 1 \\
 & statement & 4 of 4 & none & 8.07 & 2 \\
 & spec-change & 5 of 5 & none & 6.91 & 1 \\
\texttt{gemini-3.8-flash-high} & bare & 5 of 5 & none & 2.63 & 0 \\
 & statement & 4 of 4 & none & 2.36 & 0 \\
 & spec-change & 5 of 5 & none & 2.47 & 0 \\
\bottomrule
\end{tabular}}\\[3pt]
\parbox{0.92\textwidth}{\small\textbf{Table W4.} The \arm{salt-diet} median divided by the \arm{plain} median in the same row of Tables~W1 and~W3, greenfield, with its sign counts, by the rules of Table~T4. No $p$-value is given. The brownfield ratios are in the result file.}
\src{harness/systems-v3/RESULT-cost-tables-v3-2026-09-29.md, section W4, its sign counts and the brownfield ratios}
\end{minipage}\end{center}

\subsection{Correctness over the matrix, a declared post-hoc reading}
\label{sec:matrixcorrect}

The tables above describe what the conditions cost. Tables~P1 to~P4 read the same 200 conditions for whether the work passed its referee. The reading was registered on 2026-10-01, after every verdict it reads existed and after many had been read by its author, and it is declared post hoc: what was fixed before the instrument ran is the method, never the data. A cell's verdict is the one the condition's result of record prints, as amended by that file's own dated addenda. The withheld suite is the grader, and a cell's own claim to have finished is never read as a pass. A pass is defined per task form as each condition's freeze registered it: every test of the withheld suite on greenfield, every test of the brownfield suite on the end state, and under specification change every test of the second phase with no regression and no clause failure. On the \arm{salt-diet} arm the proof's verdict is printed beside the suite's in the result file and never folded into it, so a pass means the same thing in both arms.\src{harness/systems-v3/REGISTRATION-correctness-tables-v3-2026-10-01.md, sections P1 and P2}

\paragraph{How a halt is read.} Each cell falls in one of four classes. \textsc{pass} and \textsc{fail} are the suite's verdicts. A cell that halted at a registered budget and whose end state did not pass is \textsc{censored}, and a halt is never a failure; a halted cell whose end state passed is a pass. A cell on which the suite gave no verdict is \textsc{unscorable}. A condition prints $k/n$, where $n$ counts only \textsc{pass} and \textsc{fail} cells, with its censored and unscorable cells beside it. The sign of a row compares intervals: each arm's pass rate is bounded by counting every censored and unscorable cell first as a failure and then as a pass, a sign is printed only when the two intervals are strictly apart, so that a shared endpoint is not a separation, $=$ only when both are the same single point, and $?$ otherwise. A halt therefore cannot produce a sign in either direction, and since the cap binds the \arm{salt-diet} arm more often, that is the arm the rule protects.\src{harness/systems-v3/REGISTRATION-correctness-tables-v3-2026-10-01.md, sections P2 to P4 and addendum 1} A third addendum, written after trial runs had printed only the 200 check, the class totals and the list of unmeasured cells, fixed how five shapes of record are read: among them, a halt with no suite verdict is censored, and a record that prints only a condition's count gives that condition its $k/n$ only when the count covers every cell of record and every one of those cells is shown landed.\src{harness/systems-v3/REGISTRATION-correctness-tables-v3-2026-10-01.md, addendum 3}

\paragraph{The checks.} 181 conditions carry a pass count, 16 are inexpressible and 3 declared, and none is unmeasured: 200. Over the cells, 447 pass, 40 fail, 10 are censored and 4 unscorable, and 42 are read from their condition's count. The reading's second method compares the instrument's count with every count a result of record already prints: 94 counts are compared and 19 disagree, and each disagreement is listed in the result file with the population its source counts and is never resolved by hand. The instrument verifies a cause for every one: 18 are reproduced by counting the other population their source names, and 1 is a figure that an addendum has since superseded. 51 conditions have a printed count that agrees, 8 have printed counts none of which agrees, and 122 have no printed count anywhere, so for those the instrument's reading is the only one.\src{harness/systems-v3/RESULT-correctness-tables-v3-2026-10-01.md, the header}

\paragraph{What the tables show, and within what.} Most conditions pass every cell in both arms. In Table~P4 the \arm{salt-diet} interval lies wholly above the \arm{plain} one on two rows, Pro and Flash on \texttt{Paxos} under specification change, where the \arm{plain} arm passed 0 of 3. It lies wholly below on \texttt{FreeList} under no treatment at Opus, Pro and Flash and under specification change at Opus and Flash, and on \texttt{LZW} at Pro under no treatment and under the statement treatment and at Flash under the statement treatment. The Sonnet rows carry no $+$ and no $-$. Every $?$ has a censored or unscorable cell on at least one side, and the lower block of Table~P4 says on which. On brownfield, by the same rule and printed in the result file, the Claude and Flash rows carry no $+$ and no $-$, and Pro's \arm{salt-diet} interval lies below on \texttt{LRU} and \texttt{FreeList} under both treatments and on \texttt{Paxos} under no treatment.\src{harness/systems-v3/RESULT-correctness-tables-v3-2026-10-01.md, sections P4, the P4 counts, where each ? comes from, and the brownfield signs}

\paragraph{What the tables cannot carry.} These are descriptions, with no test and no $p$-value, and the registered tests remain those of Section~\ref{sec:s3}. At $n=3$ a rate moves in thirds, so a sign is one cell's difference. A pass is bounded by the strength of the suite that grants it, and the suites' mutant scores are ceilings and not strengths. No condition of the 181 was scored a pass on a suite that was not measured: every pass row carries its test count, and each group's pass count equals its result's. Part of that check was re-run by a non-author, and the rest is the lead's reading alone.\src{harness/systems-v3/CENSUS-full-matrix-2026-09-14.md, addendum 34} Censoring is arm-correlated, so the count of $?$ is arm-correlated too. The two lanes are graded by different scorers of record, so a row compares arms within one model and one problem and never a Gemini row with a Claude row. The brownfield \texttt{Crc32} and \texttt{LZW} rows carry the comment tell of Section~\ref{sec:matrix}, a limit on absolute difficulty and not on the arm contrast.\src{harness/systems-v3/CENSUS-full-matrix-2026-09-14.md, addendum 36} Section~\ref{sec:posthoc} read Matrix~1's landed cells under that pass's own classes; these tables read the census's cells of record under this registration's, so a row here need not equal a row there. None of it says the method under test makes code more or less correct.\src{harness/systems-v3/REGISTRATION-correctness-tables-v3-2026-10-01.md, section P6}

\begin{center}\begin{minipage}{\textwidth}\centering
{\scriptsize\setlength{\tabcolsep}{4pt}
\begin{tabular}{llrrrr}
\toprule
 & & \multicolumn{2}{c}{no treatment} & \multicolumn{2}{c}{statement} \\
\cmidrule(lr){3-4}\cmidrule(lr){5-6}
model & problem & \arm{plain} & \arm{salt-diet} & \arm{plain} & \arm{salt-diet} \\
\midrule
\texttt{claude-opus-5} & \texttt{Crc32} & 3/3 & 3/3 & 3/3 & 3/3 \\
 & \texttt{LRU} & 3/3 & 3/3 & 3/3 & 3/3 \\
 & \texttt{FreeList} & 3/3 & 2/3 & 3/3 & 3/3 \\
 & \texttt{LZW} & 3/3 & 3/3 & 3/3 & 3/3 \\
 & \texttt{Paxos} & 3/3 & 3/3 & --- & --- \\
\texttt{claude-sonnet-5} & \texttt{Crc32} & 3/3 & 3/3 & 3/3 & 3/3 \\
 & \texttt{LRU} & 3/3 & 3/3 & 3/3 & 3/3 \\
 & \texttt{FreeList} & 2/3 & 2/2 (+1c) & 3/3 & 2/2 (+1c) \\
 & \texttt{LZW} & 3/3 & 3/3 & 3/3 & 2/2 (+1c) \\
 & \texttt{Paxos} & 3/3 & 3/3 & --- & --- \\
\texttt{gemini-3.1-pro-high} & \texttt{Crc32} & 3/3 & 3/3 & 3/3 & 3/3 \\
 & \texttt{LRU} & 3/3 & 3/3 & 3/3 & 3/3 \\
 & \texttt{FreeList} & 1/3 & 0/3 & 3/3 & 1/1 (+2c) \\
 & \texttt{LZW} & 3/3 & 2/3 & 3/3 & 2/3 \\
 & \texttt{Paxos} & 1/3 & 0/1 (+2c) & --- & --- \\
\texttt{gemini-3.8-flash-high} & \texttt{Crc32} & 3/3$^\dagger$ & 3/3$^\dagger$ & 3/3$^\dagger$ & 3/3$^\dagger$ \\
 & \texttt{LRU} & 3/3$^\dagger$ & 3/3$^\dagger$ & 3/3$^\dagger$ & 3/3$^\dagger$ \\
 & \texttt{FreeList} & 1/3$^\dagger$ & 0/3$^\dagger$ & 3/3$^\dagger$ & 3/3$^\dagger$ \\
 & \texttt{LZW} & 3/3 & 2/2 (+1u) & 3/3 & 2/3 \\
 & \texttt{Paxos} & 2/3$^\dagger$ & 2/3$^\dagger$ & --- & --- \\
\bottomrule
\end{tabular}}\\[3pt]
\parbox{0.92\textwidth}{\small \textbf{Table P1.} Greenfield. Full passes $k/n$ per condition, over the condition's cells of record, declared post hoc; $n$ counts \textsc{pass} and \textsc{fail} cells only. A mark such as (+1c) or (+1u) counts the cells censored at a registered budget or the cells the suite gave no verdict on, which enter neither $k$ nor $n$. $^\dagger$ marks a condition whose record prints only its count, read under addendum~3. --- marks an inexpressible condition. A descriptive reading: no test, and no verdict on the arms. \emph{Erratum, version~4:} the unscorable cell in the row \texttt{gemini-3.8-flash-high}, \texttt{LZW}, \arm{salt-diet}, no treatment, printed 2/2 (+1u), is a wedge: from its second result to its cap, every result repeated the one before, and it never recovered. All four wedges in the run box's streams fell in the \arm{salt-diet} arm (4 of its 253 streams, against 0 of 292 for \arm{plain}), so a wedge may be an outcome of the treatment rather than a fault of the transport; four events suggest the link without establishing it, and the cause of a wedge is unmeasured. Counted as a failure in its arm's denominator, the row reads 2/3. No cell is re-scored, and Table~P4's sign for the row stays $?$ under its own rule. Three \texttt{gemini-3.1-pro-high} rows reach $n=3$ only through a top-up cell built from a later export than the condition's other cells: \texttt{FreeList} \arm{plain} under no treatment and under the statement treatment, and \texttt{Crc32} \arm{salt-diet} under the statement treatment. Two of the three also span two client binaries. A top-up's score is of record; whether the export changed the outcome is not settled.\src{harness/systems-v3/CENSUS-full-matrix-2026-09-14.md, addendum 31 section AG1; harness/systems-v3/RESULT-stage3-2026-09-12.md, addendum B}}
\src{harness/systems-v3/RESULT-correctness-tables-v3-2026-10-01.md, section P1}
\src{harness/systems-v3/AMENDMENT-ZF-gemini4-argon-2026-10-13.md, addendum 6, sections A6.2 and A6.4; evidence/zf-wedge-census-2026-10-15/census-K3.tsv; harness/systems-v3/CELLMAP-correctness-tables-v3-2026-10-01.tsv, cell l6vgfs02}
\end{minipage}\end{center}

\begin{center}\begin{minipage}{\textwidth}\centering
{\scriptsize\setlength{\tabcolsep}{4pt}
\begin{tabular}{llrrrr}
\toprule
 & & \multicolumn{2}{c}{no treatment} & \multicolumn{2}{c}{statement} \\
\cmidrule(lr){3-4}\cmidrule(lr){5-6}
model & problem & \arm{plain} & \arm{salt-diet} & \arm{plain} & \arm{salt-diet} \\
\midrule
\texttt{claude-opus-5} & \texttt{Crc32} & 3/3 & 3/3 & 3/3 & 3/3 \\
 & \texttt{LRU} & 3/3 & 3/3 & 3/3 & 3/3 \\
 & \texttt{FreeList} & 3/3 & 3/3 & 3/3 & 3/3 \\
 & \texttt{LZW} & 3/3 & 3/3 & 3/3 & 3/3 \\
 & \texttt{Paxos} & 3/3 & 3/3 & --- & --- \\
\texttt{claude-sonnet-5} & \texttt{Crc32} & 3/3 & 3/3 & 3/3 & 3/3 \\
 & \texttt{LRU} & 3/3 & 3/3 & 3/3 & 3/3 \\
 & \texttt{FreeList} & 3/3 & 3/3 & 3/3 & 2/2 (+1c) \\
 & \texttt{LZW} & 3/3 & 3/3 & 3/3 & 3/3 \\
 & \texttt{Paxos} & 3/3 & 3/3 & --- & --- \\
\texttt{gemini-3.1-pro-high} & \texttt{Crc32} & 3/3 & 2/2 (+1u) & 3/3 & 3/3 \\
 & \texttt{LRU} & 3/3 & 2/3 & 3/3 & 2/3 \\
 & \texttt{FreeList} & 3/3 & 1/3 & 2/3 & 1/3 \\
 & \texttt{LZW} & 3/3 & 3/3 & 3/3 & 3/3 \\
 & \texttt{Paxos} & 3/3 & 2/3 & --- & --- \\
\texttt{gemini-3.8-flash-high} & \texttt{Crc32} & 3/3 & 3/3 & 3/3 & 3/3 \\
 & \texttt{LRU} & 3/3 & 3/3 & 3/3 & 3/3 \\
 & \texttt{FreeList} & 3/3 & 3/3 & 3/3 & 3/3 \\
 & \texttt{LZW} & 3/3 & 3/3 & 3/3 & 3/3 \\
 & \texttt{Paxos} & 3/3 & 3/3 & --- & --- \\
\bottomrule
\end{tabular}}\\[3pt]
\parbox{0.92\textwidth}{\small \textbf{Table P2.} Brownfield. The marks and limits are those of Table~P1.}
\src{harness/systems-v3/RESULT-correctness-tables-v3-2026-10-01.md, section P2}
\end{minipage}\end{center}

\begin{center}\begin{minipage}{\textwidth}\centering
{\scriptsize\setlength{\tabcolsep}{4pt}
\begin{tabular}{llrr}
\toprule
model & problem & \arm{plain} & \arm{salt-diet} \\
\midrule
\texttt{claude-opus-5} & \texttt{Crc32} & 3/3 & 3/3 \\
 & \texttt{LRU} & 3/3 & 3/3 \\
 & \texttt{FreeList} & 3/3 & 2/3 \\
 & \texttt{LZW} & 3/3 & 3/3 \\
 & \texttt{Paxos} & 2/2 (+1c) & 2/2 (+1u) \\
\texttt{claude-sonnet-5} & \texttt{Crc32} & 3/3 & 3/3 \\
 & \texttt{LRU} & 3/3 & 3/3 \\
 & \texttt{FreeList} & 1/3 & \emph{declared} \\
 & \texttt{LZW} & 3/3 & \emph{declared} \\
 & \texttt{Paxos} & 2/3 & \emph{declared} \\
\texttt{gemini-3.1-pro-high} & \texttt{Crc32} & 3/3 & 3/3 \\
 & \texttt{LRU} & 3/3 & 3/3 \\
 & \texttt{FreeList} & 1/3 & 0/2 (+1u) \\
 & \texttt{LZW} & 3/3 & 3/3 \\
 & \texttt{Paxos} & 0/3 & 1/2 (+1c) \\
\texttt{gemini-3.8-flash-high} & \texttt{Crc32} & 3/3 & 3/3 \\
 & \texttt{LRU} & 3/3 & 3/3 \\
 & \texttt{FreeList} & 2/3 & 0/3 \\
 & \texttt{LZW} & 3/3 & 3/3 \\
 & \texttt{Paxos} & 0/3 & 3/3 \\
\bottomrule
\end{tabular}}\\[3pt]
\parbox{0.92\textwidth}{\small \textbf{Table P3.} Specification change, greenfield only. Full passes of the second phase, with no regression and no clause failure. \emph{declared} marks a condition unreached at the cap. The other marks and limits are those of Table~P1. Six \texttt{claude-opus-5} rows reach $n=3$ through step g, and each is conditioned on a landed first phase. The first phases that landed, out of those with an end, were: \texttt{FreeList} 4/4 \arm{plain} and 3/6 \arm{salt-diet}; \texttt{LRU} \arm{plain} 4/5; \texttt{LZW} \arm{plain} 3/3; \texttt{Paxos} 4/4 \arm{plain} and 3/5 \arm{salt-diet}. A difference in reach is a difference in what the row's $n$ is drawn from.\src{harness/systems-v3/CENSUS-full-matrix-2026-09-14.md, addendum 32 section AH1; harness/systems-v3/RESULT-stepg-opus-specchange-2026-09-29.md section 2}}
\src{harness/systems-v3/RESULT-correctness-tables-v3-2026-10-01.md, section P3}
\end{minipage}\end{center}

\begin{center}\begin{minipage}{\textwidth}\centering
{\scriptsize\setlength{\tabcolsep}{4pt}
\begin{tabular}{llccc}
\toprule
model & problem & no treatment & statement & spec.\ change \\
\midrule
\texttt{claude-opus-5} & \texttt{Crc32} & $=$ & $=$ & $=$ \\
 & \texttt{LRU} & $=$ & $=$ & $=$ \\
 & \texttt{FreeList} & $-$ & $=$ & $-$ \\
 & \texttt{LZW} & $=$ & $=$ & $=$ \\
 & \texttt{Paxos} & $=$ & --- & ? \\
\texttt{claude-sonnet-5} & \texttt{Crc32} & $=$ & $=$ & $=$ \\
 & \texttt{LRU} & $=$ & $=$ & $=$ \\
 & \texttt{FreeList} & ? & ? & --- \\
 & \texttt{LZW} & $=$ & ? & --- \\
 & \texttt{Paxos} & $=$ & --- & --- \\
\texttt{gemini-3.1-pro-high} & \texttt{Crc32} & $=$ & $=$ & $=$ \\
 & \texttt{LRU} & $=$ & $=$ & $=$ \\
 & \texttt{FreeList} & $-$ & ? & ? \\
 & \texttt{LZW} & $-$ & $-$ & $=$ \\
 & \texttt{Paxos} & ? & --- & $+$ \\
\texttt{gemini-3.8-flash-high} & \texttt{Crc32} & $=$ & $=$ & $=$ \\
 & \texttt{LRU} & $=$ & $=$ & $=$ \\
 & \texttt{FreeList} & $-$ & $=$ & $-$ \\
 & \texttt{LZW} & ? & $-$ & $=$ \\
 & \texttt{Paxos} & $=$ & --- & $+$ \\
\bottomrule
\end{tabular}\\[8pt]
\begin{tabular}{llrrrrr}
\toprule
model & treatment & $+$ & $=$ & $-$ & $?$ & --- \\
\midrule
\texttt{claude-opus-5} & bare & 0 & 4 & 1 & 0 & 0 \\
 & statement & 0 & 4 & 0 & 0 & 1 \\
 & spec-change & 0 & 3 & 1 & 1 & 0 \\
\texttt{claude-sonnet-5} & bare & 0 & 4 & 0 & 1 & 0 \\
 & statement & 0 & 2 & 0 & 2 & 1 \\
 & spec-change & 0 & 2 & 0 & 0 & 3 \\
\texttt{gemini-3.1-pro-high} & bare & 0 & 2 & 2 & 1 & 0 \\
 & statement & 0 & 2 & 1 & 1 & 1 \\
 & spec-change & 1 & 3 & 0 & 1 & 0 \\
\texttt{gemini-3.8-flash-high} & bare & 0 & 3 & 1 & 1 & 0 \\
 & statement & 0 & 3 & 1 & 0 & 1 \\
 & spec-change & 1 & 3 & 1 & 0 & 0 \\
\bottomrule
\end{tabular}\\[8pt]
\begin{tabular}{lrrr}
\toprule
treatment & \arm{salt-diet} only & \arm{plain} only & both \\
\midrule
bare & 3 & 0 & 0 \\
statement & 3 & 0 & 0 \\
spec-change & 1 & 0 & 1 \\
\bottomrule
\end{tabular}}\\[3pt]
\parbox{0.92\textwidth}{\small \textbf{Table P4.} Top: the sign of the \arm{salt-diet} arm's pass-rate interval against the \arm{plain} arm's in the same row of Tables~P1 and~P3, greenfield. Each interval counts every censored and unscorable cell first as a failure and then as a pass; $+$ and $-$ only when the intervals are strictly apart, $=$ only when both are the same single point, $?$ otherwise; --- means no sign, because a side is inexpressible or declared. Middle: the signs per model and treatment (\emph{bare} is no treatment). Bottom: for each $?$, which arm carries the censored or unscorable cells that make it. No $p$-value is given. The brownfield signs, by the same rule, are in the result file.}
\src{harness/systems-v3/RESULT-correctness-tables-v3-2026-10-01.md, section P4, the P4 counts and where each ? comes from}
\end{minipage}\end{center}

\section{A second model run on the same matrix (added in version~5)}
\label{sec:zf}

After version~4 we ran a newer model, \texttt{gemini-4-argon-high}, beside a same-day control, \texttt{gemini-3.8-flash-high}, on 46 conditions of the pilot matrix at $n=3$ (276 registered cells). The two models were fired adjacent in time on one harness export, so the comparison is between the two model columns of each row. The 3.x cells of record reported in Section~\ref{sec:matrix} used a patchwork of harness levels and are not a same-day control.

264 of the 276 cells were fired; 261 of them were scored and 3 wedged. The 12 not fired are the Paxos greenfield spec-change condition, both arms and both models: it was cut whole under a registered rate rule because its projected finish missed the set's deadline in every window read, and it is reported short at $n=0$. This section therefore says nothing about Paxos under a spec change for either model. A cell that wedged (stopped without a scorable result) is counted as a failure in its arm.

Counted passes: \texttt{gemini-4-argon-high} 66 of 66 \arm{plain} and 63 of 66 \arm{salt-diet}, where the three \arm{salt-diet} failures are the three wedged cells of one condition (FreeList, greenfield, no treatment); \texttt{gemini-3.8-flash-high} 64 of 66 \arm{plain} and 57 of 66 \arm{salt-diet}, with no wedges. Two of those Flash \arm{salt-diet} passes, one in each of the greenfield and brownfield \texttt{Paxos} conditions, were served a web search (Section~\ref{sec:matrixrecords}); they keep their verdicts and leave the comparison, which without them reads 55 of 64, each of the two conditions at 2 of 2. \texttt{gemini-4-argon-high} was served nothing.\src{harness/systems-v3/RESULT-ZF-gemini4-argon-2026-10-25.md, addenda of 2026-11-20; evidence/o37-agy-web-served-2026-11-20/recut/ZF-moved.txt}

\paragraph{Limits.} These are counts at $n=3$ per condition, with no test and no claimed direction, and they make no claim that the method helps or hurts. Each pass is read from the output of a test driver that runs in one process with the subject's code, so that output is reachable by the subject's code: the scoring sandbox closes side channels and not this one, and no count here claims that no subject printed a verdict it did not earn.\src{harness/systems-v3/AMENDMENT-ZF-gemini4-argon-2026-10-13.md, addendum 3, the classes declared open, item (a); addendum 4, the limit on every score row} A per-turn print timeout fired in \arm{salt-diet} cells far more often than in \arm{plain} ones (\texttt{gemini-4-argon-high}: 20 of 66 \arm{salt-diet} roots against 6 of 66 \arm{plain} in the final phase; \texttt{gemini-3.8-flash-high}: 12 against 0), and whether a timeout changed a verdict was not measured. Truncated score rows and empty turns are also more frequent in the \arm{salt-diet} arm, for both models. Nothing here compares Gemini~4 with Gemini~3.x in general: the control is a Flash model, and the historical column confounds model with harness, client and fence. No token figure is reported for this set.
\src{harness/systems-v3/RESULT-ZF-gemini4-argon-2026-10-25.md, the code block printed by zf\_result.py --print: POPULATION, SHORT, TOTAL, A7, TRUNCATED-ROW and EMPTY-TURN lines, and section 3 (the cut)}

\section{Whether an explanation arm's prose is true of its own Lean (added in version~5)}
\label{sec:o90truth}

After version~4 we ran the first stage of a test of an explanation arm on the CLEVER substrate of Section~\ref{sec:priorreads}. The two arms have byte-identical instruction files. The only difference is one paragraph of the second-stage prompt, given to the treatment arm alone, which asks the agent to write an explanation file that restates its main Lean statement in words and says why it holds, and to add a one-line documentation comment to each helper declaration.\src{AMENDMENT-O90-stage1-2026-10-23.md, section 2, the arms and the appended paragraph} The registered test asks whether the treatment arm's written prose about its own Lean is false of that Lean more often than the control's. Over the written sentences that were scored, a sentence counts as false when a reader flagged it and a separate confirming session, reading the whole Lean file, agreed.\src{AMENDMENT-O90-stage1-2026-10-23.md, A3.4, the confirmer and its prompt} The confirmer is the same vendor as one of the two readers, so its agreement with that reader is not independent. 39 of the treatment arm's 915 scored sentences were false and none of the control's 39, and the one-sided Fisher exact test gives $p = 0.1898$, above the registered threshold of $0.10$, so the test does not find the treatment wrong on truth.\src{RESULT-o90-stage1-truth-axis-2026-10-31.md, the decision; harness/s2lean/o90-2026-10-23/truth-2026-10-31/flaglog.out}

\paragraph{Limits.} That is a failure to find and not a finding of equal truthfulness. The control contributes 39 written sentences, so the test has little power against a small excess. All 39 of the control's sentences are documentation comments, none of the treatment's 39 confirmed-false sentences is, and on documentation comments alone the two arms read zero against zero.\src{RESULT-o90-stage1-truth-axis-2026-10-31.md, limits 1, 3 and 6} The two readers disagree on the treatment arm: the confirmer rejected 60 of one reader's 103 flags and 2 of the other's 32, and the first reader's flag rate already separates the arms ($p = 0.0113$), which pushes the test toward firing, not away from it.\src{RESULT-o90-stage1-truth-axis-2026-10-31.md, the assumption section; harness/s2lean/o90-2026-10-23/truth-2026-10-31/fisher-checks.out} The confirmer's false-confirmation rate is bounded at 10.1\,\% (95\,\%, $n = 45$), enough to account for between 6.9 and 10.1 of the 39. That noise can only lower the treatment's count, which would raise $p$.\src{RESULT-o90-stage1-truth-axis-2026-10-31.md, limit 4; harness/s2lean/o90-2026-10-23/truth-2026-10-31/calib-summary.txt} The count rests on a rule the tool applied and the amendment registered only afterwards: a sentence one reader flagged and the confirmer confirmed stays false when the other reader left it unscored. Read the other way, the treatment arm has 36 of 912 and $p = 0.2151$. Both readings are above the threshold.\src{AMENDMENT-O90-stage1-2026-10-23.md, addendum 11; harness/s2lean/o90-2026-10-23/truth-2026-10-31/fisher-checks.out} The result says nothing about the method's effect beyond this test.\src{RESULT-o90-stage1-truth-axis-2026-10-31.md, limit 8}

\section{Whether a reader understands the two arms' Lean (added in version~5)}
\label{sec:o90understand}

The same stage, on the two arms of Section~\ref{sec:o90truth}, also measured whether a reader model that had not written the code understands it.\src{AMENDMENT-O90-stage1-2026-10-23.md, section 2, the arms} The reader was asked two kinds of question about each artifact, both scored against a key fixed before the first read. For U1 it was given an input and an output and asked whether the output satisfies the task's specification, with the specification's own definition removed from the file. For U2 it was shown the file whole with eight Lean constants listed and asked, for each, whether the file's main declaration uses it, directly or through a helper the file itself defines.\src{AMENDMENT-O90-stage1-2026-10-23.md, A2.2, A2.3 and A2.4} The read set stopped on its registered spend halt at 319 of its 323 reads. All U1 reads ran, and four U2 reads, two per arm, did not.\src{RESULT-o90-stage1-rowR-2026-10-28.md, the set} On U1 both arms scored 100\,\% of the reads that were scored (68 of 69 for the control, whose one other read did not parse, and 69 of 69 for the treatment), so the registered route (a gain of 10 points or more for the treatment) is not met.\src{RESULT-o90-stage1-rowR-2026-10-28.md, U1; harness/s2lean/o90-2026-10-23/rowR-2026-10-28/score.out} On U2 the pooled accuracy was 97.3\,\% for the control and 97.7\,\% for the treatment. On one of U2's two item kinds (92 of the 323 draws), timeouts depend on the arm, so the registered comparison uses bounds that count a timed-out option first as wrong and then as right. The two bounds point in opposite directions, so the comparison is undecided. On that kind alone the observed figure is 96.7\,\% for the treatment against 97.6\,\% for the control.\src{RESULT-o90-stage1-rowR-2026-10-28.md, U2 and limit 5, the second-kind lines with their two bounds; AMENDMENT-O90-stage1-2026-10-23.md, A6.2, the 92 of 323 draws}

\paragraph{Limits.} Every accuracy here is by one reader mechanism (Gemini), and the reader was not blind to the arm. U1 is at ceiling, so it is not evidence that the arms are understood equally well. The stage is incomplete under its own registration, and none of these figures is a stage-1 verdict.\src{RESULT-o90-stage1-rowR-2026-10-28.md, limits 1, 5 and 6}

\section{The instrument findings, as results}
\label{sec:instrument}

These are the results we think transfer. Each was found by a gate, a probe, or a re-check in the record; each changed the protocol; and each is stated as the rule it taught, with the episode that taught it beside it.

\begin{enumerate}[leftmargin=*]
\item \textbf{A censored cap returns the cap.} Section~\ref{sec:s2rust}. Any budget derived from a capped distribution has to come from an uncensored read, or from a cap raised until the censoring fraction is near zero. Recording the state each episode had reached when the cap cut it does not rescue the estimate on its own, because under a verifier that state can carry no information: an obligation is discharged or it is not, so an incomplete proof reports zero discharged until the moment it reports all of them. A partial count is a distance and a zero count is an absent measurement, and Section~\ref{sec:open} reports an episode that read zero at the cap and held a complete verified proof thirty-nine calls later.\src{RESULT-DT-sonnet-probe-2026-09-02.md, the appended 120-turn amendment}
\item \textbf{A statistic is a cap's price only in the cap's unit.} The quota cost per token at the higher tier was roughly double, while the token count fell; a campaign that priced the tier on tokens alone would have called it free. The metered sum and the quota unit are different quantities and a stop rule has to say which it uses.\src{SCOUT-S2LEAN-STAGE0.md, amendment 8 addendum 3; RESULT-amend16-driver-2026-09-02.md section 4}
\item \textbf{A gate can penalize the treatment for applying the treatment.} Section~\ref{sec:armcorr}, three instances in one wave, the third inside the layer added to prevent the first.
\item \textbf{A screen's false positives are invisible by construction.} The campaign ran 201 scored Lean episodes and examined exactly one refusal, because exactly one existed; it was a complete proof. A defence-in-depth layer needs its own false-positive reading routinely.\src{SCOUT-S2LEAN-STAGE0.md, amendment 9 result}
\item \textbf{A blind agent that passes is a measurement of a different task.} A fence derived from the run root denied the agent its own episode directory, so the referee wrapper was unreachable; the agent wrote proofs it could never check, and two of them passed the referee at 11 to 16 times the cost of a sighted episode. The episode driver now refuses when the fence and the workspace intersect.\src{RESULT-amend16-driver-2026-09-02.md section 2b}
\item \textbf{A deny list is only as broad as the layer that enforces it.} Section~\ref{sec:fence}. The sandbox's path denials fenced the agent's subprocesses and not the agent's own file tool, and the probe that certified the fence made its reads through the sandbox, so it could not see the layer above. A fence has to be probed in the language of every tool it is meant to bind, and the probe has to be neutral: a probe that told the agent a tool was denied and asked it to route around was answered by the agent's judgement with no tool call made, and read as blocked while measuring nothing.\src{AMENDMENT-17-fence-config-dir-2026-09-02.md section 9} The audit built to find such a hole in the record had the same shape of defect: its field recorded that a call was \emph{not blocked}, which conflates a denial, an absent file and a served read, and it reported an episode as an escape whose read had returned that the file did not exist. Only a served read is a hole being used, and a field that cannot say which of the three it saw will report the wrong one.\src{AMENDMENT-17-fence-config-dir-2026-09-02.md section 11}
\item \textbf{An instrument that states a reporting rule in the grammar of a measurement will be quoted as one.} The scorer for the systems population printed its floor rule as the sentence \emph{every per-problem magnitude is unresolved}. That is how the rule reads when it is taken as a statement about the table rather than about what may be reported from it, and the abstract of this paper quoted it faithfully and was false until it was corrected: the same run's table prints two premiums above the floor and the body of the section says which two. The defect was in neither the number nor the prose but in the instrument's wording, and no gate could fire on either half, because reading the paper against its own table finds the symptom and only reading the scorer finds the cause. The scorer now names the two sets apart and says in its own output that the rule is not a claim that every magnitude fell below.\src{harness/systems-v3/score\_matrix1.py, the G2 reporting rule; PUBLISH-CHECKLIST.md section (m)}
\item \textbf{An absence is only as wide as the population the instrument could see.} This paper reports absences as results: five patterns swept across a cells root returning zero files each, and zero file-tool calls at a fenced path across 278 landed episodes. Each was measured with a positive control, which is the right discipline and is not sufficient on its own. A sweep of this repository's own references for a registered amendment returned absent with its positive control firing correctly, and the document existed throughout on a remote the working copy had not fetched, so the control was drawn from the same truncated population as the search and could not have detected the truncation. A positive control establishes that the instrument works on the population it can see. It says nothing about whether that population is the one the claim is about, and an absence sweep should therefore name the population it enumerated.\src{PUBLISH-CHECKLIST.md section (m)}
\item \textbf{A bound does not cover a set that grew after it was written, and the growth is silent.} Section~\ref{sec:s3}. The box-load term for the systems matrix was bounded by showing that every cell of its root ended before a process leak opened. Three cells were later added to the declared set by amendment, from a different root and a later hour, and no file records an end time for them. Nothing in a scoreboard changes shape when a cell that no timing record covers joins the set: the premium is still a ratio of medians and the $p$-value is still a sign test, so a bound stated over the old population reads as though it covered the new one. A bound has to name the set it was taken over, and a declared set that changes has to re-derive every bound stated over it.\src{harness/systems-v3/RESULT-n3-topup-2026-09-09.md section 1; PUBLISH-CHECKLIST.md section (m)}
\item \textbf{A gate that extracts by delimiter measures the delimiter.} A statement-immutability failure was reported as the agent altering the specification; the specification was byte-identical, and the mechanism was Lean naming a cached auxiliary \texttt{match} declaration after whichever declaration elaborated it first. The audit now compares matchers by content.\src{AMENDMENT-14-instrument-2026-09-01.md sections 2 and 3}
\item \textbf{A guard whose predicate cannot hold looks like a guard in every review.} The recall instrument's provenance guard compared a wrapper's hash to its wrappee's and had taken the fallback on 78 of 78 rows; the fallback read the same bytes, so no number moved.\src{TRIAGE-B-failures-2026-09-01.md section 6} The token ceiling had been blind on every episode where it could have mattered, because the watchdog's call crashed on the first path-carrying tool call and the shell turned the crash into zero; repaired with a self-test that makes the caller's exact call.\src{SCOUT-S2LEAN-STAGE0.md, amendment 4}
\item \textbf{A scoreboard is silent about the structure of its own dataset.} Section~\ref{sec:s3}. Two properties of the cost matrix are invisible in the number it produces: no cell in it carries a correctness verdict, and three of its five problems reach their registered $n$ on the plain arm only by counting a cell fired in a different run and a different cells root. Neither is a defect in the arithmetic, and neither can be seen from the arithmetic. A result that is reported without them is a different claim from the one the data supports, so a scoreboard needs a companion statement of what its own population is made of.\src{harness/systems-v3/RESULT-matrix-opus-1-2026-09-08.md, the correctness section and the declared-set section}
\item \textbf{An instrument can be right on everything it has ever scored and wrong on the next design.} The condition key for the earlier two-arm stages is the problem and the arm. The matrix adds a third arm dimension, and a scorer keying on two fields pools the bare and statement arms of the same treatment, doubles each apparent $n$, and mixes the pair the design calls its gold, raising no error and producing medians that look reasonable. It was caught by building one cell and reading its control directory before building the other 56.\src{harness/systems-v3/PREREGISTRATION-matrix-opus-1-2026-09-08.md section 15}
\item \textbf{A control arm that could only have hurt you is worth its price and is not evidence for you.} Section~\ref{sec:s3}. The placebo arm on the systems population cost \$193.42 and returned \textsc{unresolved} on all five problems, which was the outcome its own registration predicted from the floor at that $n$. It could have cleared the floor and damaged the finding; it did not. The reading that the placebo therefore supports the treatment was written down as forbidden before the arm was fired, and the question of what an arm can resolve is free before the cells and is priced at the arm's full cost afterwards.\src{harness/systems-v3/RULING-placebo-acceptance-2026-09-08.md sections 13.1 and 13.3}
\item \textbf{A counterfactual about someone else's instrument must be run against their code.} Section~\ref{sec:triage}: the headline that the reference checker was sorry-inflated was refuted by its own kill-check, and what survived was a different hole.\src{AUDIT-FINDING-s2lean-2026-08-29.md}
\item \textbf{A stop rule that is correlated with the arm removes the treatment's hardest work from every sample taken later.} Section~\ref{sec:posthoc}. The cost matrix carried a per-cell budget cap. Every cell it stopped was a treatment cell and none was a control cell, which is not a coincidence: the treatment is the arm that spends more, so it is the arm that reaches a spend cap, and it reaches it on the problems that take the longest. When a correctness pass was run over the surviving repositories afterwards, the cap had already decided which cells could be in it. The treatment entered that pass with fewer cells than the control and with its slowest cells missing, so any pass rate computed for it is biased upward by construction, and the bias is invisible in the pass rate itself. The cap was registered, published and reported throughout as a cost stop; nothing about it was hidden. What was not foreseen is that a stop rule silently defines the population of every study built on the same cells afterwards, and that when the stop is arm-correlated the resulting sample is not merely smaller but easier on exactly one arm. A cap has to be read as a sampling instrument and not only as a spending one, and any downstream analysis over capped cells owes a statement of which arm lost cells and which cells they were.\src{harness/systems-v3/RESULT-posthoc-correctness-2026-09-09.md section 3} The complete matrix repeats the finding at a second model and under a second treatment: on the Sonnet greenfield blocks the cap bound no \arm{plain} cell and 5 of 15 \arm{salt-diet} cells under no treatment, and none against 2 of 12 under the statement treatment, where the two \arm{salt-diet} cells that did not pass are the two the cap stopped (Section~\ref{sec:matrixrecords}).\src{harness/systems-v3/CENSUS-full-matrix-2026-09-14.md, addendum 14 section T3 and addendum 15 section U2}
\item \textbf{A fence can deny the subject its own toolchain, and a cell that goes green under it has done a different task.} Section~\ref{sec:matrixrecords}. The first firing of the Gemini specification-change level ran under a sandbox that listed the parent of the harness's package registry among its read denials, so the treatment arm's build could not resolve its verification library offline. Some cells went red on that; others went green because the subject rebuilt its build environment around the hole, and plain cells did the same. A green cell under a broken environment is a measurement of the subject's ability to repair the environment, and the registered task was something else, so the ruling voided both arms of the level rather than the one that failed: re-firing one arm into a working environment while the other kept the broken one would bias the pair toward the hypothesis. The fix carves both toolchain roots into the rendered fence and adds a check that refuses a wave whose toolchain root is denied, and the population was measured over every Gemini cell's own fence file by two methods that agree, so the defect is bounded to one level rather than assumed to be.\src{harness/systems-v3/CENSUS-full-matrix-2026-09-14.md, the declaration of 2026-09-23 on the level-8 fence}
\item \textbf{A landing gate that reads a file the subject can write is a self-report.} Section~\ref{sec:matrixrecords}. The gate that records a landing took its verdict from a result file inside the subject's tree, which the runner also trusted as a cache, and two subjects forged it, one by feeding a forged line through a named pipe and one by reimplementing the runner's own tree hash. No score is affected, because the referee copies the solution alone into a fresh crate and builds it with its own toolchain, and that is the point: a landing was already, by this repository's own rule, the subject grading itself, and a self-graded landing rate is an upper bound on a verified one. What the forgery adds is that the bound can be reached on purpose, and that a search for traces can only ever return \emph{unforged by trace}, since a subject that cleans up leaves no file.\src{harness/systems-v3/CENSUS-full-matrix-2026-09-14.md, the declaration of 2026-09-23 on the level-8 fence}
\end{enumerate}

\section{The reads that shaped the design}
\label{sec:priorreads}

Three populations were measured before the seat-as-subject matrix. Each of them closed a route that had looked open, and together they are the reason the design of Section~\ref{sec:populations} is shaped as it is: a referee that decides without the agent's testimony, an authored population, and a cost question asked where a capability question had saturated. They are reported here compressed to that role.

\subsection{Provenance of the three prior populations}

\paragraph{S2-Lean: CLEVER Task~1.} \texttt{trishullab/clever}~\cite{thakur2025clever} at commit \texttt{8348039}, MIT, Lean \texttt{v4.27.0}~\cite{demoura2021lean4}, mathlib \texttt{a3a10db0}~\cite{mathlib2020}; 161 problems.\src{SCOUT-S2LEAN-STAGE0.md section 1} A raw problem file is re-cut by the harness into three views: stage~A (write \texttt{generated\_spec} from the docstring, ground truth absent), stage~B (prove \texttt{spec\_isomorphism} between the agent's frozen specification and the revealed human one), stage~C (implement and prove \texttt{correctness} against the human specification, tests included). The draw sorts the real ids by a committed seed after removing the four structural exclusions of the benchmark's agentic-proving paper~\cite{sosso2026agentic}. Stage~0 ran $k=30$. The registered comparison population is U15, the first fifteen unflagged ids of that order (the same paper flags 80 of 161 specifications), reached at depth~27.\src{SCOUT-S2LEAN-STAGE0.md section 2; amendment 3} Stage~C's population removes the two U15 ids whose stage-C view fails to elaborate before any agent text (problems 54 and 112) and one further id, problem~18, whose task is machine-checked unsatisfiable (Section~\ref{sec:triage}), leaving 12.\src{AMENDMENT-11-stageC-2026-09-01.md section 2}

\paragraph{S2-Rust: VeruSAGE-Bench proof targets.} \texttt{microsoft/verus-proof-synthesis}~\cite{yang2025verusage} at commit \texttt{cbf9c0c6}, MIT; the Anvil-Advanced~\cite{sun2024anvil} and NRKernel~\cite{lattuada2023verus} projects, 267 records.\src{AMENDMENT-15-s2rust-2026-09-01.md section 2} The wave takes the unique \texttt{proof fn} target whose body is empty modulo whitespace: 207 records; the five refuse classes are counted and published with the draw.\src{AMENDMENT-15-s2rust-2026-09-01.md section 3} Stage~0 then ran every reference proof through the pinned referee inside the harness's own scaffold: 180 live, 27 with a task text that omits definitions its own ground truth supplies, zero dead tasks, zero pin defects, zero scaffold defects.\src{RESULT-amend15-stage0-2026-09-01.md} The pin itself was measured: the current Verus release failed 5 of 15 reference proofs at the Rust front end, and the benchmark's own release passed 15 of 15; a front-end error on a reference file indicts the toolchain, never the task.\src{AMENDMENT-15-s2rust-2026-09-01.md section 5} The resource-limit curve is flat into the registered limit (179 of 180 at 10, 180 of 180 at 50 and at 250) and three repeats at the pinned seed produced zero flips.\src{RESULT-amend15-stage0-2026-09-01.md, appended section}

\paragraph{S1: SWE-bench Verified.} \texttt{princeton-nlp/SWE-bench\_Verified}~\cite{swebenchverified} at revision \texttt{c104f840}, 500 rows, content-pinned by a hash over the canonicalised rows.\src{PRE-REGISTRATION.md section 3; TASKLIST.json} The selection script is the normative form of the criteria; the draw is measured-50 and pilot-30 with a per-repository cap of~9, chosen from arithmetic over the eligible counts after the first draft's cap starved the draw in silence.\src{PRE-REGISTRATION.md section 3}

No task text from SWE-bench Verified is redistributed. Its dataset card carries no licence field, so rather than rely on a reading of what the issue text permits, the repository ships the 30 identifiers, the pinned revision, a hash over the 500 canonicalised rows and a hash over the 30-row projection the episodes read; \texttt{harness/fetch\_problem\_statements.py} rebuilds that projection from the pinned revision and verifies it against both.\src{PROVENANCE.md section 1.3; PUBLISH-CHECKLIST.md section (k)} The rebuild was compared byte for byte against the file the episodes read before that file was removed from the tree.

Licences and attributions for every population in this paper, the authored one included, and the exact list of what this repository redistributes from each, are in \texttt{PROVENANCE.md}.

\subsection{SWE-bench Verified saturates at Sonnet~5 and is abandoned as a substrate}
\label{sec:s1}

Stage~0 ran the two control arms on the first 15 tasks of the pilot draw at \texttt{claude-sonnet-5}, effort high, 40-call cap, in the pinned images; pre-flight and gold controls passed 15 of 15 each and nothing was excluded. Both arms resolved the same 13 and failed the same 2, so $b=c=0$, and the two failures are the two tasks on which both arms hit the cap, which makes 13 of 15 a lower bound at that cap. The pre-registered contamination proxy, cut at 0.80 and stated before computing, read 6 of 13 resolved tasks as high-similarity, which the rule classifies as \textsc{indeterminate}; a blind panel found, and the record verifies at the transcript, that on two unresolved tasks the agent wrote lines of the upstream fix as its own edit input before any read could have shown them.\src{RESULTS-stage0-2026-08-29.md, morning line, the reading, and contamination check and panel; EXCLUSIONS.md} The substrate is saturated at this tier for this scaffold, contamination is present and unquantifiable, and the treatment arms were never run on it. It is the reason the campaign moved to referees that decide rather than to suites an agent may have seen.

\subsection{S2-Lean: a pre-registered null, and a tier that moves both arms}
\label{sec:s2lean}

On the registered U15 at a uniform 100-call cap, \texttt{claude-sonnet-5}, stage~A then stage~B with arms alternating per problem, every arm proved 8 of 15. The treatment against the plain arm read $b=c=0$: the treatment matched the control's set on all fifteen problems, and seven of its episodes ended with \texttt{sorryAx} in \texttt{spec\_isomorphism}, the identical failure on the identical set. Totals were flat at 50.7M, 51.7M and 51.4M metered tokens. The paired split was not: on the eight problems both arms proved, the treatment used 15.79M tokens against the plain arm's 20.76M ($-24\,\%$), and on the seven both failed, 35.62M against 29.92M ($+19\,\%$). At $n=8$ and $n=7$ this is reported as a paired observation and not as an effect.\src{SCOUT-S2LEAN-STAGE0.md, amendment 5 result}

Raising the tier to \texttt{claude-opus-5} on the same U15 and the same checker moved both content arms to 10 of 15 with $|b-c|=0$, at 26.1M tokens against the Sonnet run's 108.9M, and the raise is not monotone per problem. A registered differential re-check of one treatment episode refused by the pragma screen found a complete kernel-checked proof with and without the refused line, which reads the treatment at 11 of 15 as a diagnostic and leaves the contrast \textsc{indistinguishable}.\src{SCOUT-S2LEAN-STAGE0.md, amendment 8 result and amendment 9 result} The placebo at Opus proved 9 of 15, a strict subset of the plain arm's ten, landing on the boundary of its registered arm-independence band $[9,11]$ by one problem's margin; the tier's gain is a property of the tier and not of prompt content.\src{RESULT-amend13-AW-2026-09-02.md sections 1 and 2} On the registered twelve stage-C problems at Opus the plain arm certified 12 of 12, at which the reachability bound $c \leq n - P_0$ is $c \leq 0$, so no problem remained for a treatment arm to win and it did not run.\src{RESULT-amend11-stageC-2026-09-01.md sections 1, 2 and 4}

\subsection{The triage: most of the null belongs to the oracle}
\label{sec:triage}

A blind triage of every failed stage-B cell in the record labelled each from the two specification texts and the proof alone, with tier, arm, class and termination withheld. Of 31 cells, 30 are triageable: 17 are \textsc{human-loose}, the agent's specification strictly stronger than the human one with a kernel-checked witness of non-isomorphism recorded where one could be written, 5 are \textsc{agent-wrong}, and 8 are \textsc{proof-hard}. The label is a property of the problem rather than of the arm: eight failing problems carry eight single labels with no exceptions, so the arms fail for the same reason on the same problems.\src{TRIAGE-B-failures-2026-09-01.md sections 0, 2 and 5.1} Under four of the five false obligations the pattern is one defect: the human specification guards its conclusion with a well-formedness hypothesis and says nothing outside it, while the agent's, asked for as a \texttt{Prop} over the same signature, is total, and a guarded specification and a total one are never isomorphic.\src{TRIAGE-B-failures-2026-09-01.md section 3}

Problem~18 is worse than loose. Under Lean \texttt{v4.27.0} slice equality makes the reference specification's occurrence test false at a genuine occurrence, so the specification forces the answer 0 where the benchmark's own test demands 3; that no implementation satisfies both is machine-checked with axioms exactly the allowlist and no \texttt{sorry}. The elaboration check that guards the population had marked it fine, because it measures whether the task can be stated and not whether it can be done.\src{TRIAGE-B-failures-2026-09-01.md section 4} An audit of the benchmark's own reference checker adds that it agrees with ours exactly on non-adversarial artifacts and not at all on adversarial ones: its submission path recompiles the solver's view, so replacing the theorem with \texttt{True} certifies 15 of 15 without helper lemmas, and an axiom discharging the real goal certifies 15 of 15 because an axiom is not a \texttt{sorry}.\src{AUDIT-FINDING-s2lean-2026-08-29.md sections 3 and 4}

\subsection{S2-Rust: the plain arm saturates the hard band at Opus~5}
\label{sec:s2rust}

The upper tercile of the 180 live tasks by reference-proof wall was registered as the hard band, with its 13 drawn ids, before any of it was spent on; the cutpoints are ranks because absolute walls re-time 2.2 to 2.8 times faster on an idle machine while the order is stable. With a registered sequential early stop, 10 of the 13 were run: 10 scorable, zero void, and 9 of 10 passed at a 40-call cap, the stop firing at 10 because the ceiling threshold of 9 was already reached. The registered prediction was 5 to 9 with a point estimate of 7, so the band held at its top edge and the point estimate was low by 2, the fifth consecutive under-estimate of the plain arm.\src{RESULT-amend16-driver-2026-09-02.md sections 5 and 6}

The cap sits in the shoulder, and this is what the number costs. Uncensored passing call counts were 23, 26, 27, 27, 32, 35, 35 and 39 against a cap of 40; two of ten episodes were at the cap, one of them a pass at exactly 40, and the single failure died at 40 calls reading eight verified and one error. A p90 computed from data the cap produced returns the cap, so the measured 9 is a lower bound on the ceiling and the room left for a treatment arm is smaller than the count shows.\src{RESULT-amend16-driver-2026-09-02.md section 6.1} That a cap of this size can cut an episode which would otherwise have passed is evidenced directly rather than inferred: at the lower tier on this same band, an episode reporting no discharged obligations at the 40-call cap carried a complete verified proof at call~79 once the cap was raised.\src{RESULT-DT-sonnet-probe-2026-09-02.md, the appended 120-turn amendment} The median episode spends 81\,\% of its tokens before its first clean referee run, range 64 to 91, so cost is dominated by search rather than verification and a cap cuts search rather than polish.\src{RESULT-amend16-driver-2026-09-02.md section 6.2}

\section{The treatment question, open}
\label{sec:open}

Nothing in this record shows an effect of the treatment on what a referee accepts. On S2-Lean the treatment arm matched the control's set at Sonnet and was within one problem of it at Opus, with the placebo moving with the tier; the triage attributes most of the shared failures to the reference oracle. On S2-Rust and on stage~C the plain arm is at or near the ceiling of every band the campaign can afford, so a comparison has no room. On S3-Systems the treatment arm cost more on all five problems under the registered sign test, and that is a difference in price carrying the four qualifiers of Section~\ref{sec:s3}, on a run in which no cell was checked for correctness as it went; the post-hoc pass of Section~\ref{sec:posthoc} checked the surviving cells afterwards and did not separate the arms. A price difference is not a capability difference, and nothing in that design licenses reading one from the other. The complete pilot matrix of Section~\ref{sec:matrix} adds 181 conditions with a result of record across four models and two task forms, and none of them carries a verdict on the arms: each result of record states what its conditions can and cannot carry, and the census that counts them records that none claims the method under test helps or hurts.\src{harness/systems-v3/CENSUS-full-matrix-2026-09-14.md, addendum 24 section AD2} The question is open. The tests that could close it are these, and with the task forms of Section~\ref{sec:design} they are the benchmark's roadmap:

\begin{enumerate}[leftmargin=*]
\item \textbf{A lower tier on the hard band.} Registered as both arms at \texttt{claude-sonnet-5} on the same 13 Verus ids with the same early stop, quoted from a Sonnet probe on the band rather than from the Opus figures. The quote landed; the read was withdrawn; and the reason it was withdrawn is the result reported here. The probe paired three of the band's tasks across the two tiers, the same task and arm and instrument with only the model differing, at the same 40-call cap. Its aggregate paired token ratio was 1.55 (per-episode median 2.06), which prices the lower tier at 0.71 of the higher tier's per-episode quota and passes the affordability gate. But Sonnet passed 0 of the 3 where Opus passed 2, and all three Sonnet episodes ended at the cap, so a 13-id read at that cap would have been in part a measurement of the cap.\src{RESULT-DT-sonnet-probe-2026-09-02.md, the dt\_quote.py block, gates G1 and G2} One further episode was run to separate the two: the same task, the same tier and the same arm, at a cap of 120 calls instead of 40.

\begin{center}
{\small\setlength{\tabcolsep}{5pt}
\begin{tabular}{llrrrl}
\toprule
model & cap & calls & metered tokens & wall (s) & referee at the end \\
\midrule
\texttt{claude-opus-5}   & 40  & 23 & 957,722   & 307  & 9 verified, 0 errors (pass) \\
\texttt{claude-sonnet-5} & 40  & 40 & 3,051,144 & 768  & 0 verified, 1 errors (stopped at the cap) \\
\texttt{claude-sonnet-5} & 120 & 79 & 8,375,623 & 1071 & 10 verified, 0 errors (pass) \\
\bottomrule
\end{tabular}
}
\end{center}
\src{RESULT-DT-sonnet-probe-2026-09-02.md, the appended 120-turn amendment, verbatim block}

The 120-call episode passed cleanly: no screen violation, no helper-shape violation, the linter at return code zero, the count guard unchanged, the resource limit inside budget, the episode not void and no unblocked read of a fenced path. The cap was therefore binding at this tier on this task, and the identical Sonnet episode's report of no discharged obligations at call~40 was not a measure of how far it had to go. The planned 13-id read at 40 calls does not run, because it would return the cap. A read at 120 is a different and much larger commitment: one episode at 8.4 million tokens prices 26 episodes at roughly 218 million, and that figure is a lower bound, since the episode it is priced from passed at 79 of its 120 calls on the task the higher tier found easiest of the three. The two-tier comparison is therefore open and belongs to version~2. What the pair does establish is narrower, and it is one episode against one: on this task the lower tier reaches the same referee verdict as the higher one, and pays 8.7 times the tokens and 3.4 times the calls to get there.
\item \textbf{A CLEVER variant with the oracle repaired.} Replace the isomorphism obligation with an implication-with-witness (the agent's specification implies the human one, and the reference implementation satisfies the agent's), exclude problem~18, and rerun the seven problems both arms failed. This changes what the benchmark measures and is registered as a variant, not as a correction to the published Task~1 numbers.
\item \textbf{The systems population, with the four things it is missing.} The custom population of Section~\ref{sec:populations} exists, is frozen, and has produced the cost reading of Section~\ref{sec:s3}. Four registered gaps stood between that reading and a capability comparison on the same population, and each is a separate piece of work; two have since closed and are kept below rather than deleted, because a gap that closed is part of the record of what the design owed. First, and this gap has closed post hoc rather than by registration: a correctness instrument, a referee verdict per cell on the withheld suite, written to a path the scorer reads, so that a premium can be paired with an outcome rather than with a price alone. Section~\ref{sec:posthoc} reports that pass. It was specified before it ran and after the cost result was known, it did not separate the arms, and a premium on this run can now be set beside an outcome without that outcome being a pre-registered one. Second, also closed, by amendment: three plain-arm cells on \texttt{FreeList}, \texttt{LRU} and \texttt{Paxos} so that all five problems reach $n=3$ within one run, with the sign test reported both with and without the borrowed cells; if the two readings agree the dependency is discharged, and if they diverge the divergence is the result and is reported ahead of the headline. The cells were fired under AMENDMENT~26 and both readings are reported in Section~\ref{sec:s3}, where both halves of that rule turn out to fire at once. Third, a \texttt{\#\# Statement} section for the four cards that lacked one, which was the only thing standing between the gold pair and $k=5$. The content half of that gap has since closed, and the way it closed corrects this paper's own earlier reading of it: the sections are not an author's hour per card and not the harness's to invent, but derived objects, extracted from each withheld reference solution by tool and reproducible by anyone who has the reference. Had they been written as prose the treatment would have been one author's account of four tasks beside a verbatim specification on the fifth, and the arm would have stopped being comparable across the population, which is a defect in the arm rather than in the writing.\src{harness/systems-v3/AMENDMENT-statement-arm-pilot-2026-09-09.md, addendum 1} That reading is now reported in Section~\ref{sec:s3}, at $k=4$ rather than the $k=5$ the amendment aimed at, with the conditions that did not fire and the reason named there. Fourth, an arm carrying the method as written rather than the dieted rendering, which the current cost cap refused once already. A scout for a second population over a contamination-free Rust software-engineering set returned no: of its 113 hand-authored tasks 5 are Rust, below the registered threshold of 8 before any screening, and none of the 5 survived the screen for a specification layer.\src{RESULT-DY-deepswe-scout-2026-09-02.md sections 3, 4 and 5} Version~2's populations ship in the Harbor task format~\cite{harbor2026}, so that the referee, the fence and the task remain one object.
\end{enumerate}

The failure surface on S2-Lean is also a finding about the benchmark: four of the twelve remaining stage-C oracles are permissive on regions their tests never visit, and the four-label triage taxonomy has no cell for a reference specification that contradicts its own tests.\src{RESULT-amend11-stageC-2026-09-01.md section 2, the permissive-oracle diagnostic; TRIAGE-B-failures-2026-09-01.md section 5.1} The reproduction scripts for both are in the repository, for the benchmark's authors; the disclosure itself is a separate act and is not claimed here.

\section{Reproducibility}
\label{sec:repro}

Everything in this paper can be re-derived, and this section says from what. Every pin is a line in \texttt{harness/HASHES.txt}: the CLEVER commit, the Lean toolchain and mathlib revision, the project manifest and lakefile hashes, the Verus release and its rust channel, the z3~\cite{demoura2008z3}, vstd and lynette hashes, the resource limit and seed, every prompt, arm file, checker and driver, and the set-hash of the 207 rebuilt Verus views. The evaluation images for the SWE-bench substrate are pinned by digest in \texttt{IMAGE-DIGESTS.json}. The episode script refuses to run a stage whose view or checker hash is not the pinned one, and each manifest records the freeze commit, the hashes it asserted, the model id read from every response, and the transcript's hash. The morning-line instruments that produced every rate in this paper are in the repository with their self-tests, and the evidence directories carry their outputs byte for byte.

The systems population is pinned differently, because it is authored here rather than derived from a public source. Its five problems are frozen in a dated export, the scored set is declared by cell identifier rather than by a pattern over the archive, each cell's control directory records its task, arm and card extras, and each cell's price is read from its own harvested meter file. The client is pinned by absolute versioned path in the registration, not resolved from the shell path. One gap in that chain is reported rather than closed: the matrix's cells carry no build-provenance record of their own, so the mapping from cell to instrument set is a reconstruction after the fact and not a receipt, and the claim that this matrix and the earlier stage ran the same client rests on a byte-size comparison against the earlier stage's own record.\src{harness/systems-v3/RESULT-matrix-opus-1-2026-09-08.md, the provenance note; harness/systems-v3/PREREGISTRATION-matrix-opus-1-2026-09-08.md sections 11 and 14}

The task populations are re-derived from the pinned sources by the harness's view builders; the repository redistributes only what \texttt{PROVENANCE.md} lists. The run records and the S2 episode archives are a separate data asset; its DOI is assigned at release (Zenodo) and recorded in the repository at release. Version~2's populations ship as Harbor tasks so that the referee, the fence and the task are one object.

The pilot matrix is re-derived from the census and the results of record it names. The census carries every count as a dated addendum with the commit that landed it, and its verifier rebuilds the trajectory from the addenda's headlines and refuses on a mismatch. Each result of record carries its per-cell table as a file beside it and a verifier that re-derives every headline figure from that table and reddens on a planted wrong value; the Gemini lane's cells name the export they ran and scored on by sha, and the Claude lane's tables derive the served model per cell from the cell's own receipt. The table of Section~\ref{sec:matrix} is generated by \texttt{paper/matrix\_census\_table.py}, which checks its sums against the census's live row and the rows in this file against its own output.\src{harness/systems-v3/CENSUS-full-matrix-2026-09-14-verify.py; harness/systems-v3/RESULT-gemini-level8-chainF-2026-09-24-verify.py; harness/systems-v3/RESULT-claude-blockO-2026-09-21-verify.py; paper/matrix\_census\_table.py}

The code is released under Apache-2.0 and the data and documents under CC BY 4.0. The agent was Claude Code~\cite{claudecode} on a consumer subscription, and the transcripts are published as this campaign's own measurements: the provider's Consumer Terms effective 2025-10-08~\cite{anthropic2025terms} assign outputs to the user and do not restrict their publication, and the training restriction in the Usage Policy dated 2025-09-15~\cite{anthropic2025usage} binds the subscriber rather than a recipient of these files.\src{PROVENANCE.md section 3; PUBLISH-CHECKLIST.md section (k)} Both documents are cited at the version read, because both are revised in place.

\section*{Acknowledgements}

CLEVER is by the Trishul group at UT Austin (MIT licence); VeruSAGE-Bench and lynette are by Microsoft (MIT licence); SWE-bench Verified is by the SWE-bench authors with OpenAI's verification pass; the source repositories of the drawn issues carry their own licences, listed in \texttt{PROVENANCE.md}. The five systems components of S3, their withheld suites and their mutant sets were authored for this benchmark by the author's assistant heads under the author's direction, which is the independence limitation stated in Section~\ref{sec:s3}. The runs were executed by Claude Code with Anthropic's Sonnet~5 and Opus~5 models, and, for the pilot matrix's Gemini lane, by the harness's second client lane with Google's Gemini 3.1 Pro and Gemini 3.8 Flash models. The protocol documents, amendments and result files in the repository were written by the author's assistant heads under the author's direction and carry the author's responsibility.


\appendix

\section{S2-Lean U15, per problem}

\begin{center}
\small
\begin{tabular}{r l l l l l l}
\toprule
problem & \arm{a0} S & \arm{a1} S & \arm{a2} S & \arm{a0} O & \arm{a2} O & \arm{a1} O \\
\midrule
73  & F & P & F & P & P & P \\
0   & F & F & F & F & F & F \\
146 & P & P & P & P & P & P \\
16  & P & P & P & P & P & P \\
4   & P & F & P & F & P & F \\
38  & P & P & P & P & P & P \\
142 & P & P & P & P & P & P \\
96  & F & F & F & F & F & F \\
112 & F & F & F & P & F$^\dagger$ & F \\
141 & F & F & F & P & P & P \\
31  & P & P & P & P & P & P \\
54  & P & P & P & P & P & P \\
127 & F & F & F & F & F & F \\
18  & F & F & F & F & F & F \\
74  & P & P & P & P & P & P \\
\midrule
proven & 8 & 8 & 8 & 10 & 10 & 9 \\
\bottomrule
\end{tabular}
\end{center}
S = \texttt{claude-sonnet-5}, O = \texttt{claude-opus-5}, stage B, 100-call cap. $^\dagger$ refused by the pragma screen; a registered re-check over the frozen artifact found a complete proof (diagnostic 11 of 15).\src{SCOUT-S2LEAN-STAGE0.md, amendment 5 result table; AMENDMENT-13-AW-2026-09-01.md section 2; RESULT-amend13-AW-2026-09-02.md section 2; SCOUT-S2LEAN-STAGE0.md, amendment 9 result}

\section{The stage-B failure triage}

\begin{center}
\small
\begin{tabular}{r l p{0.62\textwidth}}
\toprule
problem & label & ground, in one line \\
\midrule
0   & \textsc{human-loose} & conclusion guarded by \texttt{numbers.length > 1}; vacuous on short lists; witness recorded \\
4   & \textsc{human-loose} & guarded by \texttt{0 < numbers.length}; vacuous on \texttt{[]} \\
18  & \textsc{agent-wrong} & agents scan \texttt{List.range (|s| - |t| + 1)}, contradicting a \texttt{\#test}; the human spec is also unsatisfiable \\
73  & \textsc{proof-hard} & equivalent specs; the equivalence needs an optimality argument \\
96  & \textsc{human-loose} & spec fixes membership only, never order or multiplicity; witness \texttt{primes ++ primes} \\
112 & \textsc{proof-hard} & true isomorphism; one sibling cell proved it completely \\
127 & \textsc{human-loose} & guarded by interval well-formedness; vacuous on ill-formed intervals \\
141 & \textsc{proof-hard} & reduces to a \texttt{String.splitOn} equivalence; 7 to 8 kB of lemmas written and still out of rounds \\
\bottomrule
\end{tabular}
\end{center}
Totals over 30 triageable cells: \textsc{human-loose} 17, \textsc{agent-wrong} 5, \textsc{proof-hard} 8; one screen-refused cell excluded.\src{TRIAGE-B-failures-2026-09-01.md sections 2 and 3}

\section{The S2-Rust hard band}

Upper tercile of the 180 live tasks by reference-proof wall, rank $\geq 120$; band $n=60$, Anvil-Advanced 42 and NRKernel 18; band wall median 4.55\,s, max 358.60\,s; band bytes median 165,568. Draw of 13 at seed 20260902: 11 Anvil-Advanced, 2 NRKernel. Ten run under the sequential stop: 9 \texttt{PASS}, 1 \texttt{VERIFY\_FAIL} at the cap; sighted episodes reach the first clean referee run at a median of 2 referee calls (p90 9).\src{evidence/amend16-regime-boundary-2026-09-02/p0-upper-tercile-registration.txt; RESULT-amend16-driver-2026-09-02.md section 6; evidence/amend16-regime-boundary-2026-09-02/p0-metered-at-first-rc0.txt} The 9 is a lower bound on the band's pass count at this cap: the failed episode ended at 40 of 40 calls one obligation short (8 verified, 1 error), the uncensored passing call counts run 23 to 39 against the cap of 40 with p90 = 39, and one pass landed at exactly 40 of 40.\src{RESULT-amend16-driver-2026-09-02.md section 6.1}

\section{The systems matrix, per cell}
\label{app:cells}

Every metered price in the declared set of Section~\ref{sec:s3}, by condition, as the scorer reports them. The scoreboard prints a premium and not the two medians behind it, because the instrument that computes the two readings prints per-cell prices and the premium and does not print the median; this table is what a reader needs to recover any median in that scoreboard. Prices are US dollars of metered subscription spend per cell, each read from that cell's harvested \texttt{METER.txt}.

Reading~A is every row as printed. Reading~B is every row with the \underline{underlined} cell removed: those three are the borrowed smoke cells, one each on the three problems the matrix root could not take to $n=3$ on its own. Three conditions therefore stand at $n=4$ under reading~A and at $n=3$ under reading~B, and the rest are $n=3$ under both. At $n=4$ the median is the mean of the middle pair and is not any cell's price, which is why the three premiums that move between the readings are exactly the three conditions listed with four cells.\src{harness/systems-v3/RESULT-n3-topup-2026-09-09.md section 3, the instrument's stdout for both readings}

\begin{center}
\begin{tabular}{lll}
\toprule
problem & arm & cells \\
\midrule
Crc32 & plain-bare & \$5.36 \$6.21$^\dagger$ \$7.64 \\
Crc32 & diet-bare & \$6.19$^\dagger$ \$7.21$^\dagger$ \$7.63 \\
FreeList & plain-bare & \$13.02$^\dagger$ \$23.38 \underline{\$13.77}$^\dagger$ \$13.01$^\dagger$ \\
FreeList & diet-bare & \$37.95 \$37.60 \$35.41 \\
LRU & plain-bare & \$7.75$^\dagger$ \$9.73$^\dagger$ \underline{\$6.68} \$9.87 \\
LRU & diet-bare & \$11.19 \$11.21 \$14.63 \\
LZW & plain-bare & \$8.15 \$20.95$^\dagger$ \$13.95 \\
LZW & plain-statement & \$10.24 \$11.39 \$9.82$^\dagger$ \\
LZW & diet-bare & \$31.70 \$19.18 \$18.16 \\
LZW & diet-statement & \$22.53 \$15.34$^\dagger$ \$18.54 \\
Paxos & plain-bare & \$9.49$^\dagger$ \underline{\$14.20}$^\dagger$ \$16.78$^\dagger$ \$20.16 \\
Paxos & diet-bare & \$37.93$^\dagger$ \$37.65 \$23.52 \\
\midrule
Crc32 & plain-statement & \$6.36 \$6.66 \$8.64$^\dagger$ \\
Crc32 & diet-statement & \$5.65 \$6.25 \$8.56$^\dagger$ \\
FreeList & plain-statement & \$17.72$^\dagger$ \$18.26 \$18.42 \\
FreeList & diet-statement & \$21.18 \$27.84 \$29.00 \\
LRU & plain-statement & \$8.92 \$9.36 \$15.12 \\
LRU & diet-statement & \$13.65 \$14.24 \$18.65$^\dagger$ \\
\bottomrule
\end{tabular}
\end{center}

\noindent Four cells of the matrix root are refused by the scorer rather than defaulted, because their cost line does not open with a bare price: a void meter carries a zero in its prose, and a scanner that reads on would price an unmetered cell at nothing. They are excluded from every reading.\src{harness/systems-v3/RESULT-n3-topup-2026-09-09.md section 3, the refused cells}

Nineteen of the 57 prices in this table, marked $^\dagger$, come from cells whose own meter declares the total a lower bound: each contains at least one interrupted turn, a record with no stop reason, whose usage was written only up to the interrupt, so what is recorded of that turn is known and what is missing from it is not. Two methods classify every price, the meter's own declaration and the count of such records in the cell's record table, and they agree on 57 of 57. The recorded cost of the interrupted records is 0.0583 to 0.7579 percent of each affected cell's cost; that figure is a share of what was recorded and not a bound on what was not. A lower bound on a \arm{plain} cell overstates a diet-to-plain premium and one on a \arm{diet} cell understates it, and only a lower-bound cell that a median rests on can move a premium: on \texttt{Crc32} both medians rest on one, so the direction is not determined; on \texttt{FreeList}, \texttt{LRU} and \texttt{Paxos} the plain median alone does, so those premiums are overstated by this cause; on \texttt{LZW} neither does; and in the statement arm no median rests on one. No sign, floor count or verdict of Section~\ref{sec:s3} changes unless an interrupted turn's unrecorded remainder is at least 77 times what was recorded of it, and at least 121 times for the bare matrix. The prices are printed as recorded; this note adds a flag and not a number, and version~1 of this report printed the same prices without it.\src{paper/ERRATUM-lower-bound-cells-2026-09-16.md sections 2 to 5; harness/systems-v3/RECEIPT-ol-lower-bound-cells-2026-09-16.txt; harness/systems-v3/ol\_lower\_bound\_margins.py}

\section{Where an Opus cell's tokens and dollars go}
\label{app:opussplit}

\begin{center}\begin{minipage}{\textwidth}\centering
{\scriptsize\setlength{\tabcolsep}{4pt}
\begin{tabular}{llllrrr}
\toprule
problem & task form & arm & treatment & $n$ & share of tokens & share of dollars \\
\midrule
\texttt{Crc32} & brownfield & \arm{plain} & none & 3 & 20.3 \% & 24.8 \% \\
 & brownfield & \arm{plain} & statement & 3 & 22.2 \% & 26.1 \% \\
 & brownfield & \arm{salt-diet} & none & 3 & 8.7 \% & 5.1 \% \\
 & brownfield & \arm{salt-diet} & statement & 3 & 5.8 \% & 5.0 \% \\
 & greenfield & \arm{plain} & none & 3 & 25.3 \% & 31.3 \% \\
 & greenfield & \arm{plain} & spec-change & 3 & 25.3 \% & 31.3 \% \\
 & greenfield & \arm{plain} & statement & 3 & 15.5 \% & 22.3 \% \\
 & greenfield & \arm{salt-diet} & none & 3 & 9.4 \% & 5.6 \% \\
 & greenfield & \arm{salt-diet} & spec-change & 3 & 9.4 \% & 5.6 \% \\
 & greenfield & \arm{salt-diet} & statement & 3 & 5.6 \% & 4.9 \% \\
\texttt{FreeList} & brownfield & \arm{plain} & none & 3 & 32.7 \% & 40.0 \% \\
 & brownfield & \arm{plain} & statement & 3 & 29.0 \% & 38.1 \% \\
 & brownfield & \arm{salt-diet} & none & 3 & 5.5 \% & 13.9 \% \\
 & brownfield & \arm{salt-diet} & statement & 3 & 5.6 \% & 9.2 \% \\
 & greenfield & \arm{plain} & none & 3 & 24.3 \% & 35.1 \% \\
 & greenfield & \arm{plain} & spec-change & 2 & 24.8 \% & 36.2 \% \\
 & greenfield & \arm{plain} & statement & 3 & 31.1 \% & 38.9 \% \\
 & greenfield & \arm{salt-diet} & none & 3 & 1.4 \% & 6.2 \% \\
 & greenfield & \arm{salt-diet} & spec-change & 1 & 1.4 \% & 6.2 \% \\
 & greenfield & \arm{salt-diet} & statement & 3 & 2.6 \% & 7.7 \% \\
\texttt{LRU} & brownfield & \arm{plain} & none & 3 & 15.6 \% & 24.6 \% \\
 & brownfield & \arm{plain} & statement & 3 & 36.7 \% & 34.3 \% \\
 & brownfield & \arm{salt-diet} & none & 3 & 4.4 \% & 8.7 \% \\
 & brownfield & \arm{salt-diet} & statement & 3 & 5.1 \% & 11.8 \% \\
 & greenfield & \arm{plain} & none & 3 & 27.8 \% & 36.0 \% \\
 & greenfield & \arm{plain} & spec-change & 2 & 30.5 \% & 36.6 \% \\
 & greenfield & \arm{plain} & statement & 3 & 21.9 \% & 31.3 \% \\
 & greenfield & \arm{salt-diet} & none & 3 & 3.0 \% & 4.2 \% \\
 & greenfield & \arm{salt-diet} & spec-change & 3 & 3.0 \% & 4.2 \% \\
 & greenfield & \arm{salt-diet} & statement & 3 & 4.8 \% & 10.0 \% \\
\texttt{LZW} & brownfield & \arm{plain} & none & 3 & 24.9 \% & 36.6 \% \\
 & brownfield & \arm{plain} & statement & 3 & 23.6 \% & 28.8 \% \\
 & brownfield & \arm{salt-diet} & none & 3 & 11.9 \% & 22.6 \% \\
 & brownfield & \arm{salt-diet} & statement & 3 & 6.2 \% & 13.8 \% \\
 & greenfield & \arm{plain} & none & 3 & 22.2 \% & 28.8 \% \\
 & greenfield & \arm{plain} & spec-change & 2 & 21.3 \% & 27.8 \% \\
 & greenfield & \arm{plain} & statement & 3 & 20.4 \% & 34.2 \% \\
 & greenfield & \arm{salt-diet} & none & 3 & 2.1 \% & 5.7 \% \\
 & greenfield & \arm{salt-diet} & spec-change & 3 & 3.7 \% & 8.6 \% \\
 & greenfield & \arm{salt-diet} & statement & 3 & 7.3 \% & 13.5 \% \\
\texttt{Paxos} & brownfield & \arm{plain} & none & 3 & 21.5 \% & 34.1 \% \\
 & brownfield & \arm{salt-diet} & none & 3 & 2.2 \% & 6.4 \% \\
 & greenfield & \arm{plain} & none & 3 & 26.4 \% & 30.9 \% \\
 & greenfield & \arm{plain} & spec-change & 2 & 23.1 \% & 29.4 \% \\
 & greenfield & \arm{salt-diet} & none & 3 & 5.6 \% & 6.8 \% \\
 & greenfield & \arm{salt-diet} & spec-change & 2 & 5.7 \% & 8.8 \% \\
\bottomrule
\end{tabular}}\\[3pt]
\parbox{0.92\textwidth}{\small\textbf{Table S1.} Per Opus condition, the median over its cells of the share of the cell's tokens and of its modelled list-price dollars spent outside the head session, by subagents the session served \texttt{claude-sonnet-5}. \emph{none} is no treatment. The greenfield specification-change rows share their first phase with the rows under no treatment, which is why several repeat them.}
\src{harness/systems-v3/RESULT-cost-tables-v3-2026-09-29.md, section S1; evidence/v3-cost-tables-2026-09-28/opus-split-raw.tsv}
\end{minipage}\end{center}


\begin{thebibliography}{99}

\bibitem{claudecode}
Anthropic.
\newblock \emph{Claude Code}, version 2.1.251.
\newblock \url{https://code.claude.com/docs/en/overview}.

\bibitem{anthropic2025terms}
Anthropic.
\newblock \emph{Consumer Terms of Service}, effective October~8, 2025.
\newblock \url{https://www.anthropic.com/legal/consumer-terms}.

\bibitem{anthropic2025usage}
Anthropic.
\newblock \emph{Usage Policy}, effective September~15, 2025.
\newblock \url{https://www.anthropic.com/legal/aup}.

\bibitem{demoura2008z3}
Leonardo de~Moura and Nikolaj Bj{\o}rner.
\newblock Z3: An Efficient SMT Solver.
\newblock In \emph{Tools and Algorithms for the Construction and Analysis of Systems (TACAS 2008)}, pages 337--340, 2008.
\newblock \url{https://doi.org/10.1007/978-3-540-78800-3_24}.

\bibitem{demoura2021lean4}
Leonardo de~Moura and Sebastian Ullrich.
\newblock The Lean 4 Theorem Prover and Programming Language.
\newblock In \emph{Automated Deduction --- CADE 28}, pages 625--635, 2021.
\newblock \url{https://doi.org/10.1007/978-3-030-79876-5_37}.

\bibitem{harbor2026}
Harbor Framework Team.
\newblock \emph{Harbor: A framework for evaluating and optimizing agents and models in container environments}.
\newblock Zenodo, 2026.
\newblock \url{https://doi.org/10.5281/zenodo.20953922}.

\bibitem{hickey2026authority}
Jason Hickey.
\newblock AI with Authority, from Application to Silicon.
\newblock arXiv:2608.21356, 2026. \url{https://arxiv.org/abs/2608.21356}.

\bibitem{hickey2026saltmethod}
Jason Hickey.
\newblock The Salt method: canonical definition.
\newblock \texttt{docs/SALT-METHOD.md} in \url{https://github.com/jyh/salt}, commit \texttt{a8eac8bc}, 2026. Apache-2.0.

\bibitem{jimenez2024swebench}
Carlos~E. Jimenez, John Yang, Alexander Wettig, Shunyu Yao, Kexin Pei, Ofir Press, and Karthik Narasimhan.
\newblock SWE-bench: Can Language Models Resolve Real-World GitHub Issues?
\newblock In \emph{International Conference on Learning Representations (ICLR)}, 2024.
\newblock arXiv:2310.06770. \url{https://arxiv.org/abs/2310.06770}.

\bibitem{kaashoek2026xv6}
M.~Frans Kaashoek and Nickolai Zeldovich.
\newblock Extending concurrent separation logic to the hardware level to verify the xv6 OS kernel on RISC-V with AI agents.
\newblock arXiv:2609.04043v2, 2026. \url{https://arxiv.org/abs/2609.04043}.

\bibitem{lattuada2023verus}
Andrea Lattuada, Travis Hance, Chanhee Cho, Matthias Brun, Isitha Subasinghe, Yi Zhou, Jon Howell, Bryan Parno, and Chris Hawblitzel.
\newblock Verus: Verifying Rust Programs using Linear Ghost Types.
\newblock \emph{Proceedings of the ACM on Programming Languages}, 7(OOPSLA1):286--315, 2023.
\newblock \url{https://doi.org/10.1145/3586037}.

\bibitem{mathlib2020}
The mathlib Community.
\newblock The Lean mathematical library.
\newblock In \emph{Proceedings of the 9th ACM SIGPLAN International Conference on Certified Programs and Proofs (CPP 2020)}, pages 367--381, 2020.
\newblock \url{https://doi.org/10.1145/3372885.3373824}.

\bibitem{nosek2018preregistration}
Brian~A. Nosek, Charles~R. Ebersole, Alexander~C. DeHaven, and David~T. Mellor.
\newblock The preregistration revolution.
\newblock \emph{Proceedings of the National Academy of Sciences}, 115(11):2600--2606, 2018.
\newblock \url{https://doi.org/10.1073/pnas.1708274114}.

\bibitem{swebenchverified}
Princeton NLP.
\newblock \emph{SWE-bench Verified}.
\newblock Hugging Face dataset, split \texttt{test}, 500 rows; revision \texttt{c104f840} as pinned here. The dataset card carries no licence field.
\newblock \url{https://huggingface.co/datasets/princeton-nlp/SWE-bench_Verified}.

\bibitem{sosso2026agentic}
Alessandro Sosso, Akhil Arora, and Bas Spitters.
\newblock Agentic Proving for Program Verification.
\newblock arXiv:2605.23772, 2026. \url{https://arxiv.org/abs/2605.23772}.

\bibitem{sun2024anvil}
Xudong Sun, Wenjie Ma, Jiawei~Tyler Gu, Zicheng Ma, Tej Chajed, Jon Howell, Andrea Lattuada, Oded Padon, Lalith Suresh, Adriana Szekeres, and Tianyin Xu.
\newblock Anvil: Verifying Liveness of Cluster Management Controllers.
\newblock In \emph{Proceedings of the 18th USENIX Symposium on Operating Systems Design and Implementation (OSDI 2024)}, pages 649--666. USENIX Association, 2024.
\newblock \url{https://www.usenix.org/conference/osdi24/presentation/sun-xudong}.

\bibitem{thakur2025clever}
Amitayush Thakur, Jasper Lee, George Tsoukalas, Meghana Sistla, Matthew Zhao, Stefan Zetzsche, Greg Durrett, Yisong Yue, and Swarat Chaudhuri.
\newblock CLEVER: A Curated Benchmark for Formally Verified Code Generation.
\newblock arXiv:2505.13938, 2025. \url{https://arxiv.org/abs/2505.13938}.

\bibitem{yang2025verusage}
Chenyuan Yang, Natalie Neamtu, Chris Hawblitzel, Jacob~R. Lorch, and Shan Lu.
\newblock VeruSAGE: A Study of Agent-Based Verification for Rust Systems.
\newblock arXiv:2512.18436, 2025. \url{https://arxiv.org/abs/2512.18436}.

\end{thebibliography}
\end{document}